# The Mu2e Calorimeter Final Technical Design Report

N.Atanov[a], V. Baranov[a], J. Budagov[a], S.Ceravolo[b], F. Cervelli[e], F. Colao[b], M. Cordelli [b],
G. Corradi[b], E. Dané[b], Yu.I. Davydov[a], S. Di Falco[e], S. Donati[e;g], E. Diociaiuti[b;j], R. Donghia[b;k],
B. Echenard[c], K. Flood[c], S. Giovannella[b], V. Glagolev[a], F. Grancagnolo[l], F. Happacher[b],
D.G. Hitlin[c], M. Martini[b;d], S. Miscetti[b], T. Miyashita[c], L. Morescalchi[e;f], P. Murat[h], D. Pasciuto[e],
G. Pezzullo[e], F. Porter [c], T. Radicioni[e], F. Raffaelli[e], M. Ricci[b;d], A. Saputi[b], I. Sarra[b],
F.Spinella[e], D. Tagnani[b;k],G. Tassielli[l], V. Tereshchenko[a], Z. Usubov[a], R.Y. Zhu[c]

a. Joint Institute for Nuclear Research, Dubna, Russia
b Laboratori Nazionali di Frascati dell'INFN, Frascati, Italy
c California Institute of Technology, Pasadena, United States
d Università "Guglielmo Marconi", Roma, Italy
e INFN Sezione di Pisa, Pisa, Italy
f Dipartimento di Fisica dell'Università di Siena, Siena, Italy
g Dipartimento di Fisica dell'Università di Pisa, Pisa, Italy
h Fermi National Laboratory, Batavia, Illinois, USA
i INFN Sezione di Lecce, Lecce, Italy
j Dipartimento di Fisica dell'Università di Roma Tor Vergata, Rome, Italy
k Dipartimento di Fisica dell'Università degli Studi Roma Tre, Rome, Italy

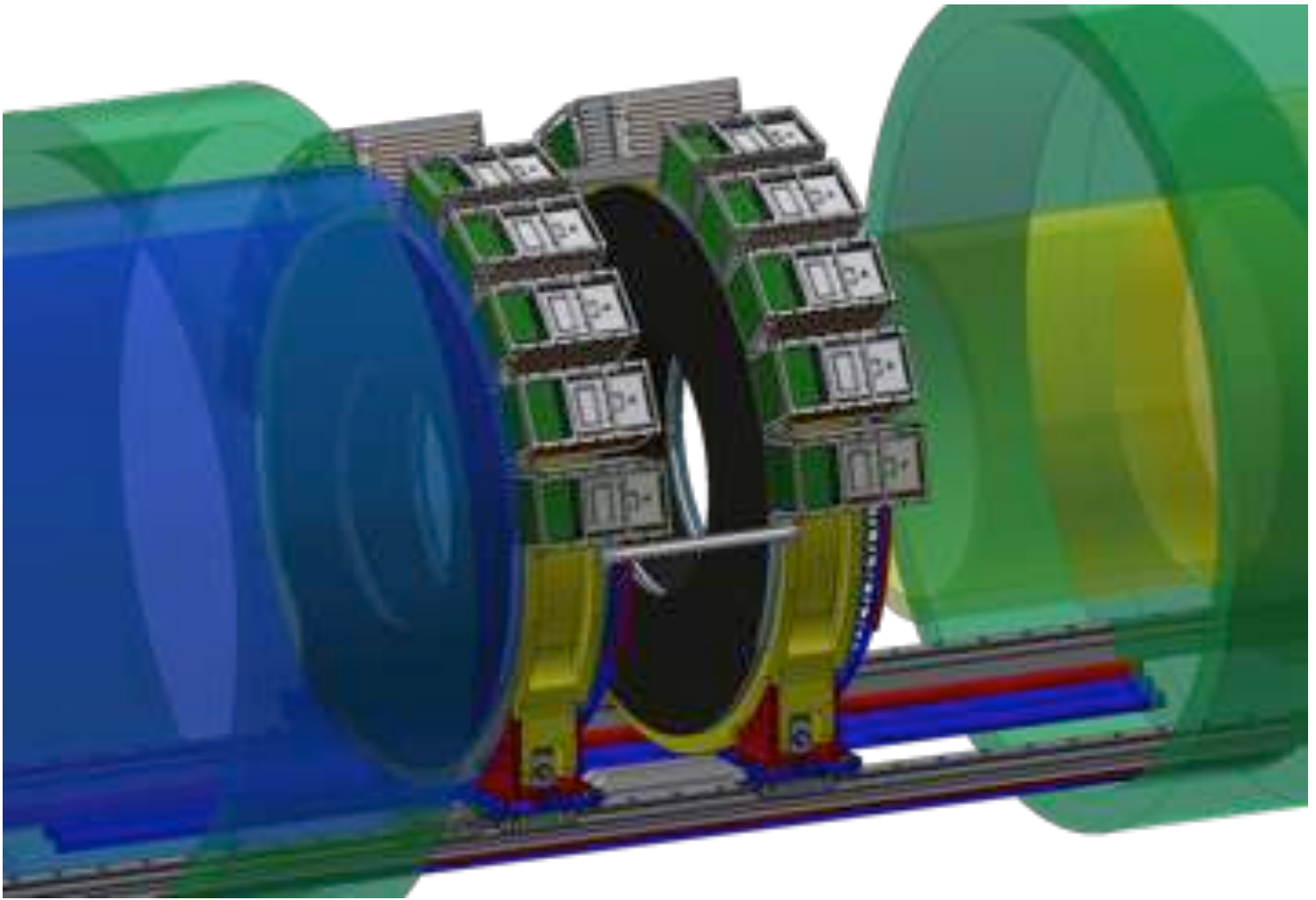

# 1  The Mu2e calorimeter concept and design

The design of the Mu2e detector [1-1] is driven by the need to reject backgrounds to a level consistent with single event sensitivity for $\mu \rightarrow e$ conversion of the order of $3\times10^{-17}$. The calorimeter system is a vital link in the chain of background rejection, arising from particles reconstructed as signal electron by the tracker system only. Such background could be produced by cosmic ray muons not vetoed by the CRV system producing a track that looks coming from the target with a momentum in the 105 MeV acceptance window. The primary purpose of the Mu2e calorimeter is to provide a second set of measurements that complement the information of the tracker (e.g. particle identification) and enable us to reject these backgrounds. A calorimeter with a coarse energy resolution and good time resolution can achieve such task.

## 1.1 Requirements

The requirements for the calorimeter have been documented by the Mu2e collaboration [1-2]. The primary functions are to provide energy, position and timing information to confirm that events reconstructed by the tracker system are indeed conversion electrons while performing a powerful $\mu/e$ particle identification. In addition, it should provide a seed for the track search and a high level trigger independent of the tracker system. As detailed in the subsequent sections, this leads to the following requirements for 100 MeV electrons:

- An energy resolution of O (5%).
- A timing resolution better than ~ 0.5 ns.
- A position resolution better than 1 cm to allow comparison of the position of the energy deposit to the extrapolated trajectory of a reconstructed track.
- The calorimeter must operate in the evacuated bore of the Detector solenoid (DS) at $10^{-4}$ Torr and stand the unique, high-rate Mu2e environment.
- The calorimeter must maintain its functionality for radiation exposures up to 15 Gy/crystal/year and for a neutron flux equivalent to $10^{11}$ n$_{1MeV\ eq}$ /cm$^2$/year.

The resolutions needed are indeed achievable in the energy range of interest. The energy resolution of a crystal calorimeter complements, but is not competitive with, that of the tracking detector. The Mu2e calorimeter group, [1-3], as well as other experiments operating in a similar energy regime [1-4], has achieved an energy resolution of 5% at 100 MeV.

For real tracks, activity in the tracker and in the calorimeter will be correlated in time. The time resolution of the calorimeter should be better than the time resolution of extrapolated tracks from the tracker, estimated to be of ~1 nanosecond.  A calorimeter timing resolution of 0.5 ns, when combined in quadrature with the tracker resolution, gives a negligible contribution to the total timing resolution, and can be easily achieved. A good timing will help resolving pileup issues related to the high background rate present. The calorimeter timing information can be used by the cluster reconstruction algorithm in several ways. For the cluster reconstruction itself, good time resolution helps in the connection/rejection of cells to the cluster and in the cluster merging.

In the 100 MeV energy regime, a total absorption calorimeter employing a homogeneous continuous medium can meet the resolution requirement. This could be either a liquid such as xenon, or a scintillating crystal; we have chosen the latter. Three types of crystals have been considered in detail for the Mu2e calorimeter: lutetium-yttrium oxyorthosilicate (LYSO), barium fluoride (BaF$_2$) and un-doped Cesium Iodine (CsI). The final selected design for the Mu2e calorimeter uses an array of CsI crystals arranged in two annular disks, as shown in Fig. 1-1.left. Electrons following helical trajectories spiral into the front faces of the crystals, as shown in Fig.1-1.right. The crystals are of parallelepiped shape, with a transversal dimension of 34x34 mm$^2$ and a length of 200 mm; there are a total of 1348 crystals. Each crystal is read out by two large-area silicon photomultipliers (SiPM); solid-state photo-detectors are required because the calorimeter resides in a 1 T magnetic field. Front-end electronics is mounted on the rear of each disk, while voltage distribution, slow controls and digitizer electronics are mounted in crates located on top of each disk. A laser flasher system provides light to each crystal for relative calibration and monitoring purposes. A circulating liquid radioactive source system provides absolute calibration and an energy scale. The crystals are supported by a lightweight carbon fiber support structure in the inner region and by an aluminum support in the outer region. Each of these components is discussed in the other sections. In the next subsections, we will explain slightly more the physics requirements.

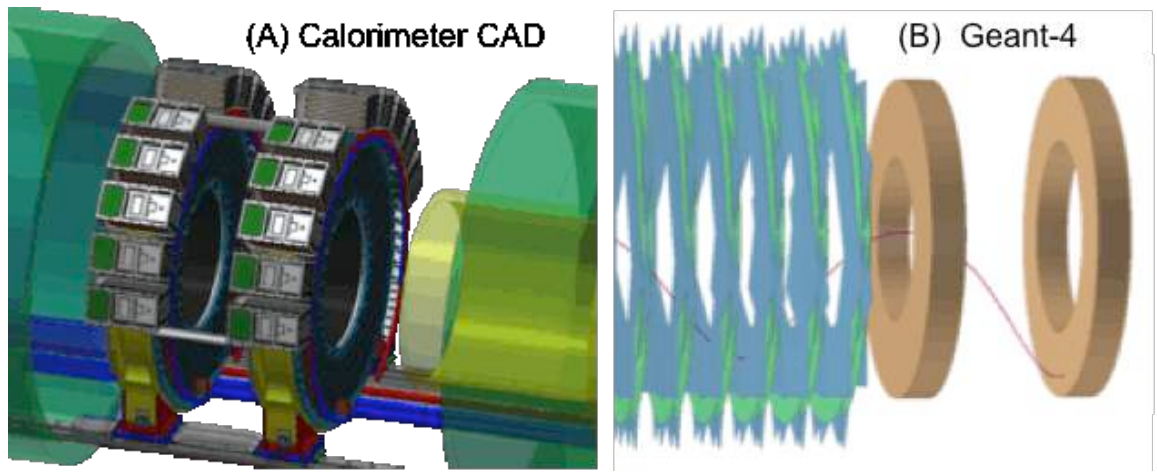

Figure 1-1 *The Mu2e calorimeter consists of an array of CsI crystals arranged in two annular disks (left). Electrons spiral into the upstream faces (right).*

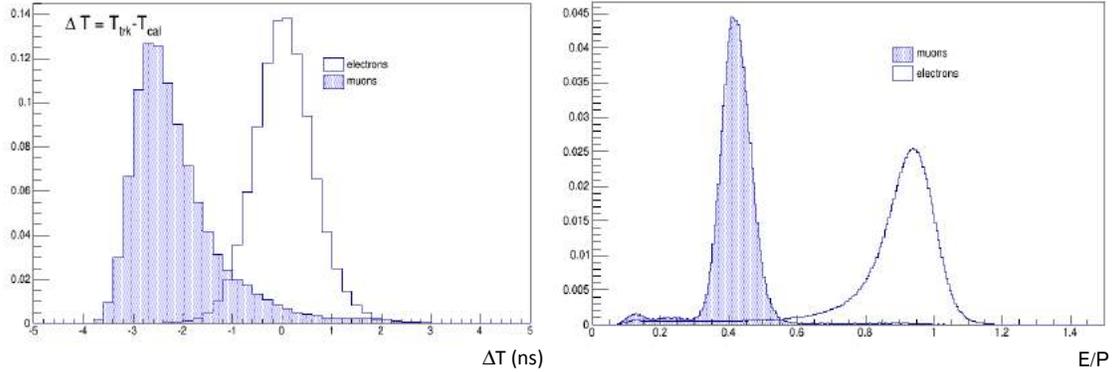

Figure 1-2 *Particle ID variables: (left) Time differences between extrapolated track and arrival time on the calorimeter, (right) E/P.*

## 1.2 Particle Identification and Muon Rejection

Cosmic rays generate two distinct categories of background events: muons trapped in the magnetic field of the Detector Solenoid and electrons produced in a cosmic muon interaction with detector material. According to recent studies of the cosmic background, after 3 years of data taking one could expect to get O(1) events in which negative cosmic muons with $103.5 < P < 105$ MeV/c enter the detector bypassing the CRV counters and surviving all tracking analysis cuts. To keep the total background from cosmics at a level below 0.2 events, a muon rejection of 200 is required to make these fakes fully negligible. Timing and dE/dx information from the Mu2e tracker allows for limited PID capabilities [1-5]. However for a muon rejection factor of 200, the efficiency of the electron identification based on the tracker-only information could be 50% or even below. The addition of energy and timing measurements from the Mu2e calorimeter (see Fig. 1.2) provide critical information to efficiently separate electrons and muons in the detector. The calorimeter acceptance has been optimized such that ~ 92 % of conversion electron (CE) events with tracks passing "Set C" quality cuts have a calorimeter cluster with E > 10 MeV produced by the conversion electron. A reconstructed CE candidate event is therefore required to have a calorimeter cluster, pointed to by the track. A track-cluster matching $\chi_{match}^2 = (\Delta U/\sigma_U)^2 + (\Delta V/\sigma_V)^2 + (\Delta T/\sigma_T)^2$ is defined, where $\Delta U$ and $\Delta V$ are the track-to-cluster coordinate residuals in directions parallel and orthogonal to the track, and $\Delta T$ is the difference between the track time extrapolated to the calorimeter and the reconstructed cluster time. The estimated resolutions are $\sigma_U = 1.5$ cm, $\sigma_T = 0.8$ cm, and $\sigma_T = 0.5$ ns. For the background occupancy level expected in the Mu2e detector during the data taking, a requirement $\chi_{match}^2 < 100$ is 98% efficient for the expected CE signal. Events are also required to be consistent with the electron hypothesis such that they have

| $\Delta T$ | < 3 ns and E(cluster)/P(track)<1.15. After the cleanup cuts, the log likelihoods of the electron and muon hypotheses are defined: $\ln L_{e,m} = \ln P_{e,m}(\Delta t) + \ln P_{e,m}(E/P)$, where $P_{e,m}(\Delta t)$ and $P_{e,m}(E/P)$ are $\Delta t$ and E/P probability density distributions for electrons and muons correspondingly. These distributions are shown in [1-6]. A ratio of the likelihoods of the two hypotheses $\ln (L_e/L_m) = \ln L_e - \ln L_m$ determines the most likely particle mass assignment. Figure 1.3 (left) shows the muon rejection factor plotted vs the CE identification efficiency for different background levels: CE only, CE plus nominal expected background, CE plus two times the expected background. For the nominal background expectation and muon rejection factor of 200, the electron identification efficiency is (96.5 ± 0.1)%. This number includes the geometrical acceptance and efficiency of all cuts and demonstrates a high efficiency of the PID procedure. Figure 1.3 (right) shows dependence of the electron identification efficiency for different values of the calorimeter energy and time resolution in the range $0.02 < \sigma_E/E < 0.2$ and $0.05 < \sigma_T < 1$ ns. The value of the muon rejection factor is fixed at 200. One can see that in the expected operational range, $\sigma_E/E < 0.1$ and $\sigma_T < 0.5$ ns, the PID is robust with respect to the calorimeter resolution, with the electron identification efficiency variations below 2% in this region of parameter space.

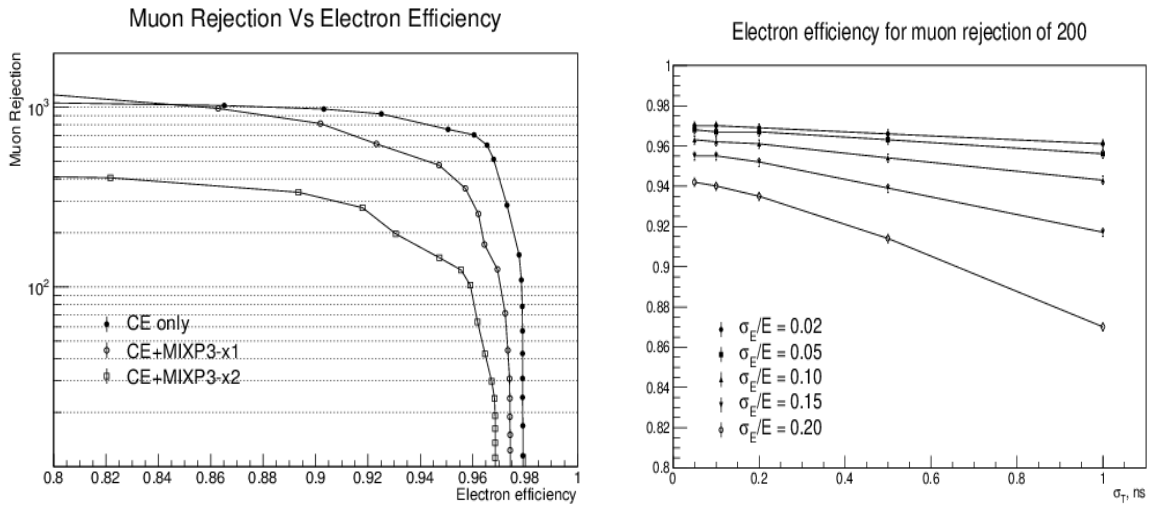

Figure 1-3 *(left) PID efficiency for CE vs. muon rejection for different background levels: no background, expected background, twice the expected background; (right) PID efficiency for CE for muon rejection factor of 200 and different assumptions about the calorimeter energy and timing resolution.*

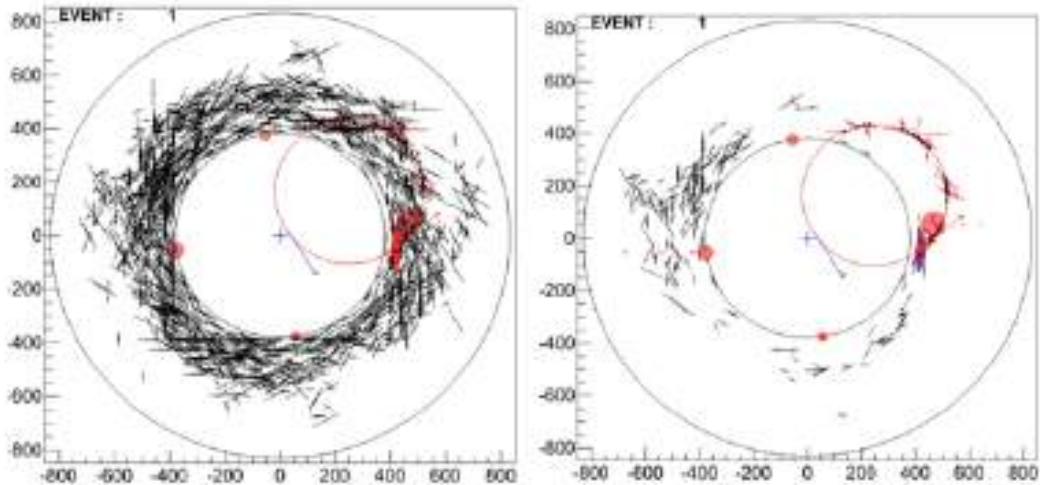

Figure 1-4  *Distribution of the hits in the tracker before (left) and after (right) the application of a timing window based on timing information in the calorimeter. The situation for the pattern recognition is dramatically improved*

## 1.3 CalTrk: seeding of tracking by calorimeter hits

In addition to improved background rejection, the calorimeter provides a robust approach to track reconstruction. Mu2e does not have an "event time"; all straw hits reconstructed within a micro-bunch therefore have to be considered by the track-finding algorithm and the track time is reconstructed as a track fit parameter. The standalone Mu2e track reconstruction attempts to find the 100 ns time slice within the microbunch with the maximum number of hits in it, and uses those hits to find a track. In the presence of the correlated in-time background produced by δ-electrons, such an approach relies strongly on the δ-electron hits being identified and excluded before execution of the track reconstruction, which at present uses a neural network-based procedure. A cluster produced by a track and reconstructed in the calorimeter can be used as a seed for the track finding. In Fig. 1.4, the distribution of the hits in the tracker before the application of a timing window of ± 40 ns on  the difference between the straw hit times and the calorimeter cluster time, $\Delta t_{HITS}$,  is shown. This simple cut cleans up the hits not related to real track and increase substantially the signal/noise ration, thus allowing to reconstruct tracks that are missed by the standalone tracking algorithm. This  calorimeter-driven track finding improves also the overall track finding efficiency. More details on the implementation of this in the final detector can be found in sec. 2.3  and in  [1-7].

## 1.4 Use of the Calorimeter for triggering

The calorimeter system can also generate a fast trigger for the experiment that is independent of the tracker. This trigger will take the form of an offline HLT/L3-like filter that can be used after streaming the events to the online computing farm, but before storing data on disk. The DAQ will read events from the tracking and calorimeter digitizers at a maximum throughput of 20 GByte/sec and the online farm will be able to fully reconstruct nearly all the streamed data. The calorimeter filter should be able to process the data in the online farm with the requirement of rejecting the background by a factor > 200. The most important aspect of this filter is that it is fully independent of the tracker, with completely different systematics due to environmental backgrounds. First studies show that, at the required level of rejection, the standalone calorimeter trigger will reach efficiency of 60-70 % that is more than enough to create samples for an unbiased estimate of the tracking trigger efficiency. An additional trigger filter based on tracking and on the application of the CalTrk method is also being developed. A more detailed discussion of simulated performance with the final detector can be found in sec. 2.4.

***Summary of Calorimeter System Parameters***

In order to simplify the reading of the next chapters, we have collected in Table 1 all the [parameter and specific numbers needed for an overview summary of the Mu2e calorimeter system.

Table 1 Summary of calorimeter parameters.

| | |
|---|---|
| Number of Disks | 2 |
| Disk  Inner and Outer Radius | 374 mm, 660 mm |
| Crystal Type, density, $X_0$, $R_M$ | CsI, 4.9 g/cm$^3$, 2.0 cm, 3.0 cm |
| Crystal Shape | Parallelepiped |
| Crystal Length | 200 mm |
| Crystal Transversal Area | 34x34 mm$^2$ |
| Total number of crystals Disk 1+2 | 1348 |
| Single crystal weight | 1.14 Kg |
| Total scintillation mass | 1540 kG |
| Number of SiPM/crystal | 2 |
| SiPM transverse dimension active area | 12x18 mm$^2$ |
| Total number of SiPMs | 2696 |
| Total number of LV/HV boards | 136 |
| Total Number of Digitizers | 136 |
| Total number of preamplifiers | 2696 |
| Power Dissipation AMP-HV | 480 mW x 2696 = 1294 W |
| Power Dissipation LV/HV |  4 W  x 136 =  544 W |
| Power Dissipation Digitizer | 31 W x 136 = 4100 W |
| Distance between disks | 700 mm |

# Chapter 1 References

# 2    Simulation and performances

## 2.1 Geometry optimization

The baseline calorimeter design for Mu2e consists of two annular disks separated by approximately a half-wavelength of the typical conversion electron helical trajectory. This configuration minimizes the number of low-energy particles that intersect the calorimeter, while maintaining excellent signal efficiency. The crystals have parallelepiped shape, with a transversal dimension of 34x34 mm$^2$ and a length of 200 mm, approximately 10 $X_0$. The disk inner and outer radii, the relative separation between the disks as well as the crystal dimensions were chosen to maximize the efficiency while minimizing the dead material inside the disks.

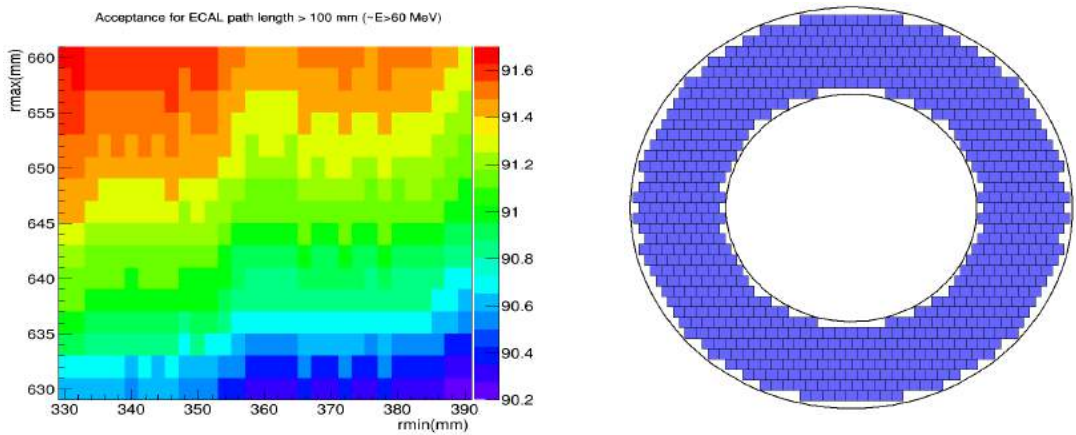

Figure 2.1 *Calorimeter efficiency for detecting a conversion electron as a function of the disk inner and outer radii with simplified algorithms (left); Crystal layout for a crystal size of 34 mm with disk radii of 374 mm and 660 mm (right).*

The dimensions of the disks were the first issue to be addressed. A simplified algorithm to estimate the signal efficiency as a function of the disk dimensions is used to select promising configurations, as shown in Figure 2.1 (left). The separation between the disks is set to 70 cm, corresponding approximately to a half-wavelength. A full simulation is then performed to confirm the results and select the final geometry: a crystal size of 34 mm with disk inner and outer radii of 374 mm and 660 mm, respectively. This solution is robust against small variations of the crystal size. The distance between the disks is then re-evaluated, confirming the value of 70 cm as optimal. The position of the disk with respect to the tracker has a negligible impact on the efficiency, as expected from translational invariance. Finally, the crystal length is studied, and a value of 20 cm, corresponding to approximately 10 $X_0$, is chosen. This geometry ensures sufficient space to mount the readout at the back of the crystals while maintaining efficiency and limiting the number of readout channels. The crystal layout is shown in Figure 2.1 (right).

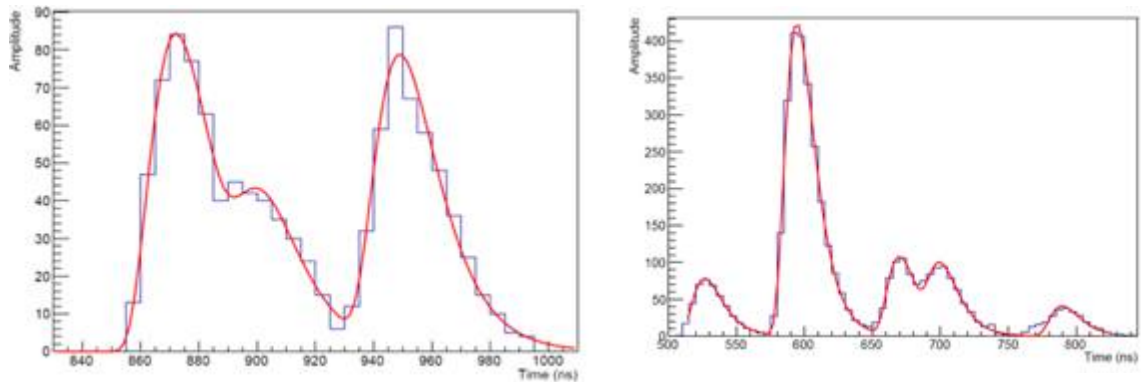

Figure 2.2 *Example of fits to the digitized waveform to extract individual crystal hits.*

## 2.2      Simulated performances

Event reconstruction in the calorimeter proceeds in several stages. The interaction of the incident particle with the crystals is first simulated by GEANT4, recording the energy, position and time of each interaction. Each energy deposit is converted into photons, taking into account corrections from Birks' law and non-uniformities in the crystal longitudinal response. The response of each SiPM is then simulated, including photo-statistics fluctuations (average photo-electron yield of 30 p.e./MeV) and the related electronic noise (150 keV equivalent), to form fully digitized waveforms. The individual hits are extracted by fits to the waveforms, including the possibility of multiple pile-up contributions, to form crystal hits. Examples of fit are shown in Figure 2.2. The pile-up identification significantly improves the resolution compared to simpler algorithms considering only the total signal in a given time window. The association of reconstructed hits to incoming simulated particles is performed at this stage.

These hits are finally used to form calorimeter clusters. The clustering [2.1] algorithm starts by taking the crystal hit with the largest energy as a seed, and adds all simply connected hits within a time window of ± 10 ns and a threshold in energy of 3 times the electronic noise. Hits are defined as connected if they can be reached through a series of adjacent hits. The procedure is repeated until all crystals hits are assigned to clusters. Additional low-energy deposits that are disconnected from the main cluster are recovered by dedicated algorithms. These fragments are usually produced by the shower, or by low-energy photons emitted by incident particles. Recovering these split-off deposits significantly improves the energy resolution.

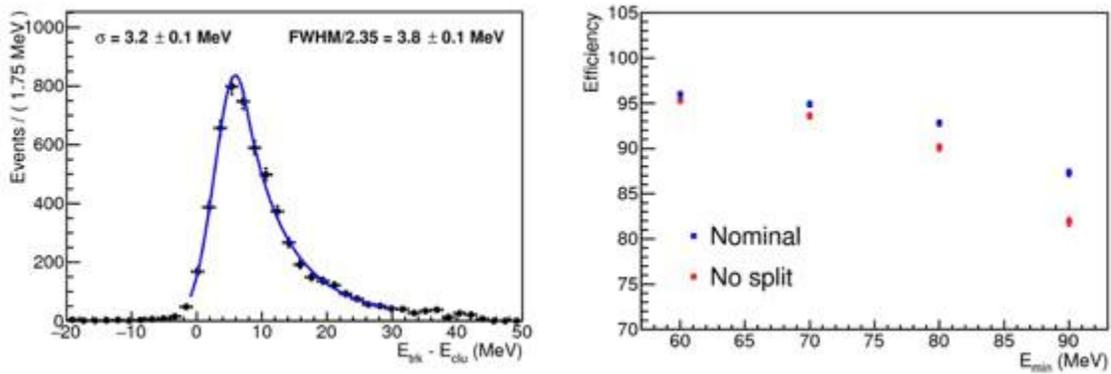

Figure 2.3. *Calorimeter energy resolution, measured as the difference between the conversion electron energy corrected for energy loss in the tracker and the cluster energy. The distribution is fit with a Crystal Ball function (left). Calorimeter efficiency for detection of good signal tracks first found in the tracker, as a function of the energy deposited in the calorimeter, with or without applying the cluster split-off recovery algorithm (right).*

The energy resolution is estimated by simulating conversion electrons distributed at random in the stopping target foils, together with the expected neutron, photon and DIO backgrounds. The distribution of the difference between the true signal electron energy obtained by simulation and the reconstructed cluster energy is plotted in Figure 2.3 (left). This variable accounts for the energy lost by the electron before hitting the calorimeter. The low-side tail is due to background pile-up with the cluster. The distribution is fit with a Crystal Ball function to extract the resolution. A full width at half-maximum (FWHM) of 8.9 ± 0.1 MeV is observed, corresponding to a resolution of FWHM/2.35 = 3.8 ± 0.1 MeV.  The dependence on the light yield and the longitudinal response  non-uniformity (LRU) are shown in Figure 2.4 (left) and (right) respectively. A LRU below 5% and a light yield above 20 to 30 p.e/MeV are sufficient to achieve good resolution. Developments to improve the simulation of the apparatus are planned in the near future.

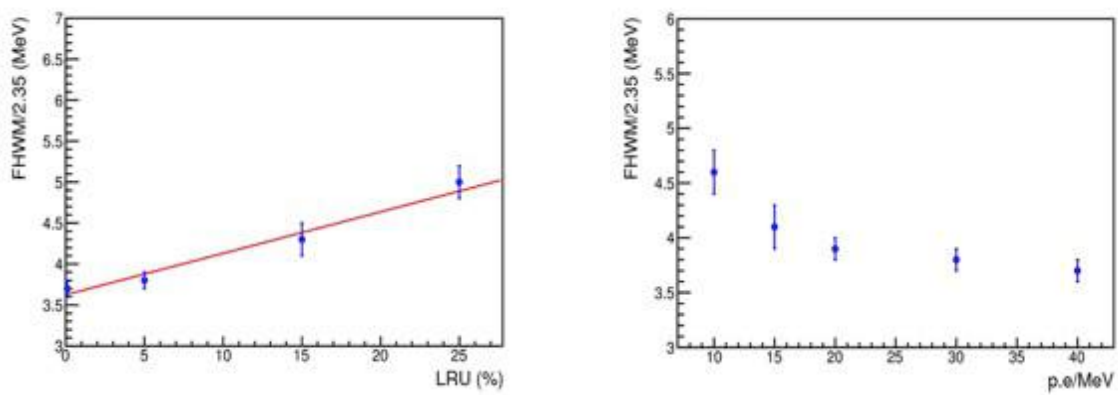

Figure 2.4 *Resolution as a function of the longitudinal response non-uniformity LRU (left) and the light yield in p.e./MeV (right).*

The relative reconstruction efficiency is evaluated by comparing the number of events containing a reconstructed signal track and a cluster above a given energy threshold to the number of events containing a reconstructed signal track. The results are displayed in Figure 2.3 (right), together with the efficiency obtained without applying the split-off recovery algorithm. As expected, the split-off recovery algorithm shows clear gain at high energies. Efficiency around 95% is achieved for cluster energies above 60 MeV. The inefficiency is mostly due to conversion electrons passing through the hole of both disks. The decrease in efficiency at higher cluster energies arise from leakage (conversion electron hitting near the inner edge of the calorimeter) or unassociated split-off deposits.

Determining the track position at the surface of the calorimeter is usually performed by standard algorithm using the calorimeter cluster centroid, the track direction and corrections to account for the non-zero interaction depth in the calorimeter. However, these corrections can be difficult to parametrize, as they depend on multiple factor. To remove these dependences, we adopted a newer approach, based on boosted decision trees. The cluster energy pattern, energies and positions of the 25 crystals centered around the most energetic energy deposit, and well as the track incident angles are used as input variables of the BDT to predict the track position. All correlations between the input parameters are automatically included. The distribution of the difference between the predicted and actual position of the track at the surface of the calorimeter are plotted in Figure 2.5 A coordinate resolution of about 6 mm can be achieved.

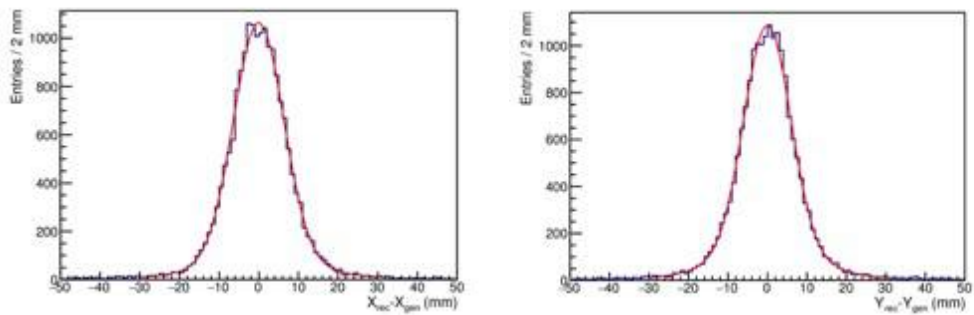

Figure 2.5 *The distribution of the difference between the predicted and actual position of the track at the surface of the calorimeter projected along the x (left) and y (right) coordinates. The distribution are fit with a sum of two Gaussian functions.*

## 2.3 Calorimeter-driven Track Finding

The Mu2e track reconstruction has several specific features. First, a CE makes 2-3 full turns moving inside the tracker, so a track consisting of several loops has to be reconstructed. This topology is very different from the typical tracks reconstructed in most HEP experiments, where "extra" loops are often discarded. In addition, as the muon nuclear capture or a muon decay happen at an unspecified time, the electron production time is also unknown. It is easy to show that the reconstructed track parameters depend on the particle timing. Let's call $T_0$ the time when the particle crosses the middle of the tracker. The $T_0$ is used as input to determine the drift radii of the straw hits. Wrong calculation of the drift radii or a mis-assignment of the drift direction impacts the reconstruction performance. For each track hit, $T_0$ explicitly enters the calculation of the hit drift radius: $r_{\text{drift}} = v_{\text{drift}} \cdot (t_{\text{measured}} - T_0 - t_{\text{flight}})$, where $v_{drift}$ is the drift velocity, and $t_{flight}$ is the particle's time of flight from the middle of the tracker to the corresponding straw. So the track hit coordinates depend on $T_0$, and $T_0$ becomes an additional parameter to be determined from the fit. The Mu2e track reconstruction proceeds in two main steps: (i) first a track search provides a pattern of straw hits consistent with a track candidate and (ii) then a Kalman-based track fitter performs the final reconstruction. The track finding uses two algorithms: a standalone algorithm, and a calorimeter-driven algorithm. The standalone algorithm relies only on the tracker information to perform a helix search. The calorimeter-seeded track search [2.2] is a specialized algorithm, optimized to search for electron tracks generated in the stopping target that produced clusters in the calorimeter. The algorithm uses the reconstructed time and position of the calorimeter cluster to search for a track pointing to that cluster. This procedure improves the track search and makes the global track reconstruction more efficient and robust with respect to the expected background level.

The pattern recognition proceeds in two stages. At the first stage the calorimeter cluster time and position are used to preselect the hits to be considered by the pattern recognition.

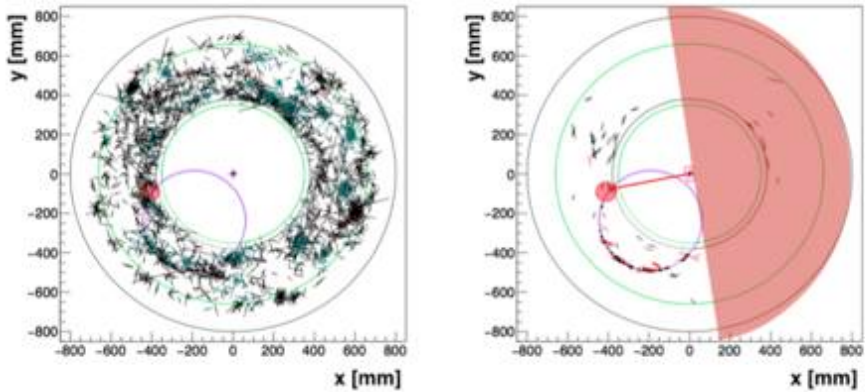

Figure 2.6: *Transverse view of an event display for a CE event with background hits included, with (right) and without (left) application of the calorimeter pre-selection. The black crosses represent the straw hits, the red bullets the calorimeter clusters, and the red circle the CE trajectory.*

The cluster time defines a time window of about 80 ns within a micro-bunch, thus improving the relative fraction of straw hits produced by a CE by a factor of 20. Furthermore, the knowledge of the cluster position allows to improve the background rejection. The graded magnetic field, between the stopping target and the tracker, acts as a lens and focuses the electrons so that CEs within the tracker geometric acceptance have a transverse momentum in the rather narrow range of 75-86 MeV/c. So in the XY plane, hits produced by a CE are contained within a semicircle centered on the cluster. The requirement that the hit azimuthal angle $\phi$ is contained within $\pi/2$ from the cluster: $|\phi_{hit} - \phi_{cluster}| < \pi/2$ reduces the background by an additional factor of two. As introduced in sec.1.3, Fig. 2.6 shows how this preliminary selection acts on a typical event where the CE is overlaid with all the expected backgrounds. At the second stage the calorimeter cluster position is used to direct the search of the hits consistent with the helical particle trajectory in the tracker. A Kalman filter is used to perform a preliminary fit through the straw tube wires and then re-perform the fit using the drift radii calculated combining the calorimeter timing measurement and the trajectory returned by the previous step. A detailed description can be found on [2.3],[2.4].

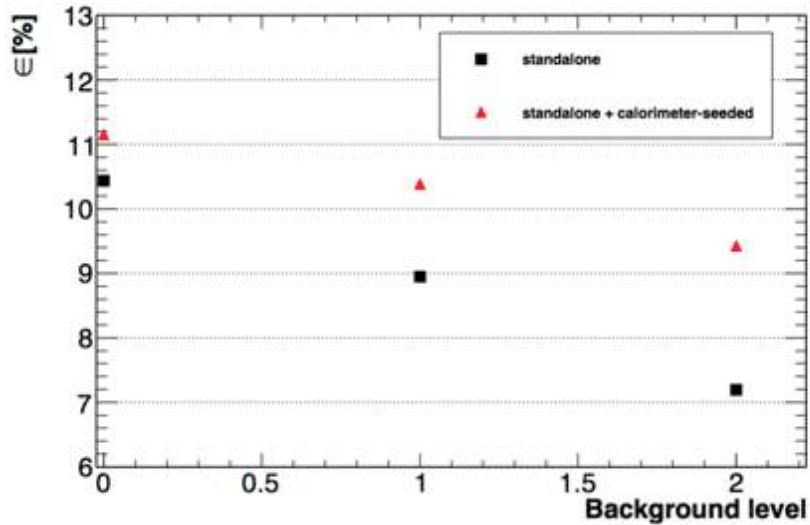

Figure 2.7: *Track reconstruction efficiency versus the background level. Black dots: standalone track reconstruction, red dots: standalone plus calorimeter-seeded track reconstruction.*

Impact of the calorimeter-seeded track search on the track reconstruction efficiency has been evaluated using simulated CE events, with tracks of reconstructed momentum $p >$ 100 MeV/c that pass the track quality cuts. Fig. 2.7 shows the track reconstruction efficiency as a function of the expected background level with and without inclusion of the calorimeter-seeded algorithm in the pattern recognition. Compared to the standalone algorithm, when including the calorimeter-seeded track-finding the number of events with the reconstructed CE tracks increases by about 15%. Moreover, the same figure shows that the calorimeter-seeded pattern recognition makes the track reconstruction more robust in scenarios with higher background levels.

## 2.4 The Calorimeter Triggers

The Mu2e trigger consists of a set of software filters which run on the server farms of the Data Acquisition (DAQ) system and are used to select the events[1] to be written to tape. These filters can be divided in two main categories: *on-spill,* when the proton beam is directed to Mu2e, and *off-spill,* when the proton beam is directed to Nova or is off [2.5].

*On-spill* filters are used to select with the highest possible efficiency the conversion electron events while keeping the global rate below 2 kHz and the total bandwidth below 0.7 GB/s [2.6]. The global trigger is the OR combination of a tracker trigger, a dual tracker-calorimeter trigger and a standalone calorimeter trigger[2]. The global background rejection factor must be better than 100.

*Off-spill* filters are used to select calibration events. In particular, for the calorimeter, there will be a cosmic trigger and a set of tagged triggers used for the calibration source, the laser calibration and the electronics pulses.

Finally, special on-spill triggers will be needed to select the DIO electrons and the electrons coming from the pion decay at rest during dedicated calibration runs at low beam intensity and reduced magnetic field. Their structure will reproduce the one of the standard on-spill triggers with a proper tuning of the parameters.

### Dual calorimeter-tracker trigger

The tracker standalone trigger [2.7] is expected to have excellent performances on the events with conversion electron tracks reconstructed using only the tracker information: a signal efficiency close to 100% and a background rejection factor of ~ 400 for nominal background conditions.

Calorimeter information can be used to simplify the tracker pattern recognition [4], improving the global track reconstruction efficiency and making it less sensitive to the background conditions. The dual calorimeter-tracker trigger uses the calorimeter information to select also the tracks recovered by the calorimeter seeded pattern recognition. In nominal background conditions, its efficiency on all the conversion electron reconstructed tracks is 92%, and the background rejection factor is ~200.

### Calorimeter standalone trigger

The calorimeter standalone trigger [2.8] is completely independent on tracker and can then be used to study the tracker trigger and track reconstruction efficiency.

---

1       A 'Mu2e event' includes the information of all the Mu2e sub-detectors in the 1.7 μs time window between two micro-bunches. This definition holds also in case of spill-off triggers.

2       On-spill filters may also include a fraction of protons and low momentum electrons used for tracker alignment and energy calibration and minimum bias downscaled events.

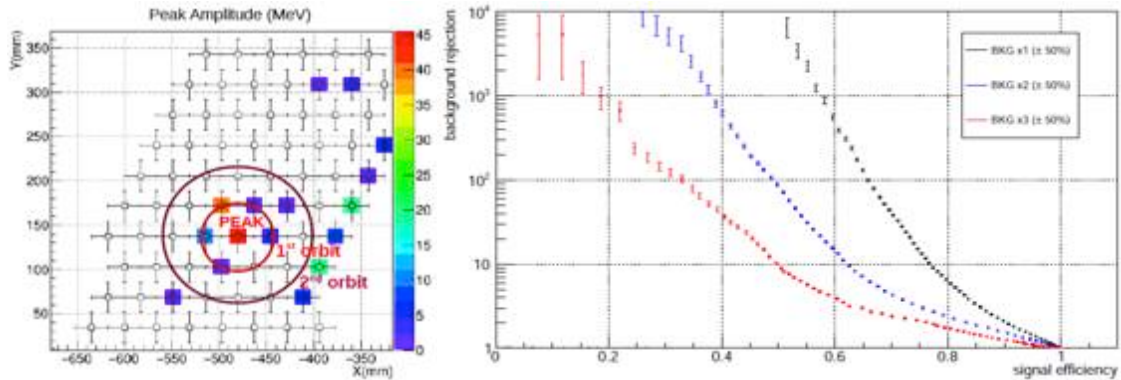

Figure 2.8: (left) Main ingredients of the standalone trigger algorithm: the cluster seed crystal exceeding the energy threshold and the adjacent crystals. (right) BDT classifier conversion electron efficiency, normalized to the reconstructable tracker tracks and background rejection for different background levels. ('bkg x1' means the current Mu2e expected background).

The calorimeter standalone trigger algorithm is based on a multivariate boosted decision tree (BDT) classifier that uses the single calorimeter crystals energy, time and position, looking for conversion electron-like electromagnetic showers around the crystals exceeding an energy threshold (Fig. 2.8.left). The BDT has been trained on a Monte Carlo sample of conversion electrons and background produced for the Mu2e CD3 analysis. The result, shown in Fig. 2.8.right, is an efficiency of ~60% for the wanted background rejection factor of 400. If the background is increased by a factor of 3 the efficiency goes down to 25%.

***Calorimeter cosmic trigger***

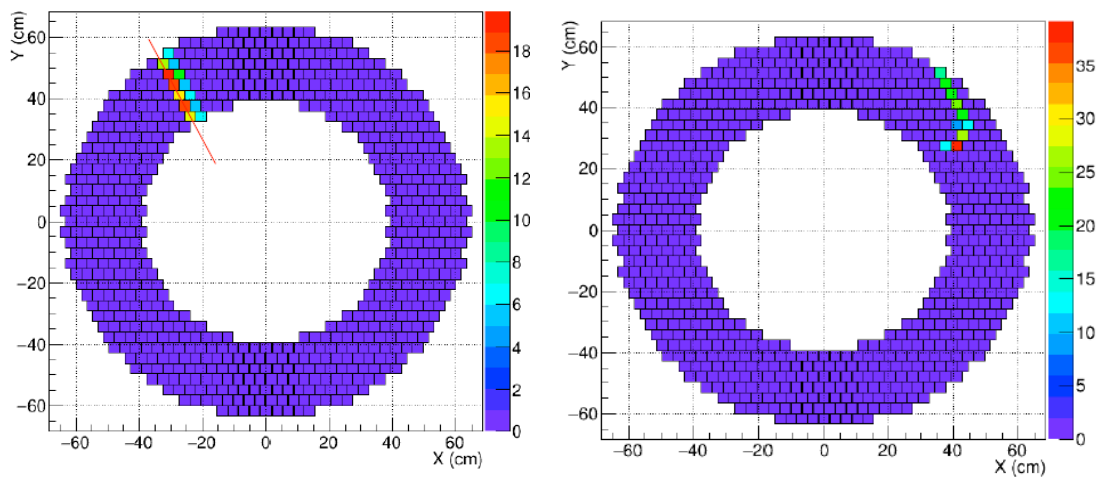

Figure 2.9: Examples of cosmic muon events selected (left) and rejected (right) by the calorimeter cosmic muon trigger. The color scale refers to the reconstructed crystal energy deposit in MeV.

A calorimeter cosmic trigger [2-9] is needed to write on tape the events in which a cosmic minimum ionizing muon crosses the calorimeter. These events are used to perform the "in-situ" calorimeter cosmic calibration that includes energy equalization, energy scale monitoring, time offsets alignment and time resolution monitoring.

The calorimeter cosmic trigger algorithm requires at least four rows with at least one crystal with an energy deposit corresponding to the crossing of a minimum ionizing muon (E~20 MeV) and, to select the high momentum particles, rejects the trajectories with a bad $\chi^2$ of a linear fit (Fig. 2.9).

The expected trigger rate is ~35 Hz and the corresponding bandwidth is ~18 kB/s, well within the allocated budget.

***Calorimeter tagged triggers***

Dedicated triggers will be formed to acquire calibration events given by the calibration source, the laser pulsing of the crystals and electronics pulsing. These trigger algorithms will simply select the events tagged as calorimeter calibration events by the DAQ system. The average requested bandwidth depends on the frequency of the calibration runs. Nonetheless, also in the worst case, assuming to have a laser pulse each spill off period, i.e. once each 1.33 s, the requested bandwidth is ~65 kB/s that is negligible with respect to the available budget.

## 2.4 Test beam performances

To test the performance of un-doped CsI readout with UV extended SiPMs, a dedicated test beam was carried out during April 2015 at the Beam Test Facility (BTF) in Frascati (Italy) where time and energy measurements have been performed using a low energy electron beam, in the energy range [80,120] MeV. The calorimeter prototype consisted of nine $3 \times 3 \times 20$ cm$^3$ un-doped CsI crystals wrapped in 150 μm of Tyvek , arranged into a $3 \times 3$ matrix. Out of the nine crystals, two were produced by Filar OptoMaterials, while the remaining 7 came from ISMA (Ukraine). Each crystal was previously tested with a $^{22}$Na source to determine its light output (LO) and longitudinal response uniformity (LRU), with the results [2.10]:

- a LO of about 90 pe/MeV, when read out with a UV-extended PMT R2059 by Hamamatsu coupled through an air gap;
- a LRU corresponding to a LO variation at both ends of the crystals less than ±6%.

Each crystal was coupled to a large area $12 \times 12$ mm$^2$ SPL TSV SiPM (MPPC) from Hamamatsu by means of the Rhodorsil 7 silicon paste. TSV stands for "through silicon via" and indicates a new technique used for building the SiPM that is characterized by a lower noise and a higher fill-factor. The operating voltage was set at 55 V for each MPPC, about 3 V above the breakdown voltage, corresponding to an average gain of $1.3 \times 10^6$ and a photon detection efficiency, PDE, of about 25 % at 300 nm. Each MPPC is composed of an array of 16 single SiPMs, each one read out with its own anode. A prototype front-end electronics (FEE) board was developed to form an analog sum of the pulses from several anodes. This board provides also a local HV regulation and an amplification by a factor of 8. Photosensor signals coming from the crystals and from a pair of scintillating counters used for triggering were read out with 12 bit, 250 Msps waveform digitizer boards, V1720 from CAEN.

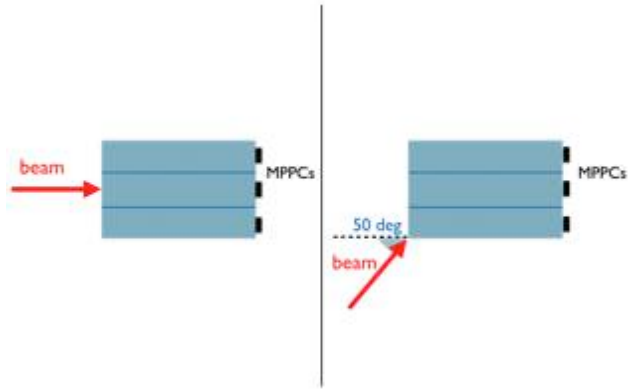

Figure 2.10: *Beam test configurations: beam normal to the prototype front face (left) and beam at 50° with respect to the normal of the prototype surface (right).*

The coincidence of the signals from two 5(L) × 1(W) × 2(T) cm$^3$ plastic scintillating counters, crossed at 90°, was used for triggering on the incoming beam. In addition, another coincidence of the signals from two 10(L) × 30(W) × 4(T) cm$^3$ scintillating counters, one above and one below the array, provided a cosmic ray trigger.  Two configurations, schematically shown in Figure 2.10, were studied during the test:

    i.    beam at 0° with respect to the prototype front face, defined as the side opposite to the photosensors;

    ii.   beam at 50° with respect to the prototype surface.

Configuration (ii) was motivated by the fact that the expected average incidence angle of a signal electron in Mu2e is about 50°. In the following sub-section, we summarize the main results obtained. A full description can be found in ref [2.10].

**Measurement of time resolution**

Assuming a constant pulse shape, the best accuracy is achieved by setting the signal time at a threshold corresponding to a constant fraction of the pulse height. Pivotal for this procedure, usually called digital constant fraction (DCF), are the choices of: fit function, fit range and threshold. The fit function that provided the best time resolution was the asymmetric log-normal function. Waveforms corresponding to signals from the CsI crystals convoluted with the SiPM and prototype pre-amplifier response function have a leading edge of about 25 ns, and a total width of about 300 ns, thus allowing to perform the fit only on the signal region. The DCF threshold, used to determine the reconstructed time, has been optimized using the data taken with a 80 MeV electron beam at 0° incidence angle. Fig. 2.11 shows that optimized value of the DCF threshold is obtained at 5% of the maximum pulse height.

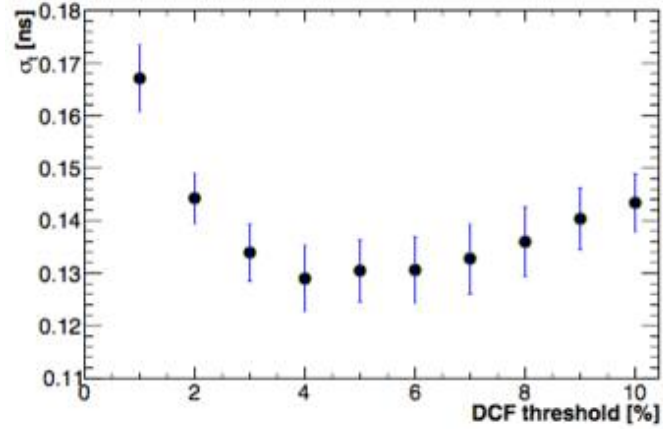

Figure 2.11 *Time resolution using 80 MeV electron beam as a function of the DCF threshold.*

Fig. 2.12 shows an example of a fit to a CsI crystal waveform.

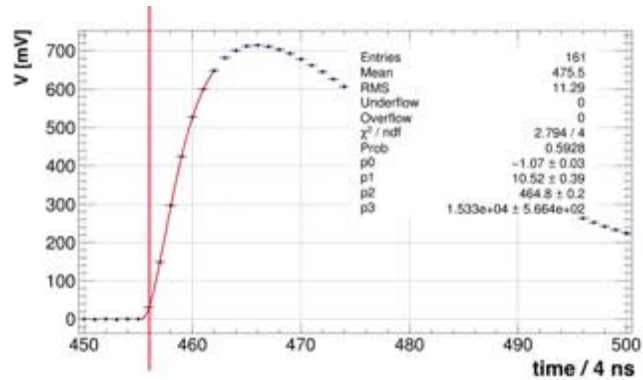

Figure 2.12 *Example of a fit to a waveform.*

The time resolution was measured using three different methods:

1.  using only the crystal with the largest energy deposition;
2.  using the energy-weighted mean time of all crystals in the matrix:

$$t_{matrix} = \sum_{i,j} (t_{crystal(i,j)} \cdot E_{i,j})/E_{tot}, \quad E_{tot} = \sum_{i,j} E_{i,j}$$

3.  using two neighboring crystals with similar energy deposition.

The first two techniques require an external time reference $t_{scint}$. No external time reference is needed for the third one. Methods 1 and 2 were used with both beam configurations: at 0° and 50°. Method 3 was used only for the runs with the beam at 50°, because these were the only ones where neighboring crystals with reconstructed energies larger than 10 MeV were present. The configuration at 0° represents the simplest one from the point of view of the analysis, providing a helpful handle for the development of the time reconstruction method.

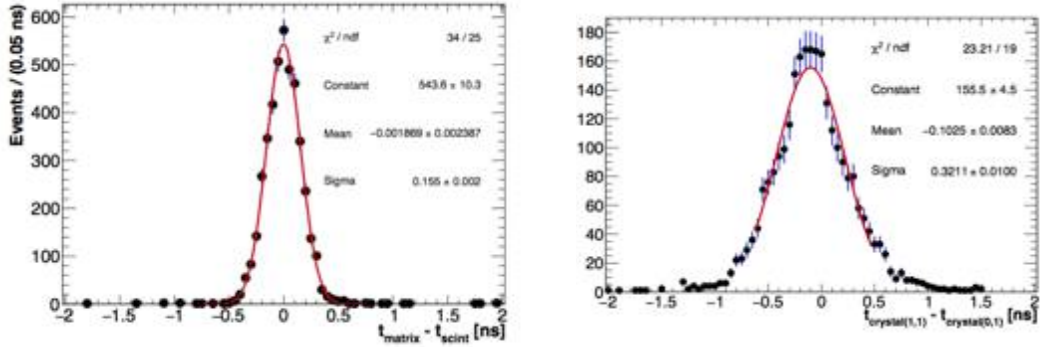

Figure. 2.13: *Distribution of time residuals between: (left) $t_{matrix}$ and $t_{scint}$ for the run at 100 MeV with the beam normal to the prototype, (right) between neighboring crystals for the run with cosmic rays.*

Fig. 2.13.left shows an example of the distribution of time residual between $t_{matrix}$ and $t_{scint}$ for the 100 MeV run. A Gaussian fit to the same distribution shows a standard deviation of about 150 ps, so that removing the contribution of the $t_{scint}$ jitter, the resulting time resolution is of about 110 ps. Data taken in the tilted configurations were combined to apply Method 3. As MIPs crossing 3 cm of CsI crystal, on average, deposit about 20 MeV of energy, cosmic muons allow a measurement of the time resolution in an energy range below the limits of the BTF. Only events where the cosmic ray crosses the central column of the prototype were selected, and the "neighboring crystals" technique was used to measure the time resolution. With a total of three crystals in the central column, there are two independent pairs of neighboring crystals that were used to measure the time resolution with Method 3. Fig.2.13.right shows as an example the time residuals for one of the two pair. Distributions of time residual were fit to a Gaussian function. Then, the time resolution is quoted using the average of the standard deviations resulting from the two fits. Assuming the resolution in all channels to be the same, the resulting average is divided by a factor √2: $\sigma_t$ ~ 250 ps. All results plotted as a function of the energy are summarized in Fig. 2.14. A clear trend of the timing resolution dependence on energy is shown. The timing resolution ranges from about 250 ps at 22 MeV to about 120 ps in the energy range above 50 MeV. The timing resolutions evaluated with different methods in the same energy range are consistent. Furthermore, in the same energy range, the time resolution using Meth. 2 is slightly worse when the beam impacts at 50° (violet triangles).

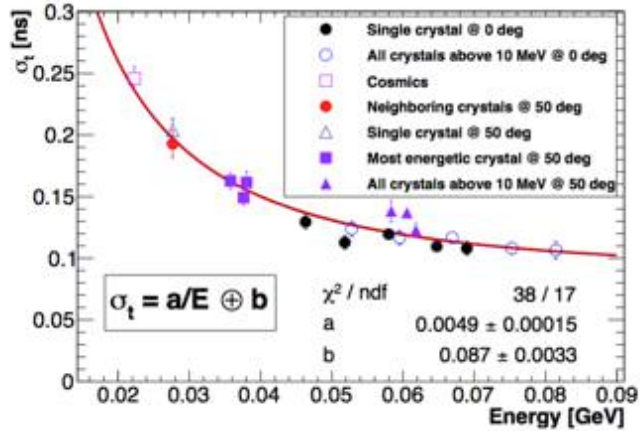

Figure 2.14: *Time resolution summary plot.*

Fluctuations of the shower development could result in additional time jitter between the signals from different crystals and might be partially responsible for this discrepancy. In principle the time resolution $\sigma_t$ depends on the undoped CsI light emission characteristics according to the following formula: $\sigma_t = a/E \oplus b$, where $a$ is proportional to the emission time constant of the undoped CsI, and $b$ represents the additional contribute from the readout electronics. The fit of the data to this function (see Fig. 2.14) shows a good agreement between the data and the fitted function.

**Measurement of energy resolution**

The active volume of the calorimeter prototype was $9 \times 9 \times 20$ cm$^3$ and corresponded to $\sim (1.3 \ R_{Moliere})^2 \times (10 \ X_0)$. Due to the small dimensions, the transverse and longitudinal leakages impact significantly the energy response.

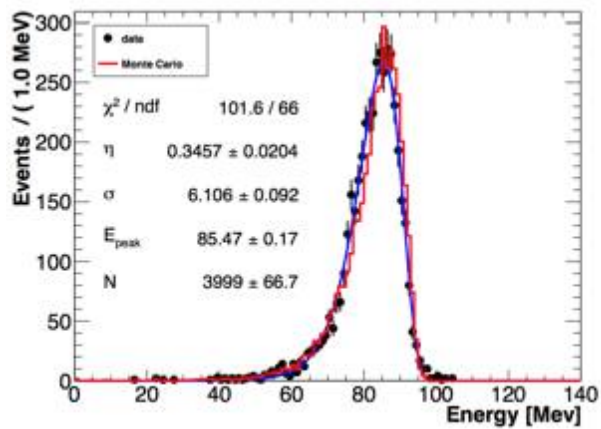

Figure 2.15 : *Distribution of reconstructed energy obtained from the data overlaid with the Monte Carlo for the run with beam energy of 90 MeV. Blue line represents a fit to the data with a log-normal function.*

Fig. 2.15 shows the distribution of the total energy deposition obtained from data taken at a beam energy of 90 MeV and 0 deg incidence angle compared with the Monte Carlo. The same Figure shows also a typical fit with a log-normal function to the data. The σ of this fit was used to evaluate the resolution.

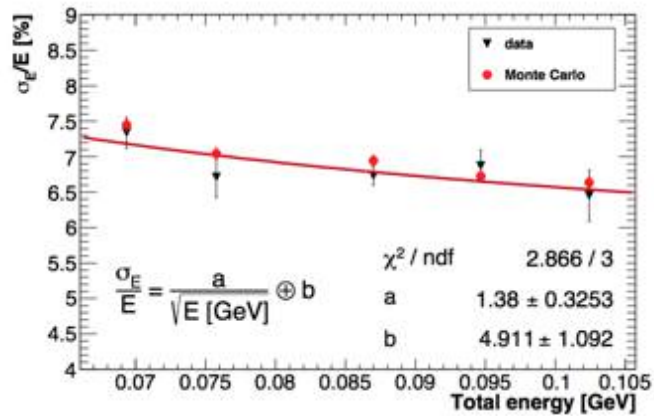

Figure 2.16: *Energy resolution obtained from the data (black) taken at 0 deg compared with the Monte Carlo (red).*

Fig. 2.16 shows the measured energy resolution as a function of the total energy reconstructed in the prototype, with the simulation results superimposed. Within the uncertainties, data and Monte Carlo distributions are in agreement. The measured energy resolution varies from 7.4% to 6.5% in the energy range [70, 102] MeV.

# Chapter 2 References

# 3   The Mu2e calorimeter crystals

## 3.1 Crystal Choice

At the start of the Mu2e project, the crystal considered for the calorimeter was lead tungstate ($PbWO_4$). The low light output and high temperature coefficient required running the calorimeter at -25$^o$C with very tight tolerances on the temperature stability. In addition, its dose rate depended radiation damage requires continuous precision monitoring, causing a difficult calibration problem. At the time of the Mu2e CDR, $PbWO_4$ had been replaced with lutetium-yttrium oxy-ortho-silicate (LYSO) crystals [3-1] LYSO is an excellent match to the problem at hand: it has a high light output, a small Molière radius, a fast scintillation decay time, excellent radiation hardness, and a scintillation spectrum that is well-matched to readout by large-area avalanche photodiodes (APDs) of the type employed in the CMS and PANDA experiments. LYSO was also the preferred option for the KLOE-2 upgrade and the proposed SuperB calorimeter. LYSO crystals are commercially available from Saint-Gobain, SICCAS (Shanghai Institute of Ceramics), SIPAT (Sichuan Institute of Piezoelectric and Acousto-optic Technology), Zecotek and other producers. Following a R&D program at Caltech in cooperation with BOET (Zecotek), SICCAS and SIPAT, large size LYSO crystals of high quality have been produced for HEP applications [3-2]. The $Lu_2O_3$ salt price, however, has increased by more than factor of three since 2013, making the cost of a LYSO calorimeter unaffordable. The only HEP experiment baselined with large size LYSO crystals is COMET at J-PARC.  Following that conclusion, we have looked for cost-effective fast crystals such as barium fluoride ($BaF_2$) and undoped cesium iodine (CsI) crystals for the Mu2e calorimeter baseline.

Table 3.1 Comparison of crystal properties for $BaF_2$, LYSO, CsI and $PbWO_4$.

| Crystal | $BaF_2$ | LYSO | CsI | $PbWO_4$ |
|---|---|---|---|---|
| Density (g/cm$^3$) | 4.89 | 7.28 | 4.51 | 8.28 |
| Radiation length (cm) $X_0$ | 2.03 | 1.14 | 1.86 | 0.9 |
| Molière radius (cm) Rm | 3.10 | 2.07 | 3.57 | 2.0 |
| Interaction length (cm) | 30.7 | 20.9 | 39.3 | 20.7 |
| $dE/dx$ (MeV/cm) | 6.5 | 10.0 | 5.56 | 13.0 |
| Refractive Index at $\lambda_{max}$ | 1.50 | 1.82 | 1.95 | 2.20 |
| Peak luminescence (nm) | 220, 300 | 402 | 310 | 420 |
| Decay time $\tau$ (ns) | 0.9, 650 | 40 | 26 | 30, 10 |
| Light yield (compared to NaI(Tl)) (%) | 4.1, 36 | 85 | 3.6 | 0.3, 0.1 |
| Light yield variation with temperature (% / °C) | 0.1, -1.9 | -0.2 | -1.4 | -2.5 |
| Hygroscopicity | None | None | Slight | None |

Tab. 3.1 shows a comparison of $BaF_2$, LYSO, undoped CsI and $PbWO_4$ [3-3]. Several points are worth discussing. Both the fast component of $BaF_2$ and CsI have about 5% light of LYSO, but are ten times of $PbWO_4$. They have a substantially larger Molière radius and radiation length, which are disadvantages. The emission of both LYSO and $PbWO_4$ is peaked at 420 nm, a wavelength matches well with APD readout, while that from both $BaF_2$ and undoped CsI is at a much high frequency. $BaF_2$ is featured with a very fast scintillation component peaked at 220 nm with sub-ns decay time, which is very useful in background rejection, providing effective compensation for the larger shower size. It may also be possible to use its bright slow component peaked at 300 nm with 650 ns decay time when event rates are not too high. The slow scintillation component in $BaF_2$, however, presents a significant pile-up effect, which needs to be reduced to accommodate high event rates by spectroscopically selective readout and selective doping in $BaF_2$ to suppress the slow scintillation light. Photomultiplier tubes with quartz windows and solar-blind photocathodes are well-matched to the fast scintillation component of $BaF_2$, but will not work in the field of the detector solenoid. Micro-channel plate PMTs are at present too expensive despite spinoffs from the LAPPD project being pursued. Our main thrusts thus were to develop solar-bind solid-state photosensors in the form of APDs or SiPMs and modify $BaF_2$ crystals by selective doping with rare earth, such as La, Ce and Y [3-4]. Such developments were followed but took too much time, so that a very fast $BaF_2$ calorimeter, with solar blind readout, is now being pursued for future Mu2e upgrade.

Because of its low melting point and raw material cost, undoped CsI is a low-cost crystal scintillator available commercially. As reported in chapter 2, we have demonstrated, both with simulation and first experimental tests, that this scintillator is a good match to  latest generation of Silicon Photomultipliers and is able to well satisfy the Mu2e calorimeter requirement. In this section, we report performance and radiation hardness of commercially available undoped CsI crystals, based upon which CsI crystals have been baselined for the Mu2e experiment since 2016. The Mu2e project described in this report will concentrate on a fast CsI crystal calorimeter.

## 3.2 Transmission and Wrapping Material

As shown in Table 3.1, undoped CsI crystals are slightly hygroscopic, and have an emission peak at 310 nm with fast decay time of about 30 ns. Its UV emission requires photodetectors with UV extensive response, which is discussed in the photodetector section of this report (chapt. 4). Its soft and slightly hygroscopic nature causes uncertainty in surface control. Fig. 3.1 (Left) shows large divergence in the longitudinal transmission spectra measured for undoped CsI crystal samples from several vendors. Also shown in this plot is the emission spectrum (blue dashed lines) and the numerical values of the emission weighted longitudinal transmittance (EWLT). The large diversity of EWLT indicates that longitudinal transmission of CsI can not be used to represent crystal's optical quality. We thus concentrate on light output measurement for CsI quality control.

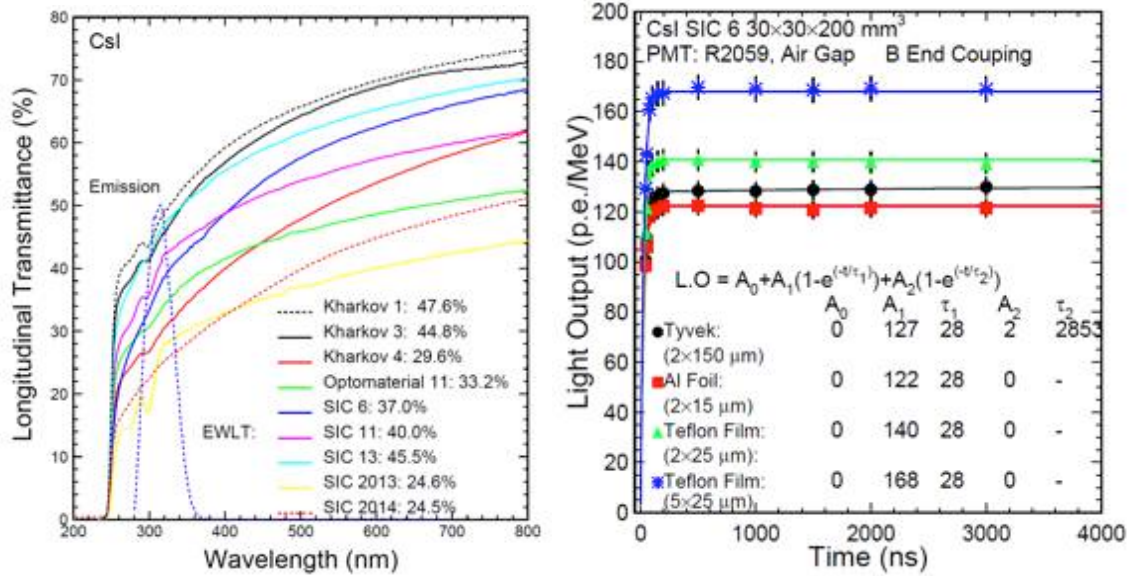

Figure 3.1. *The longitudinal transmission and scintillation emission (dashed lines) spectra (Left) and the light output (Right) are shown as a function of wavelength and integration time respectively for nine undoped CsI samples of about 20 cm long from several vendors and different wrappings.*

Fig. 3.1 (Right) shows light output as a function of integration time for a 20 cm long CsI sample SIC-6 with different wrapping materials. While the overall decay time of this CsI sample can be fit well to a single component with a fast decay time of 28 ns, the amplitude of its light output depends on the wrapping materials used. PTFE Teflon film is found to be the best wrapping material for the UV light of CsI. For good reproducibility and tenability for the light response uniformity, two layers of Tyvek paper of 150 μm thick are chosen as the wrapping material for CsI quality control.

## 3.3 Light Output, Energy Resolution and Light Response Uniformity

Light output, energy resolution and light response uniformity are important parameters for CsI quality control. Fig. 3.2 shows pulse height spectra measured by a Hamamatsu R2059 PMT with a bi-alkali cathode and a quartz window and with an air gap coupled to alternative end A (Left) and B (Right) of the sample SIC CsI-6. A total of seven spectra are measured for each coupling end by aiming Na-22 γ-rays at seven positions evenly distributed along the crystal axis. The numerical values of the FWHM energy resolution (E.R.) of the γ-ray peak are also shown in the figure. The light output and the FWHM energy resolution of the sample are defined as the average of the seven points for each coupling end. The light response uniformity (LRU) is defined as the RMS value of the seven light output data in percent to the average.

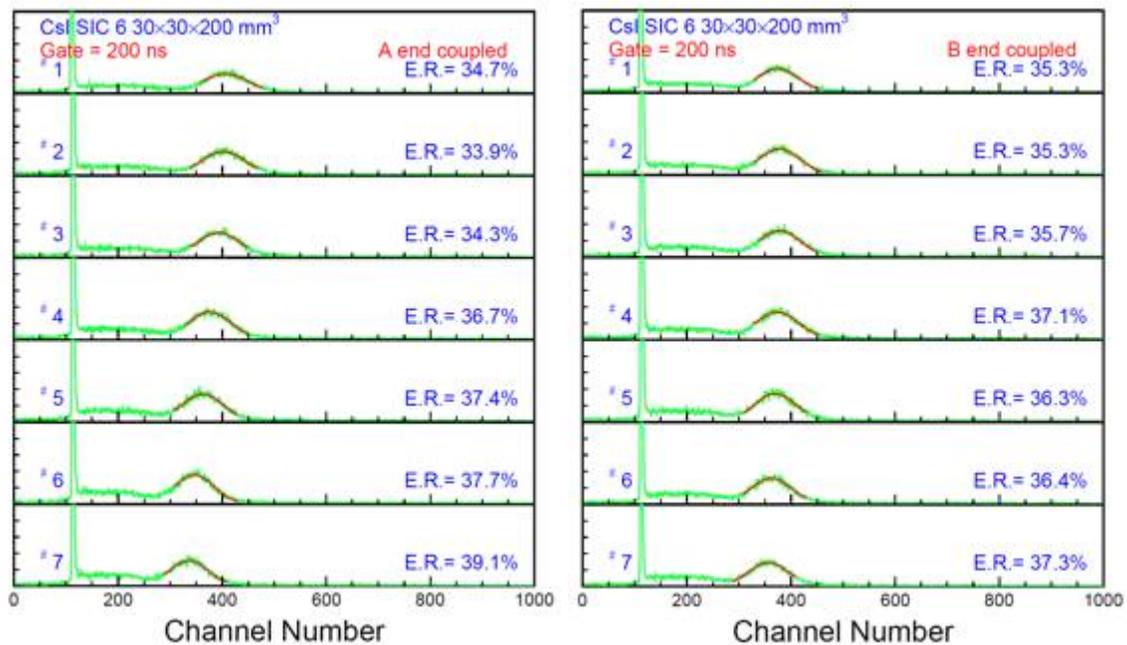

**Figure 3.2.** *The pulse height spectra measured by a Hamamatsu R2059 PMT with air gap coupled to alternative ends (A: Left and B: Right) for the sample SIC CsI-6.*

Fig. 3.3 (Left) summarizes the light output in 200 ns gate (top), FWHM energy resolution (middle) and light response uniformity (LRU, bottom) for undoped CsI crystals from different vendors. Also shown in the plots are the Mu2e specifications (pink dashed lines) used for quality control for the pre-production CsI crystals that will be used for Module-0 (chapt. 9) construction, for electron beam test and for ranking technically the different vendors. It is clear that most samples satisfy the specifications.

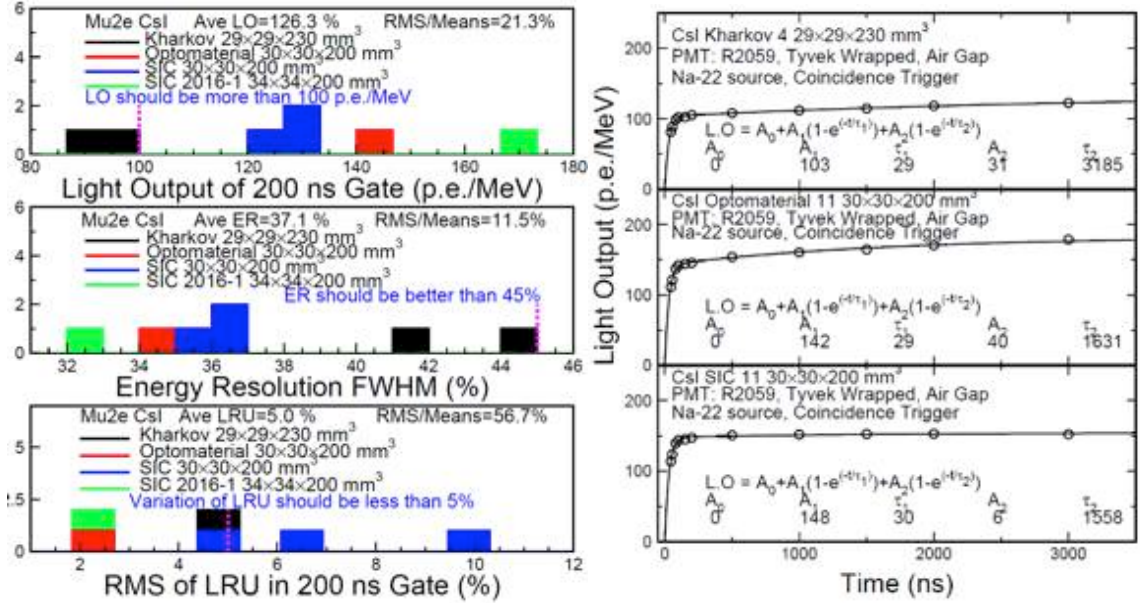

Figure 3.3 *Left: Summary of light output in 200 ns gate (top), FWHM energy resolution of Na-22 peak (middle) and light response uniformity (bottom) for pure CsI crystals from three vendors. Also shown in the plots are the Mu2e specifications (pink dashed lines). Right: Light output is shown as a function of the integration gate for CsI crystal samples from three vendors.*

## 3.4  Signal Shape: Fast/Total Ratio

In addition to the fast scintillation, commercially available undoped CsI crystals may have slow scintillation component caused by impurities or defects. Fig. 3.3 (Right) shows light output as a function of integration gate for three crystals from different vendors. In addition to the fast scintillation component some of the commercially available CsI crystals show increased light output for increased integration gate width, indicating a sample-dependent slow scintillation component with decay time of a few µs, which would cause harmful pileup effect so need to be controlled.

The top three plots of Fig. 3.4 (Left) show light output values measured at seven points evenly distributed along the crystal axis with 100, 200 and 3,000 ns gate respectively for the sample Kharkov 4. The bottom plot of Fig. 3.4 shows the Fast/Total (F/T) ratios for 100 and 200 ns gate versus 3000 ns gate. The average F/T ratio of these seven points is 73% and 75% respectively, indicating significant slow scintillation. Fig. 3.4 (Right) summarizes the F/T ratio of LO(200)/LO(3000) for all samples. We notice that the sample SIC 6 has almost no slow scintillation component, indicating that the slow scintillation component may be reduced or eliminated by optimizing raw material and crystal growth. Also shown in Fig. 3.4 (Right) is the Mu2e specification of 75% (Dashed lines). It is clear that all samples satisfy this specification.

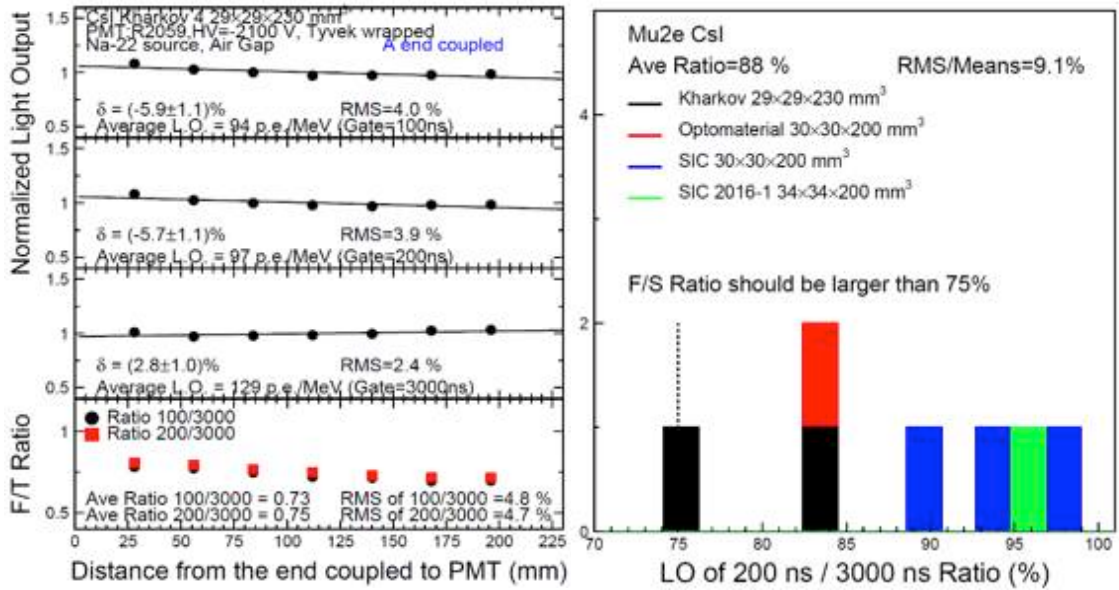

Figure 3.4. *Left: Light output (top three) measured at seven points with 100, 200 and 3000 ns integration gates respectively, and the F/T ratio (bottom) of LO(100 ns) and LO(200 ns) versus LO(3000 ns). Right: Summary of the F/T ratio of LO(200 ns)/LO(3000 ns).*

The slow scintillation in undoped CsI is found to be peaked at 450 nm, so can be eliminated by inserting a band-pass filter [3-5]. This is also highly correlated to the radiation induced readout noise, discussed in section 3.6 and ref. [3-5]. Reducing slow scintillation component thus will also reduce the radiation induced readout noise.

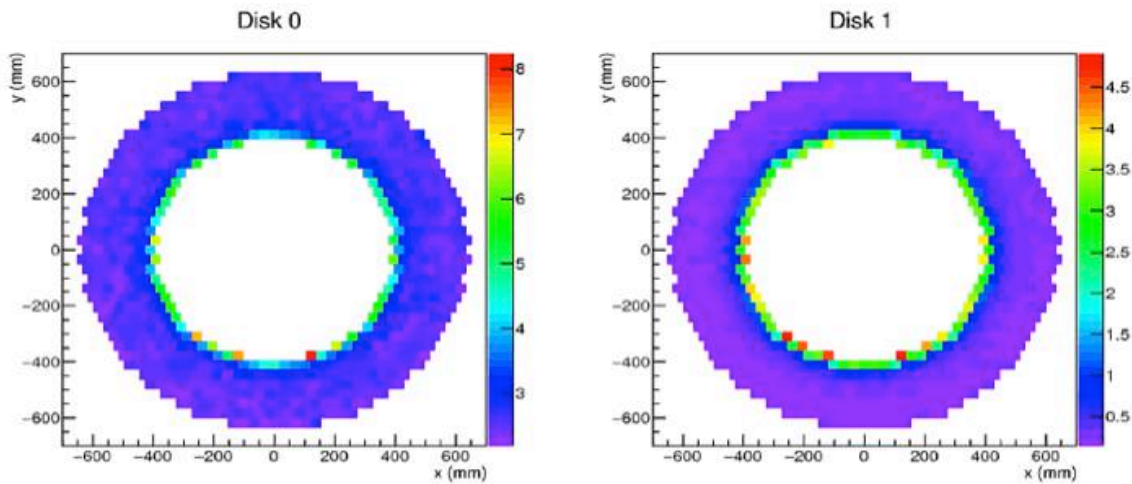

Figure 3.5 *Distribution of the Radiation dose (krad/year) in the front face of the two disks.*

## 3.5 Radiation Hardness of Undoped CsI Crystals

All known crystals suffer from radiation damage. There are three possible damage effects in crystal scintillators: (1) damage to the scintillation-mechanism, (2) radiation-induced phosphorescence and (3) radiation-induced absorption [3-6]. A damaged scintillation mechanism would reduce scintillation light yield and light output, and may also change the light response uniformity along the crystal if the dose profile is not uniform along the crystal. Radiation-induced phosphorescence (commonly called afterglow) causes an increased dark current in photo-detectors, and thus an increased readout noise. Radiation-induced absorption reduces light attenuation length and thus light output. It may also change light response uniformity if light attenuation length is shorter than twice the crystal length.

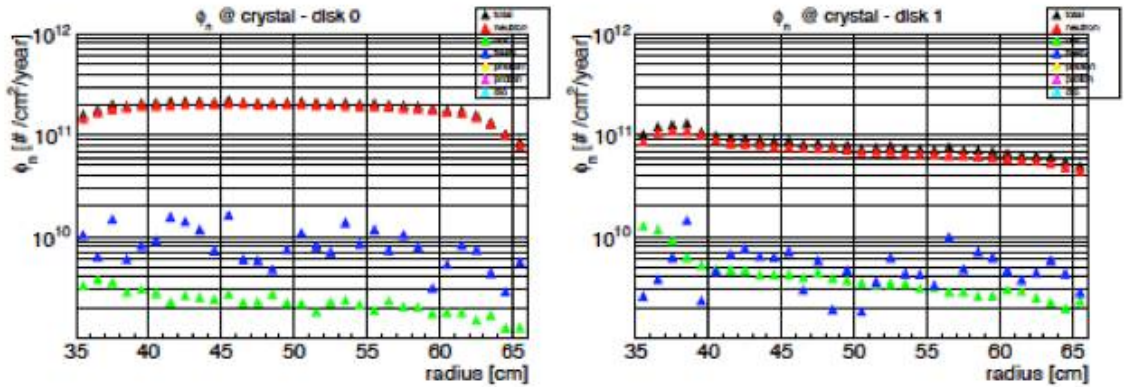

Figure 3.6 *Left: The neutron fluence (in n/cm²/year) expected in the front (Left) and back (Right) CsI disks are shown as a function the radius where the crystals are located.*

No scintillation mechanism damage was observed in crystals listed in Tab.3.1. The main radiation damage in these crystals is radiation-induced absorption, or color-center formation. Radiation-induced color centers may recover at the application temperature through color-center annihilation, leading to a dose-rate dependent damage [3-6]. If so, a precision light monitoring system is mandatory to follow variations of crystal transparency *in situ*. Radiation-induced absorption in undoped CsI crystals does not recover at room temperature, so is not dose-rate dependent [3-7]. Radiation-induced transmission and light output losses in CsI crystals were measured. Thermal annealing and optical bleaching were also found not effective for CsI [3-8]. Radiation damage study for CsI is thus a costly exercise, since crystal samples after testing are unusable. It was also found that radiation damage in 20-cm long pure CsI crystals showed no saturation, with light output loss of 70 - 80% after 1 Mrad [3-9].

The Mu2e radiation environment was calculated [3.10] by using a GEANT based full simulation for the Mu2e detector. Contributions from particles produced in beam flash, electrons from muons decaying in orbit, neutrons, protons, and photons are included in this calculation. Fig. 3.5 shows the expected ionization dose deposited per crystal per year as a function of the position for the front (left) and back (right) disk, respectively. Fig. 3.6 shows the neutron fluence as a function of the radius for the fron (Left) and back (Right) disks respectively. The average dose is about 3 and 0.5 krad/year in the front and back disk respectively with 5 krad for the innermost crystals in the front disk. The neutron fluence is dominated by neutrons coming from muon capture in the target and is reasonably flat on the calorimeter surface with an average of $2 \times 10^{11}$ ($1 \times 10^{11}$ ) n/cm$^2$/year for the front (back) disk. Assuming 230 days' run ($2 \times 10^7$ sec) each year, the hottest crystals would face the following radiation environment: 1 rad/h and $1.0 \times 10^4$ n/cm$^2$/s. The average crystal will see 0.6 rad/h. If we assume a factor of 3 safety on the simulation calculation, we will use in the following as a guidance: 1.8 rad/h for the dose and $3 \times 10^4$ n/cm$^2$/s for the neutron fluence.

Fig. 3.7 shows the longitudinal transmission (Left) and light response uniformity (Right) for a pure CsI sample of $5 \times 5 \times 30$ cm$^3$ grown at SIC in 2013 and irradiated up to 1 Mrad. Its light output was measured by using a R2059 PMT with a bi-alkali cathode and a quartz window. Also shown in the left plot is the emission spectrum (blue dashes) and the corresponding EWLT values. Radiation damage at a level of about 60% and 80% was observed respectively in EWLT and light output after an integrated dose of 1 Mrad. The damage, however, shows no saturation up to 1 Mrad, indicating a high density of defects in this crystal. The result of this measurement is consistent with the data obtained twenty years ago for 20-cm long pure CsI crystal samples from Kharkov [3-9].

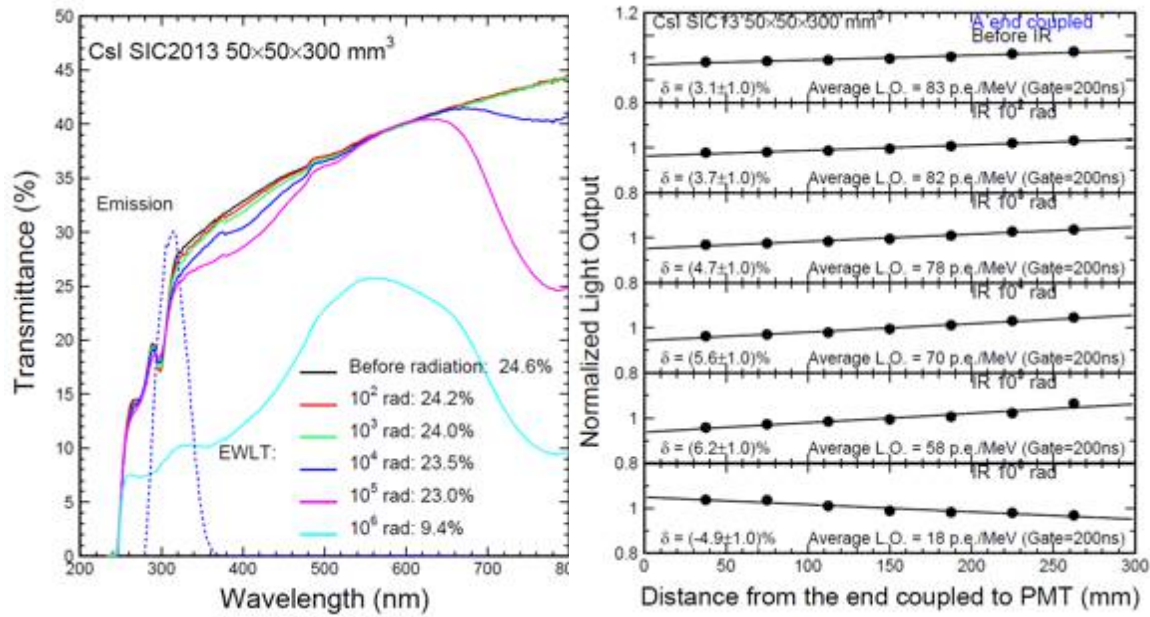

Figure 3.7 *Longitudinal transmittance spectra (left) and light response uniformities (right) are shown as a function of the integrated dose up to 1 Mrad for a CsI sample SIC 2013 of 5 × 5 x 30 cm³.*

In Fig. 3.8 (left), the normalized emission weighted longitudinal transmission (EWLT) and light output (LO) as a function of integrated dose for undoped CsI crystals from various vendors, showing very consistent light output losses up to 10 krad as compared to the 30 cm long sample SIC 2013 [38]. We notice the divergence becomes large after 100 krad. Fig. 3.8 (Right) shows a summary of the normalized light output loss measured in 200 ns integration time for undoped CsI crystals from various vendors after 100 krad of ionization dose. Also shown in the plot is the Mu2e radiation damage specification (dashed lines). It is clear that most CsI crystals from the industry satisfy the Mu2e specification.

We also investigated neutron induced radiation damage in both transmittance and light output up to $10^{12}$ n/cm² with no obvious damage in all CsI samples from different vendors, indicating that the main damage source for the Mu2e CsI crystals is ionization dose. Since radiation damage in CsI crystals is caused by oxygen contamination [3-7], it is expected that an R&D program aiming at reducing oxygen contamination could improve crystal quality and reduce the level of radiation damage in CsI. The quality of CsI can be improved through systematic R&D programs aimed at reducing oxygen contamination during crystal growth. A close collaboration with crystal growers might be followed for such an effort but it is not mandatory for the Mu2e needs and time scale.

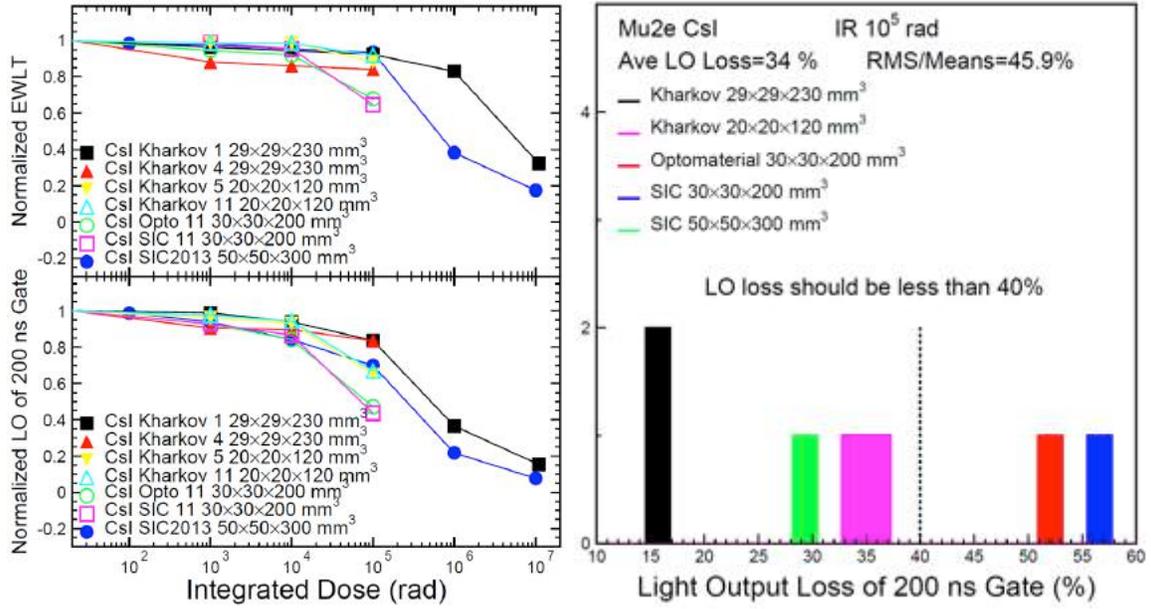

Figure 3.8 *Left: Normalized emission weighted longitudinal transmission (EWLT, top) and light output (LO, bottom) are shown as a function of integrated dose for undoped CsI crystals from various vendors. Right: Summary of the normalized light output loss measured in 200 ns integration time for undoped CsI crystals from different vendors after 100 krad.*

## 3.6    Ionization Dose and Neutrons Induced Photocurrent and Noise

One consequence of radiation damage in crystals is radiation induced photocurrent and readout noise. Assuming 230 days' run ($2 \times 10^7$ sec) each year, the hottest crystals would have the following radiation environment: 1 rad/h for ionization dose and $1.0 \times 10^4$ n/cm²/s for neutrons. Radiation-induced induced photocurrent in undoped CsI crystals was measured by using a Hamamatsu R2059 PMT coupled directly to the sample in a radiation environment.

We define an "F" factor as the radiation induced photoelectron numbers per second, determined by measuring the photocurrent of the PMT.

$$F = \frac{\frac{Photocurrent}{Charge_{electron} \times Gain_{PMT}}}{Dose\ rate_{\gamma-ray}\ or\ Flux_{neutron}} \quad (1)$$

The energy equivalent radiation induced readout noise (σ) is defined as the standard deviation of the photoelectron number (Q) in the readout gate:

$$\frac{\sqrt{Q}}{LO} \quad MeV \quad (2)$$

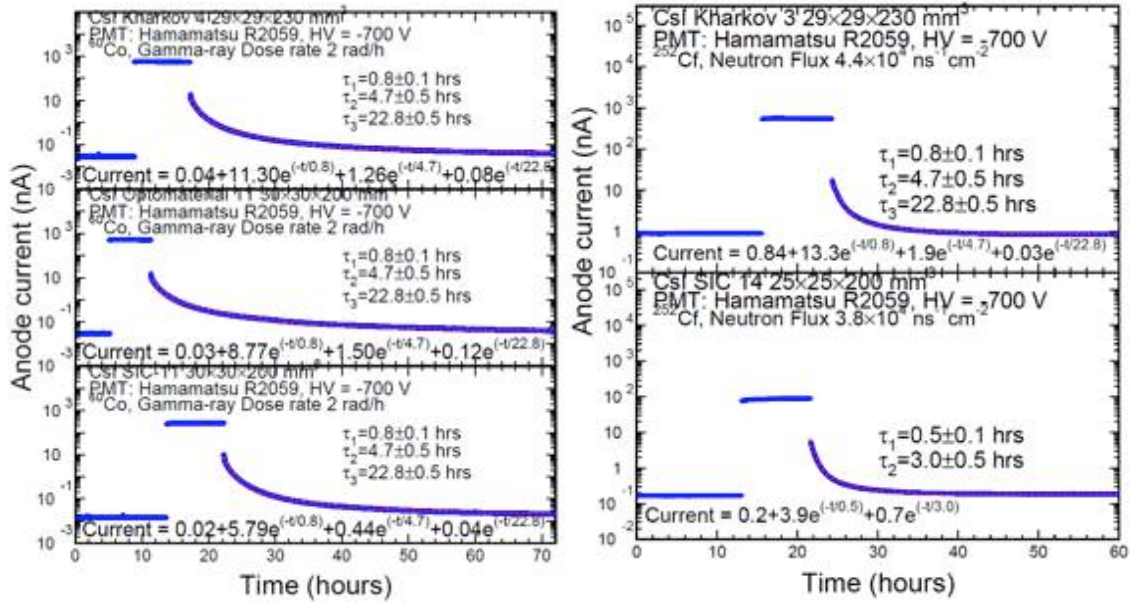

Figure 3.9 *History of photocurrent measured before, during and after γ-ray (Left) and neutron (Right) irradiation respectively for three and two undoped CsI samples, where afterglows were fitted to up to three time constants.*

Fig. 3.9 shows histories of photocurrent, measured before, during and after irradiation under a dose rate of 2 rad/h (Left) and a neutron flux of about $4\times10^4$ n/cm²/s (Right) respectively for undoped CsI crystal samples. The photocurrent in plateau was used to extract radiation-induced readout noise. We note that the decay of the γ-ray and neutron induced afterglows fit well with three time constants: 0.8, 4.7 and 22.8 hours for undoped CsI crystals with slow component, and two time constants of 0.5 and 3 hours for sample without slow component.

Table 3.2 Ionization Dose Induced Readout Noise in CsI Crystals under 1.8 rad/h

| Sample | Dimensions (cm³) | LO of 200 ns Gate (p.e./MeV) | Dark Current (nA)* | Photo current @2 rad/h (nA)* | F (p.e./s/rad/hr)* | σ (MeV)* |
|---|---|---|---|---|---|---|
| Kharkov 4 | 2.9×2.9×23 | 96 | 0.035 | 679 | 6.68E+09 | 5.11E-01 |
| Opto 11 | 3×3×20 | 140 | 0.039 | 663 | 6.53E+09 | 3.46E-01 |
| SIC 6 | 3×3×20 | 125 | 0.015 | 296 | 2.91E+09 | 2.59E-01 |
| SIC 11 | 3×3×20 | 128 | 0.018 | 330 | 3.25E+09 | 2.67E-01 |
| SIC 13 | 3×3×20 | 130 | 0.026 | 461 | 4.54E+09 | 3.11E-01 |

Table 3.3 Neutron Induced Noise in CsI Crystals under $3\times10^4$ n/cm$^2$/s

| Sample | Dimensions (cm$^3$) | LO of 200 ns Gate (p.e./MeV) | Dark Current (nA)* | Photo current (nA)* | F (p.e./n/cm) * | σ (MeV)* |
|---|---|---|---|---|---|---|
| Kharkov 3 | 2.9×2.9×23 | 88 | 0.08 | 112 | 5.01E+04 | 3.3E-01 |
| SIC 2014 | 2.5×2.5×20 | 140 | 0.013 | 165 | 8.68E+04 | 2.8E-01 |

Table 3.2 and 3.3 summarize ionization dose and neutron induced photocurrent and energy equivalent readout noise (σ) respectively in undoped CsI crystals under an ionization dose rate of 1.8 rad/h and a neutron flux of $3\times10^4$ n/cm$^2$/s. The dark current, radiation induced photocurrent, F and σ are normalized to the Mu2e CsI dimension of 3.4×3.4×20 cm$^3$. It is clear that the readout noises induced by both γ-rays and neutrons are much smaller than 1 MeV, required by the Mu2e experiment. We also notice that the readout noise induced by γ-rays is much larger than that from neutrons even with γ-ray background. CsI quality control on this aspect thus can be carried out with ionization dose only.

## 3.7 Technical Specifications for Undoped CsI Crystals

Listed below are technical specifications defined according to measured data for undoped CsI crystal samples from Kharkov (Ukraine), Opto Materials (Italy) and SICCAS (China).

- Crystal lateral dimension: ±100 μ, length: ±200 μ;
- Scintillation properties are measured by a bi-alkali PMT with an air gap coupled to the crystal wrapped by two layers of Tyvek paper of 150 μm;
- Light output: > 100 p.e./MeV with 200 ns integration gate;
- FWHM Energy resolution: < 45% for Na-22 peak;
- Fast (200 ns)/Total (3000 ns) Ratio: > 75%;
- Light response uniformity (LRU): < 5%;
- Radiation hardness:
  - Normalized light output after 10/100 krad > 85/60%; and
  - Radiation Induced Readout noise @1.8 rad/h: < 0.6 MeV.

# Chapter 3 References

# 4    The Mu2e Calorimeter Photosensors

The calorimeter requirements translates in a series of technical specification for the photo sensors that are summarized in the following list: (i) a good photon detection efficiency (PDE) of above 25%, for wavelengths around 310 nm to well match the light emitted by the un-doped CsI crystals; (ii) a large light collection area that in combination with (i) provides a light yield of above 20 p.e./MeV; (iii) a fast rise time; (iv) a narrow signal width to improve pileup rejection; (v) a high gain and (vii) the capability of surviving in presence of 1 Tesla magnetic field, operating in vacuum and in the harsh Mu2e radiation environment. Our solution to all of this is an array of large area UV extended Silicon Photomultipliers (SiPM).

A "custom" modular SiPM layout has been chosen to enlarge the active sensor area and maximize the number of collected photoelectrons from the crystal. The simulation work for optimization of granularity, acceptance and cost suggested to increase the crystal dimension from 30x30 to 34x34 mm$^2$, to accommodate two arrays of 2x3 monolithic 6x6 mm$^2$ UV extended SiPMs. This was impossible for the previous crystal dimension due to problems related to the space needed for inserting cooling lines and SiPM holders. This

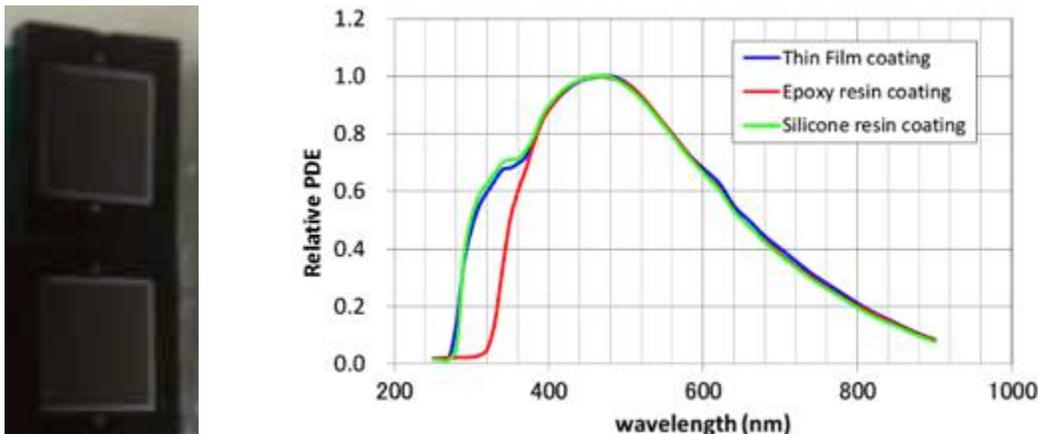

Figure 4.1: *(left) picture of a monolithic 6x6 Hamamatsu SiPM, (right) comparison of relative PDE between Epoxy, SPL and TFL coatings.*

SiPM dimension granted an area collection efficiency of ~ 19 % per sensor. Assuming a PDE > 25%, and a light yield of at least 100 pe/MeV with a 2" UV extended PMT (c.f.r. sec.3.7), we estimate to collect more than 18 p.e./MeV with a single photosensor when coupled with an air-gap to the crystal. This possibility of not using grease or glue is a great advantage since it reduces the problems related to outgassing. A second SiPM, with its own independent electronics chain, is used to read out a crystal to ensure redundancy. The choice of a 6x6 mm$^2$ monolithic SiPM was driven by the practical consideration that this format was common to many of the available commercial producers.

| Pixel pitch [$\mu$m] | 50 |
|---|---|
| Effective photosensitive area [mm] | $6.0 \times 6.0$ |
| Number of pixel | 14400 |
| Window material | Silicon resin |
| Gain (at 25°) | $1.7 \times 10^6$ |

Table 4-1: *Characteristics of the UV extended 6x6 mm² SiPM from Hamamatsu with a Silicon Protection layer: 13360-6050CS*

## 4.1    The UV extended SiPMs

To well match the wavelength of the light emitted by the CsI crystals, the SiPMs have to be extended in the UV region. As described in ref. [4-1], many efforts have been developed in the last years to extend the sensitivity to the short wavelength region. Good results have been obtained down to 280-300 nm, working on the conversion surfaces: for the Hamamatsu devices the basic change, with respect to a standard blue or green SiPM, is the coating on the protection cover. Indeed, as shown in Fig. 4.1 right, Hamamatsu presents at least two UV extended devices: one with a Silicon Protection Layer (SPL) and one with a Thin Film Layer (TFL). Both of them improve efficiency in the UV region, when compared to the ones with standard Epoxy resin. Other producers as SensL or FBK have also obtained good results using similar approaches.

In Tab. 4.1, we report the characteristics of the first UV extended 6x6 mm² SPL SiPMs we have tested from Hamamatsu. The Gain is above $10^6$, PDE is excellent, and the pixel size is of around 50 μm.

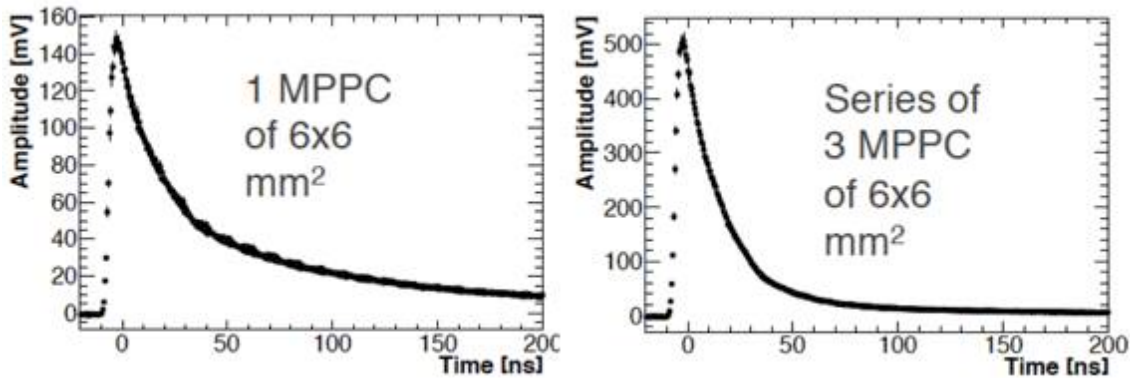

Figure 4.2 Left: *Signal shapes for 1 single 6x6 mm² SiPMs from Hamamatsu when illuminated with a 50 ps wide laser pulse. SiPMs output are without preampification and terminated on 50 Ohm. Right: Signal shape for the series of 3 SiPMs of 6x6 mm².*

### 4.3  The series configuration and the Mu2e "custom" package

A configuration in series has been chosen to overcome the issues related to the parallel connection which might affect the energy and time measurements. Indeed, the very large capacitance of a parallel connection results in an increased noise, signal rise time and width. When connecting SiPMs in a series configuration, the bias of each SiPM is regulated by the common dark current, $I_{dark}$. A reasonable equalization and similarity on the I-V curves of the SiPMs in a series, at the level of $\pm$ 15%, reduces the possible spread in the operating voltages when biased in common. Differently from the parallel configuration, where the signal becomes wider, the pulse shape of a series of SiPMs results narrower than that of a single SiPM due to the reduction of the total sensor capacitance.

This is shown in Fig. 4.2 where in the left (right) plot the signal shape for a single SiPM (the series of 3 SiPMs) is presented. A clear reduction in the width can be appreciated. The reduction of the overall width helps improving the pileup discrimination capability. The rise time of the series remains as good as the one of a single SiPM thus not modifying the time resolution performances with respect to the parallel configuration.

The design of the Mu2e custom SiPMs is reported in Fig. 4.3 and it is organized as the parallel of 2 series of 3 6x6 mm$^2$ monolithic UV extended SiPMs. In the following we refer to the array as SiPM or sensor and to the 6x6 mm$^2$ monolithic as cell. A common bias voltage point is provided for all cells. A full description of the sensor technical specifications can be found in ref [4-2]. These specifications were used for an international competitive bid done by INFN to rank, with a first pre-production, the technical capability of each producer. At the first step of the tender, three producers were selected: Hamamatsu photonics (Japan), SensL (Ireland) and Advansid (Italy). Each of this firm has been invited to produce 50 prototypes following the design shown above and the strict set of specifications of ref [2]. All prototypes are arrived and are under test at the moment of writing. A Quality Assurance program is under way at our test station in Pisa (sec. 8.3.1) to check gain, I-V curve and PDE. On a smaller sample, test of radiation hardness properties and MTTF will also be carried out as explained in the following sections. In Fig. 4.4, the pictures for the first prototypes of Hamamatsu (top), SensL (center) and Advansid (bottom) are shown.

Figure 4.3: *Mu2e SiPM technical drawing.*

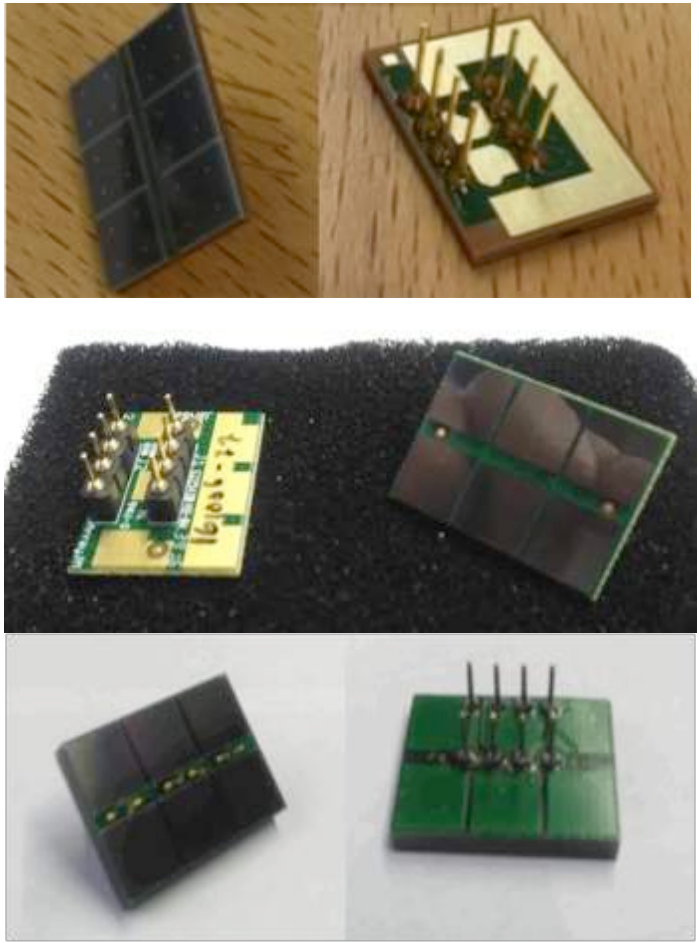

Figure 4.4 *Pictures of Hamamatsu (top), SensL (center) and Advansid (bottom) Mu2e SiPMs.*

### 4.4 Charge Resolution studies

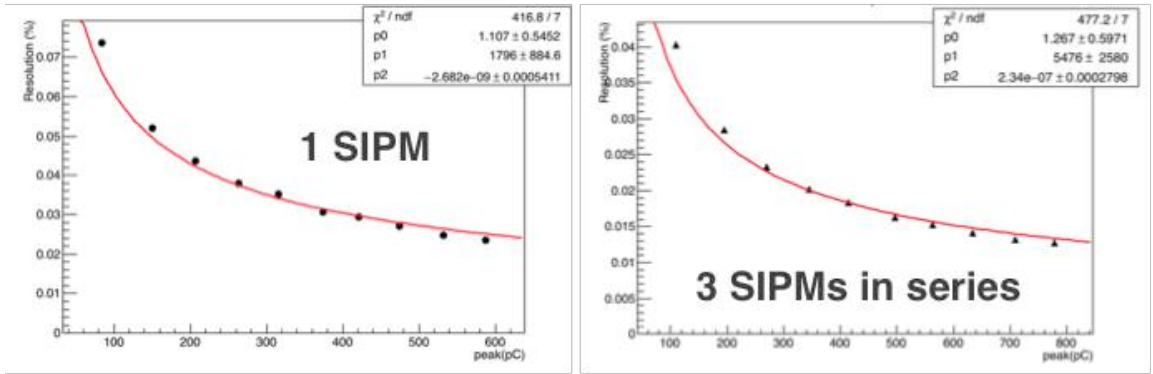

Figure 4.5: *Charge resolution of 3 SiPMs in series configuration (right) with respect to a single SIPM (left) when illuminated uniformly with a UV LED.*

The Mu2e SiPM is a 2x3 array of individual 6x6 mm$^2$ monolithic UV extended SiPMs (cell). Each series has an independent bias voltage while the anodic signals of the two series is summed together. The operation voltage applied to each cell in the series is determined by the common $I_{dark}$ and the single I-V curve. To prove that a not perfect setting of the Vop of each cell does not affect the overall charge resolution, we have carried out a direct test on this parameter. Our first test has compared the charge resolution observed in a SiPM series with respect to the one of a cell. Both the SiPM series and the single cell were illuminated uniformly with a LED. SiPMs were mounted on a PCB support, inserted inside a black box and illuminated by a UV Led through a polaroid filter. In Fig. 4.5, the distributions of resolution for a SiPM cell (left) and for a series of three SiPMs (right) are shown as a function of the charge peak. The distributions have been fitted using the following formula: $\sigma_Q/Q = \sqrt{(Fano/Npe + \sigma_{noise}^2)}$. This can be rewritten as: $\sigma_Q/Q = \sqrt{(P_0/(P_1/Q_{max}) + P_2)}$, where $P_0$ represents the Fano factor (or excess noise), $P_1$ is the number of photoelectrons, Npe measured at $Q_{max}$, $P_2$ describes the noise factor and $Q_{max}$ has been imposed equal to 600 pC. From the fit parameters, we see that the Fano factor is compatible with one and the noise is negligible. The number of photoelectrons estimated from the fit to the resolution for the SiPM in series is Npe(series)=5476. This is three times larger than that obtained for a single cell, Npe(1 cell)=1796. This demonstrates that the resolution is dominated by Poisson fluctuation of the collected light and no other effects, such as voltage or gain fluctuation, contribute.

***An extreme case: series made of two not irradiated SiPMs and an irradiated one***
In this section we quantify the effect on resolution related to the irradiation-induced increase of SiPM leakage current (c.f.r. sec.4.5). Let us consider the extreme case where only one of the three cells has been irradiated, increasing its leakage current of a factor of 2000 with respect to the not irradiated ones. Given that the chosen bias has to be the one corresponding to Idark = 0.5 uA, the irradiate SiPM will work with a (Vop - Vbr) ~0 Volt (see figure 4.6), so that its response will be negligible. This does not affect the response of the other two cells, that will keep working at the right overvoltage (Vop-Vbr=+3V).

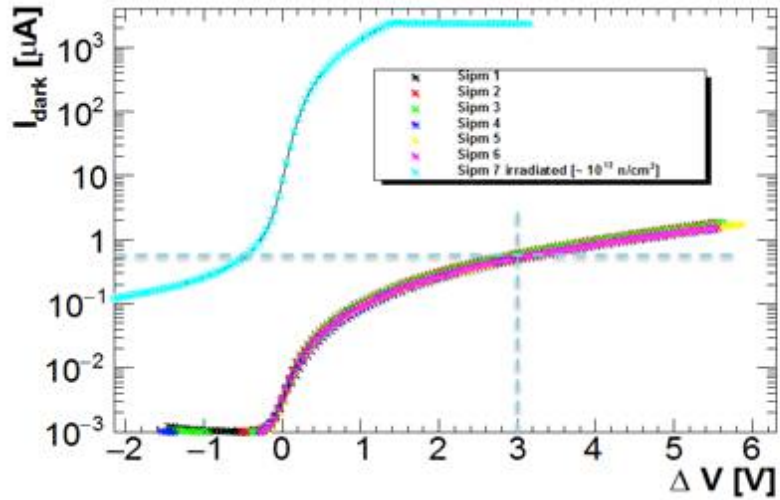

Figure 4.6: *Measured Idark vs (Vop − Vbr) for the 6x6 mm$^2$ MPPC. With the magenta curve the same measurement for an irradiated MPPC with ∼ 10$^{12}$ n/cm$^2$ of 1 MeV energy.*

To prove this, the response of the series connection of three SiPMs to a blue laser was studied in these two configurations:

1. Three MPPC not irradiated;
2. Two MPPC not irradiated and the one irradiated at Dresden up to a fluence of ∼ 10$^{12}$ n/cm$^2$ of 1 MeV energy.

We operated the arrays at constant current by fixing the operational point at 0.54 uA. This working-point corresponded to (Vop − Vbr) ∼ 0 for the irradiated SiPM, so that we biased the series without irradiated SiPMs (1) to Vop=166.4 V and the series with the irradiated SiPM (2) to Vop=162.7 V. In Fig. 4.7, the charge distributions of these two series are reported. Taking into account the ratio between the peaks of the charge distribution between SiPMs series (2) and series (1), we obtain R=Q(w_irr) / Q(w/o irr) =0.66 = 2/3. This result is compatible with the hypothesis (Vop − Vbr) ∼ 0 for the irradiated SiPM.

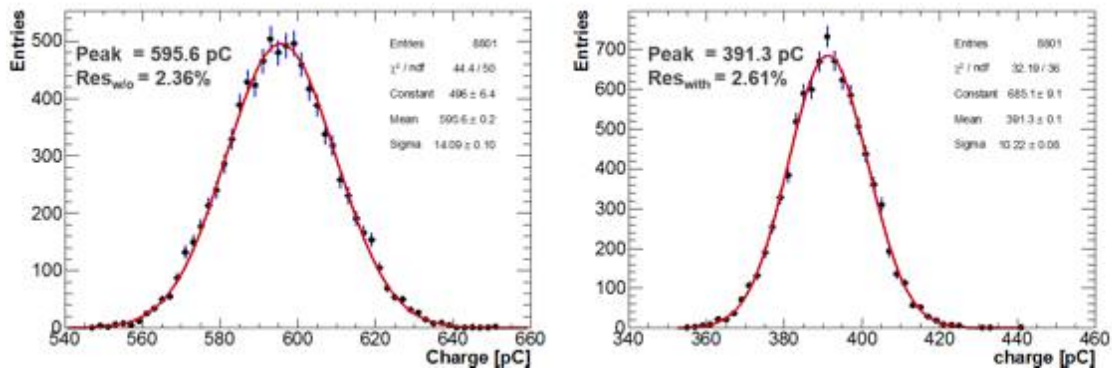

Figure 4.7: *Charge distributions for the series configurations with three not irradiated MPPCs (left) and two not irradiated MPPCs plus the one irradiated at Dresden with ~ 10^{12} n/cm^2 of 1 MeV energy.*

## 4.5    Irradiation tests with neutrons and dose

The Mu2e calorimeter must operate and survive in a high radiation environment. Simulation studies estimated that the sensors have to withstand an equivalent neutron fluency, at 1 MeV energy, of ~ $4 \times 10^{11}$ n/cm$^2$ and a Total Ionizing Dose (TID) of 20 krad [4-3]. These values assume 3 years of running, a safety factor of 3 and are calculated in the hottest irradiated region, i.e. in the innermost ring of the first disk. For this reason, we have tested our sensors before and after irradiation to measure the variation of the leakage current and of their response (i.e. the product gain x PDE).

A first irradiation campaigns was carried out in 2015, where different models of SiPMs have been tested with different protection covers: two from Hamamatsu, and one from FBK. More details on these tests can be found in ref. [4-4]. In this summary, we show only the measurements done with TID and with 1 MeV neutrons at HZDR, Dresden in 2016.  The general scheme of the experimental set-up used is shown in Fig. 4.8.

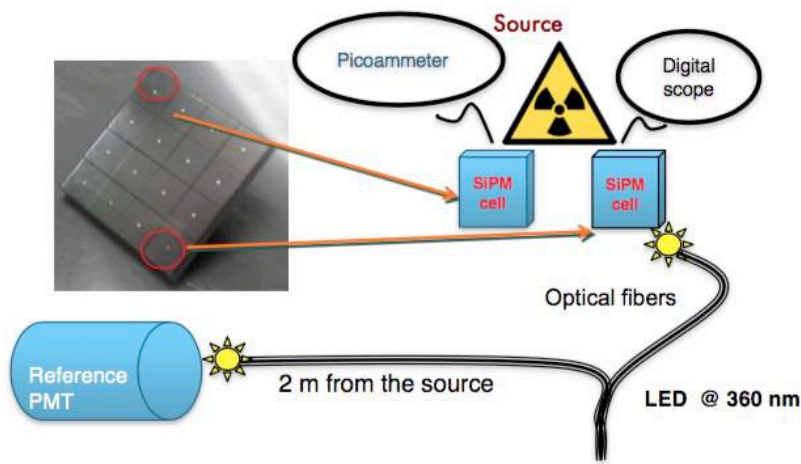

Figure 4.8: *General scheme of the setup used during the 2015/2016 irradiation campaigns.*

For the SiPM under test we acquired both the dark current with a Keithley picoammeter and the response to a fixed UV LED pulse, i.e. the gain x PDE, with a digital scope. In order to precisely monitor the input LED light, a UV photomultiplier has been illuminated with the same LED pulse by means of a split fiber and placed in a safe region located 2 m far away from the source.

### Dose irradiation test

Irradiation tests with a ionization dose have been performed with a strong $^{60}$Co source at the ENEA Calliope facility, Bracciano, Italy, where we could reach a dose from 200 to 100 rad/h at about 5 m distance from the source core. A small 3x3 mm$^2$ SPL SiPM was irradiated with these photons for three days until absorbing a total dose of ~ 20 krad. The dose effect on the SiPM performance is negligible both in term of leakage current and signal amplitude. As shown in Fig. 4.9, the leakage current before the irradiation was of ~ 0.1 uA, increasing to ~ 0.6 uA at the beginning of the irradiation, as due to the Compton effect on the SiPM active surface. In three days of irradiation the current increased to ~ 0.15 uA, thus practically doubling the initial dark current. The signal amplitude remained

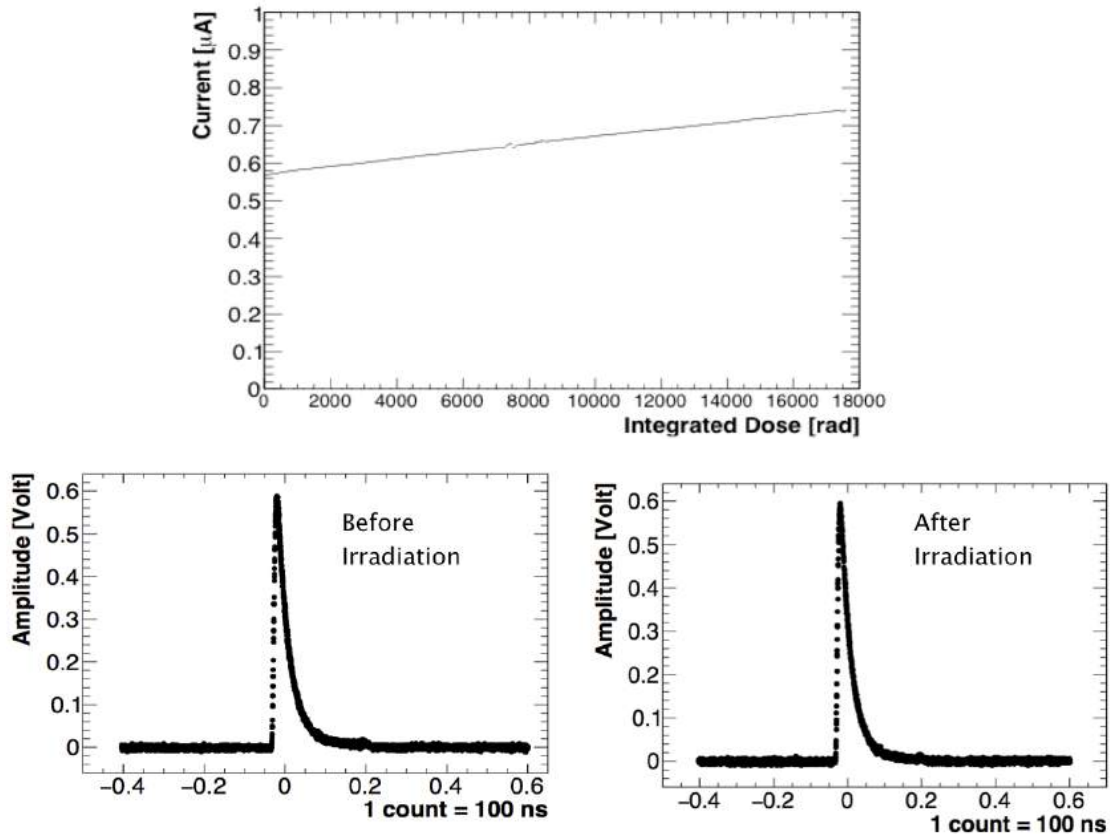

Figure 4.9: *Top: leakage current of a SPL SiPM as a function of the integrated dose. Bottom: SiPM amplitude before (left) and after (right) the dose irradiation.*

unchanged.

***Neutron irradiation tests***

During 2016, an independent irradiation test has been performed at the EPOS source (HZDR, Dresden) with 1 MeV neutrons. The experimental setup used is reported in Fig. 4.10: a 6x6 mm$^2$ SPL MPPC was located over the source in a place where no dose was present and with the active area positioned parallel to the incoming neutron flux. The SiPM was monitored both in current and in charge while illuminated with an UV LED. To maintain the SiPM temperature as stable as possible, the irradiated SiPM was connected to a Peltier cell, with the hot side glued to a cooling system. The SiPM temperature was monitored using a PT1000 sensor. To control and monitor the Hamamatsu devices, the same experimental setup reported in Fig. 4.8 was used.

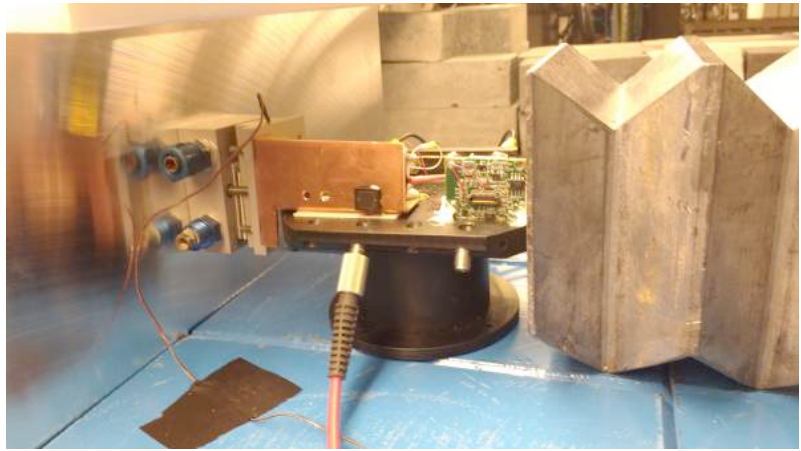

Figure 4.10: *Experimental set up used at HZDR center.*

The SiPM was biased at Vop=54.9 Volt. The total neutron fluency absorbed by the SiPM in five days was larger than 4x10$^{11}$ n/cm$^2$ of 1 MeV equivalent energy, that is three times the flux expected in the hottest region in 3 years of running. In Fig. 4.11, the irradiation results are reported as a function of the integrated flux. The signal peak decreased from ~ 650 mV to ~ 400 mV, as still due to the residual temperature variation. A rising behaviour of the SiPM leakage current, from 60 uA up to 12 mA, is instead clearly visible.

***Leakage current dependence from the temperature***

In order to study the SiPM properties as a function of its temperature, the SiPM irradiated at EPOS and one not irradiated SiPM have been tested in a vacuum chamber at 10$^{-4}$ mbar and cooled by means of a cascade of two Peltier cells. The SiPM temperature was monitored by a PT1000 sensor.

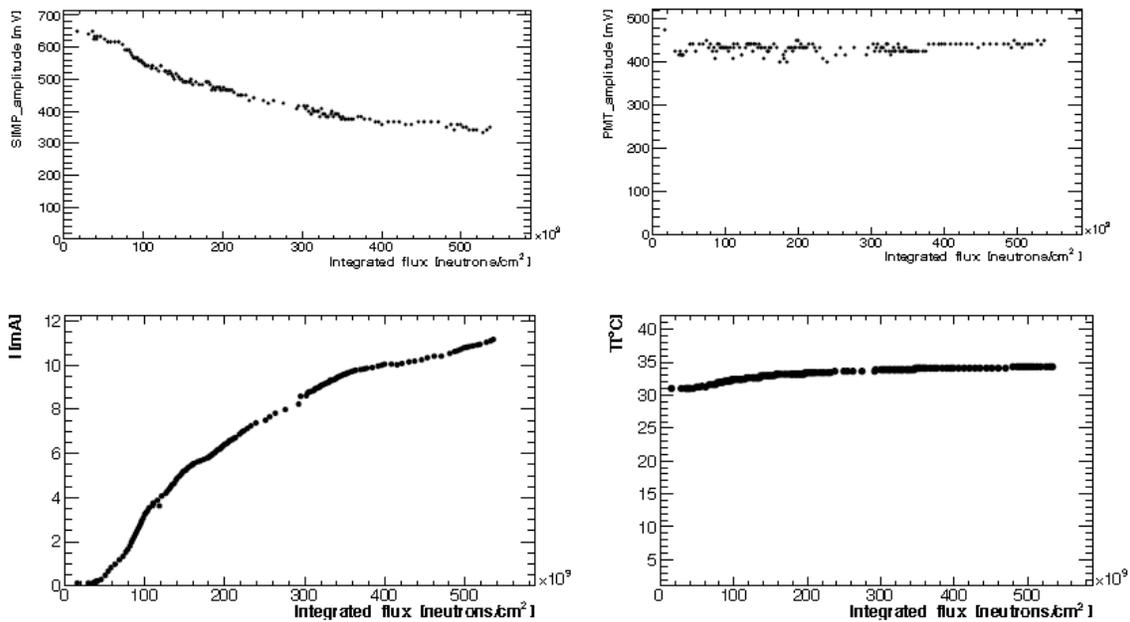

Figure 4.11: *Top: variation of SiPM (left) and (PMT) signal amplitude during the irradiation test at HZDR, in Dresden. Bottom: SiPM leakage current (left) and SiPM temperature (right) as a function of the integrated flux.*

A UV LED was illuminating directly the sample inside the vacuum chamber.

Using a pico-ammeter, the dark current was recorded and plotted as a function of the temperature. In order to maintain the gain constant during the test, the SiPM was illuminated with a LED and the peak of the signal pulse height acquired with a digital scope. When varying the temperature, we adjusted the operation voltage by keeping constant the signal amplitude. Results are shown in 4.12. The shape of the two distributions is similar but the dark current for the irradiated SiPM is larger of at least three orders of magnitude with respect to the one not-irradiated.

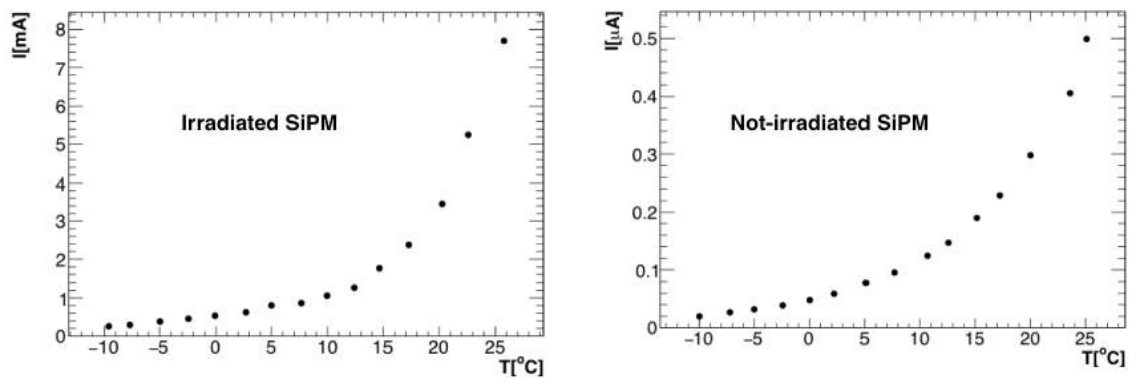

Figure 4.12: *Dependency of the SiPM dark current on the device temperature for: (left) an irradiated and (right) a not-irradiated SiPM.*

The bias supply of the front-end electronics requires the current of each channel to be smaller than 2 mA, so that it is necessary to run the SiPM at T ≤ 0 °C.

Since, as shown in Fig. 4.12, at 0 °C the current of a single 6x6 mm$^2$ SiPM reaches ~ 1 mA, and a serial connection will have the same current, we expect that in the parallel configuration of the two series a total current Id ~ 2 mA will flow, in agreement with our requirements.

## 4.6      Requirements and determination of MTTF for the Mu2e SiPMs

As explained in the introduction, one basic calorimeter requirement is to operate for at least one year with the whole apparatus inside the Detector Solenoid, without reducing performance. In this paragraph, we discuss the reliability of the Mu2e custom SiPMs. To represent their reliability, we calculate, by simulation, the needed Mean Time To Failure (MTTF) and we then describe the method developed to quantify it. The MTTF is defined as the mean time to the first failure under specified experimental conditions Evaluation of MTTF requirements.

Each calorimeter crystal is read by two identical Mu2e SiPMs, called Left and Right in the following. Each SiPM is connected to its own independent electronic chain in such a way that one sensor alone is able to satisfy the response and resolution requirements. To express the reliability as an MTTF value, we have developed an ad-hoc simulation. We assumed to be in the flat region of the sensor lifetime "bath-tube" distribution i.e. to have a constant dead rate. An exponential distribution with meantime, τ_sipm, describes the probability of survival of a single SiPM. As a starting example, we have set τ_sipm to 10$^6$ hours and simulated 1000 fake experiments. For each of these trials, we have extracted the time of dying, τ_dead, for 1360 Left and Right channels. If the value extracted is smaller than a limit set to 5000 (15000) hours, i.e. equivalent to 1 (3) years of running, we declare the SiPM "dead". The number of "dead-SiPMs" is shown in Fig. 4.13 left (right) for 1 (3) years of running. This can be expressed in words by saying that, for an assumed MTTF of 10$^6$ hours, we will have in average 2.6 (7.5) dead SiPms, with an r.m.s. of 1.4 (2.6), for the Left or Right side in 1(3) year(s) of running.

We use the same simulation to understand how the MTTF improves when we calculate the number of dead SiPM for the Left and Right side simultaneously in the same channel. The improvement is really large. In Fig. 4.14 left (right) the number of channels with both Left and Right SiPMs dead is shown for 1 (3) years of running; the average is reduced to 0.53 (0.79) with an r.m.s. of 0.16 (0.55), respectively. From this simulation, we have derived the MTTF needed to obtain a 95% (99%) probability of survival having at most one "double-dead" element in three years run. We obtained that we need an MTTF of 0.75 (1.5) x 10$^6$ hours for a 95 % (99%) survival probability.

At the moment of writing, the Mu2e calorimeter group is designing the front end electronics and the sensor cooling system to keep the SIPM to a running temperature

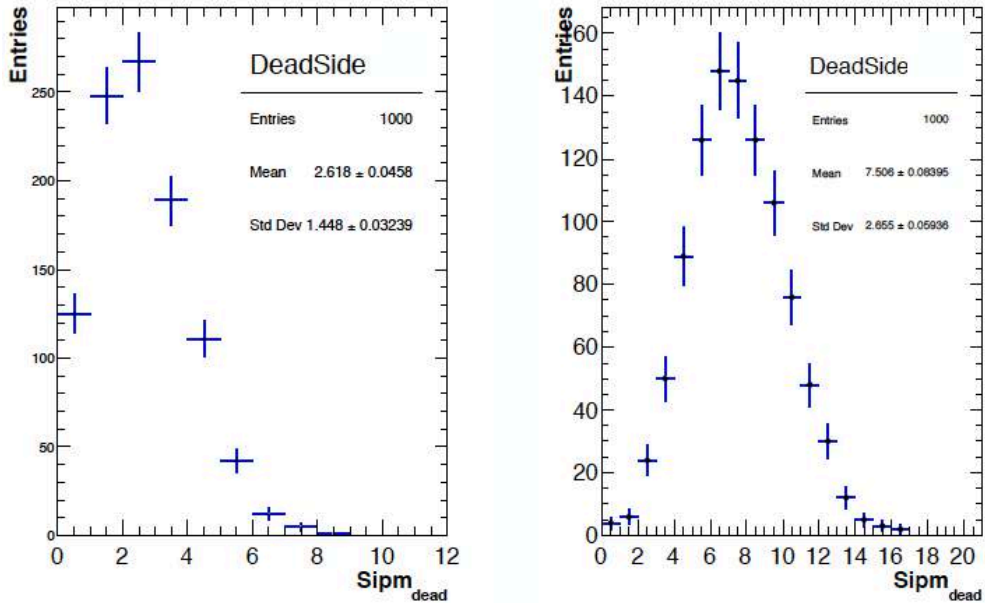

Figure 4.13: *Simulated number of dead SiPMs per crystal after one year (left) and 3 years (right) of running.*

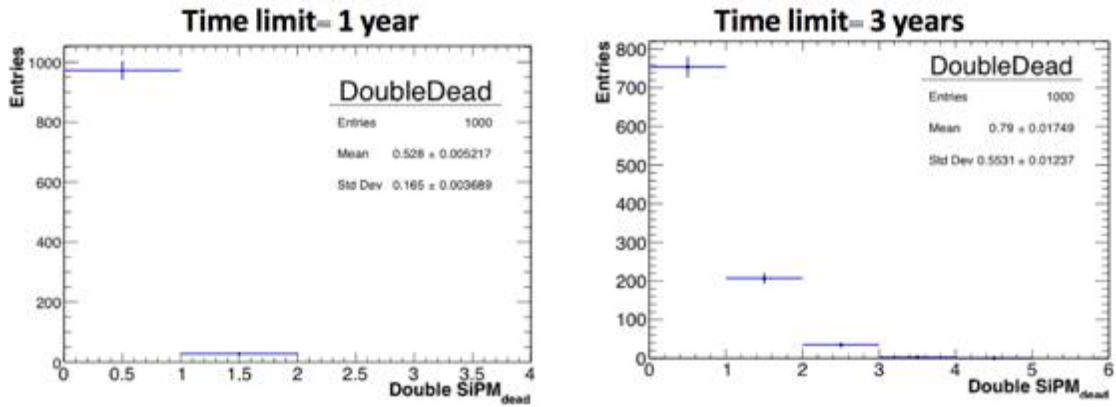

Figure 4.14: *Simulated number of double dead SiPMs per crystal after one year (left) and 3 years (right) of running.*

between -10 °C and 0 °C.

### Determination of MTTF

In literature, very high values of MTTF are reported for large area SiPMs. Ref [4-5] shows an MTTF estimate of $3 \times 10^5$ hours for running $6 \times 6$ mm$^2$ SiPMs from Hamamatsu at temperature of 25 °C. Since, we are going to run with the sensors cooled down to 0 °C, this estimate is improved with the Acceleration Factor (AF) calculated using the Arrhenius equation: $\text{AF} = e^{Ea/k(1/T\_use \ -1/T\_eval)}$, where $E_a = 0.7 \ eV$ is the Silicon activation energy of the failure mode, $k = 8.617 \times 10^{-5}\ eV/°K$ is the Boltzmann constant, $T_{use} = 273\ °K$ is the running temperature during the experiment and $T_{evaluation} = 300\ °K$ is the temperature used in the MTTF evaluation. This corresponds to an AF of 14. Scaling down also for the ratio between the SiPM area of ref.xx (6x6 mm$^2$) and the Mu2e SiPM area (6 devices of 6x6 mm$^2$), we estimate an MTTF of ~ 1x10$^6$ hours.

Since the SiPMs we are used are custom, we have developed our own control station to evaluate the MTTF of our sensors. The test is in progress at LNF (Italy) and involves 15 pre-production SiPMs (Silicon Photo Multipliers) from the three different vendors: Hamamatsu, SensL and Advansid-FBK. The stress-test is performed by keeping the SiPMs in a light tight box at 50 °C for three months. The dark current is measured once a day by a picoammeter and the response to a blue led is acquired every two minutes. Assuming no deads, it is expected to reach an MTTF value of 0.6 x10$^6$ hours. This value is evaluated as follows: MTTF> 0.5×N$_{hours}$×AF×N$_{SiPM}$, where N$_{hours}$= 2190, N$_{SiPM}$=5 and the Acceleration Factor (AF) estimated 9by the Arrhenius equation, with T_use=273 K and T_stress = 323 K is 140.

### 1. Description of the set up

A sketch of the box used for the MTTF is reported in Fig. 4.15.

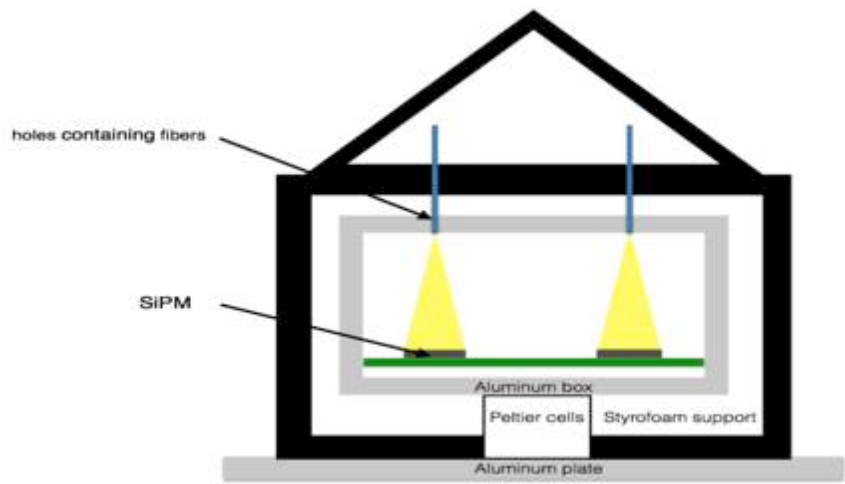

Figure 4.15: *sketch of the MTTF box used in LNF.*

The 15 SiPMs are biased at three different voltage supply, the operation voltage at 25 °C is V$_{op}$= 163.5 V for the Hamamatsu, V$_{op}$=82.5 V for the SensL, and V$_{op}$=88.5 V for Advansid-FBK. In Fig. 4.16, the pictures of the setup and of the board used to mount and polarize the SiPMs is shown. The circuit board is inserted inside an aluminum board in thermal contact with a system of two Peltier cells, allowing to maintain the temperature stable at 50°C. The temperature is constantly monitored by means of a PT1000 sensor. To illuminate the SiPMs the aluminum box presents 15 holes to insert the optical fibers, which diffuse the blue led light. The LED is driven by a pulse generator supplying a 10 V pulse of 100 ns length. The aluminum box is surrounded by a Styrofoam support avoiding heat dispersion.

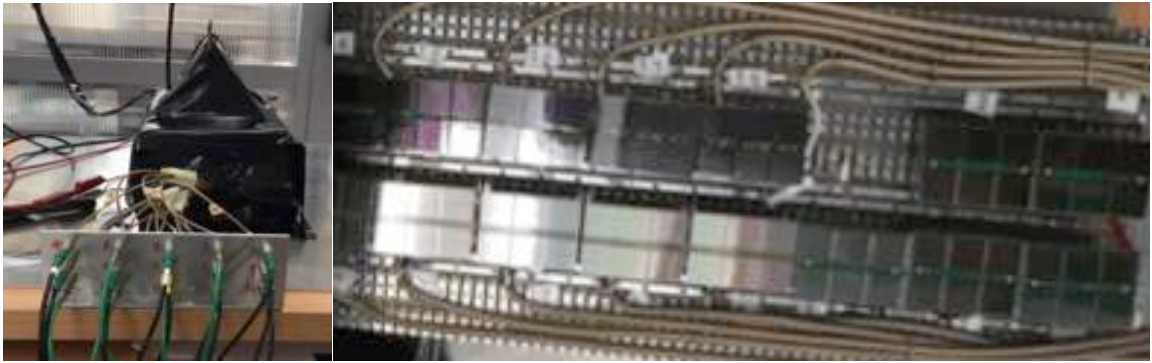

Figure 4.16: *(Left) Experimental setup used for the MTTF test. (Right) Zoom of the box*

In Fig. 4.17, the distributions of the response to a LED pulse for five of the SiPMs under test are shown. We have started running the system at 25 °C with the bias voltage set at the Vop provided by the vendors. We have then slowly increased the temperature while adjusting the voltage to keep the same response. In Fig. 4.18, the percentage variation of the response observed for all SiPMs under test for a temperature variation of +25 °C is shown. In Fig. 4.19 the behavior of the response along the time for the 5 SensL sensors under test is also shown for a running time of 430 hours. The dip observed at the beginning of the run is related to an intervention to fix better the optical fibers inside the supporting holes. No serious changes in current are observed for the overall period of test.

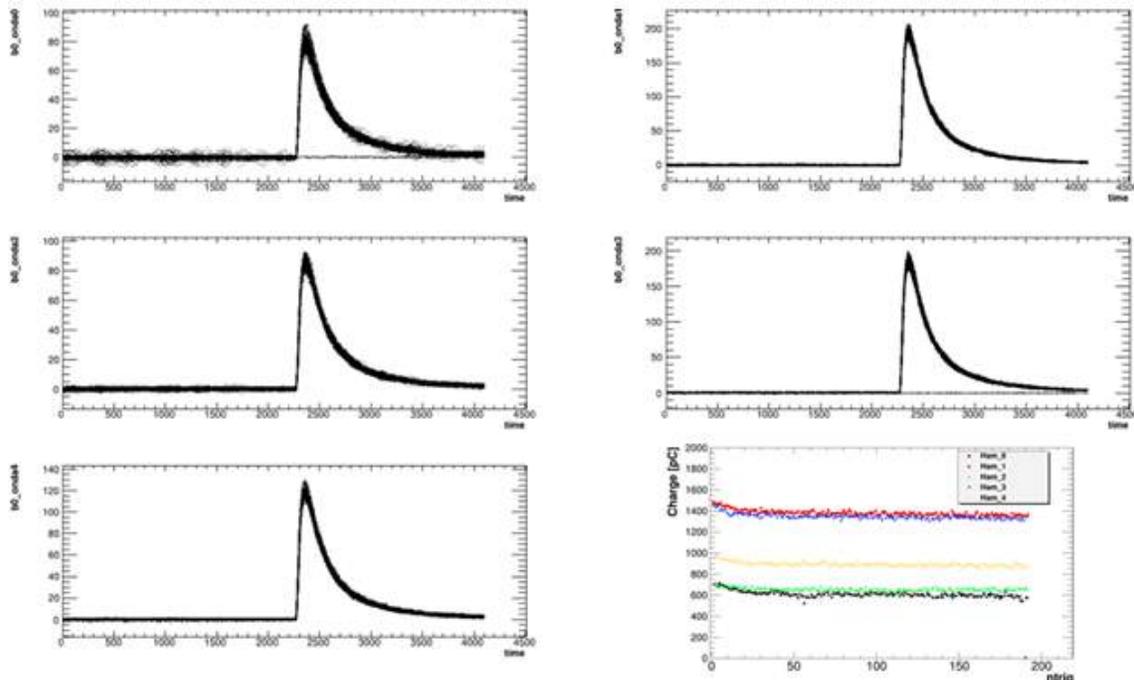

Figure 4.17: *Signal amplitude reported in mV and charge of the 5 SiPM of Hamamatsu at 25 °C*

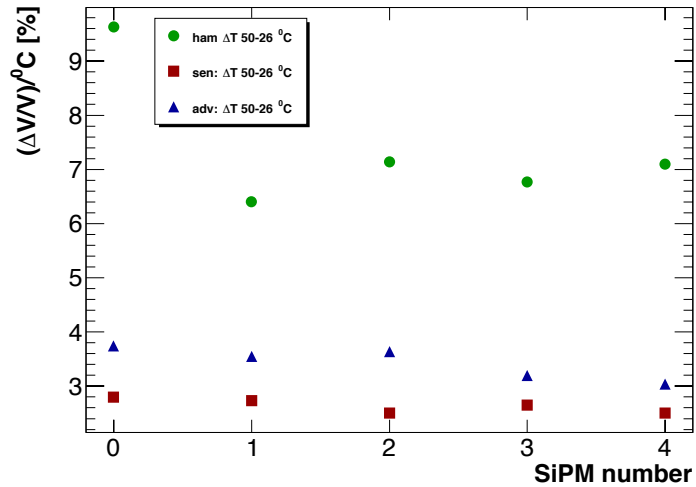

Figure 4.18: *Percentage variation of the response observed for all SiPMs under test for a temperature variation of +25 °C.*

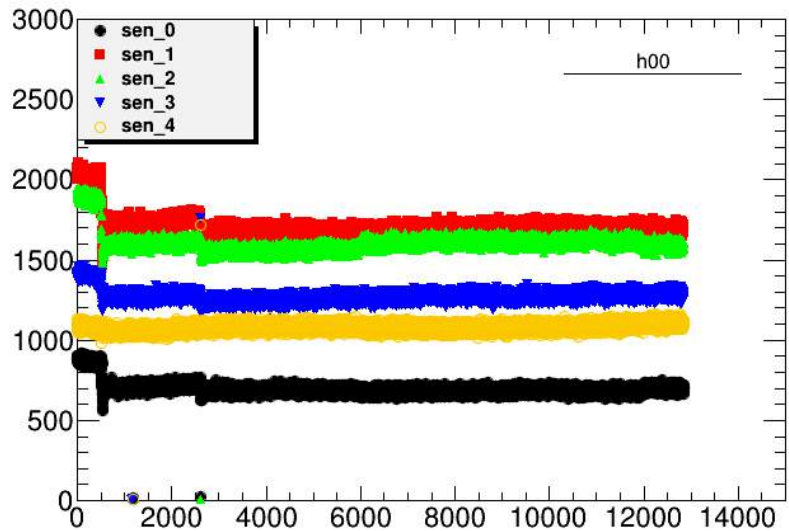

Figure 4.19: *Charge of the 5 SensL SiPMs as a function of the trigger number (1 triggered pulse each 2 minutes).*

# Chapter 4 References

# 5   The Mu2e Calorimeter electronics

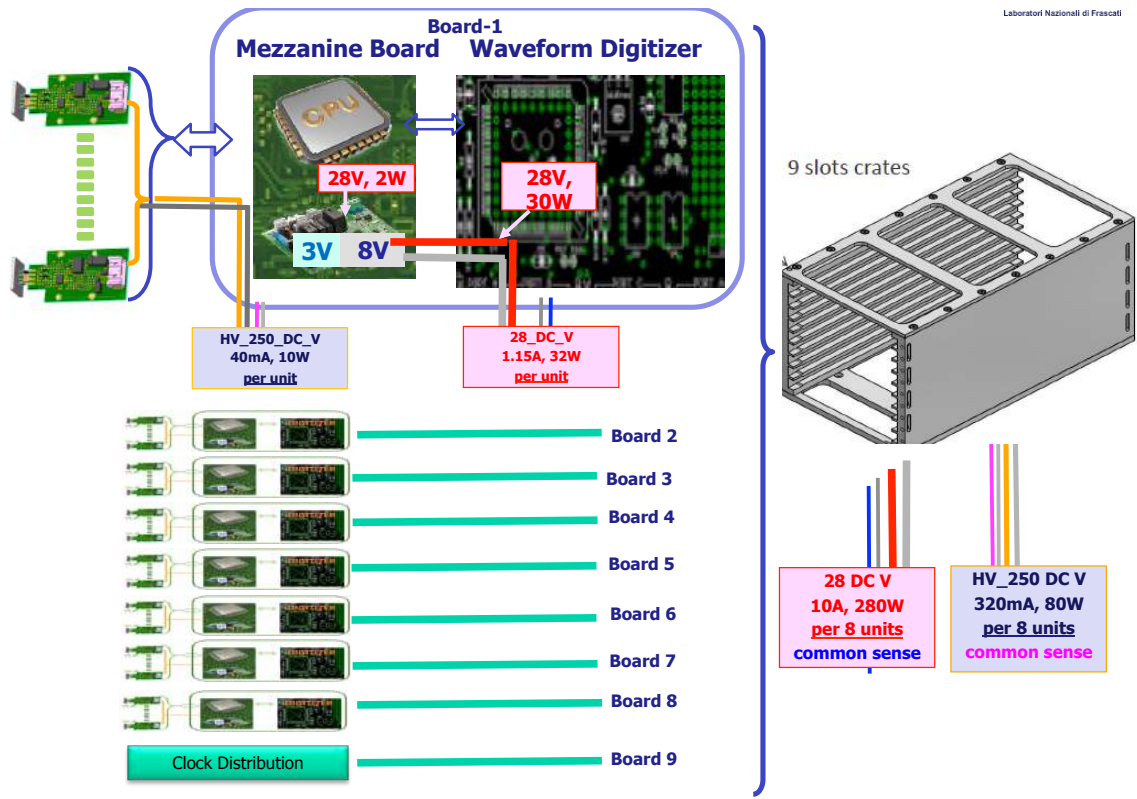

**Figure 5.1** *Overall schematic of the calorimeter electronics. The drawing represents the distribution of the electronics and the connections to one crate. Each crate can host up to 8 MB+WD boards and 1 clock distribution board.*

## 5.1   Overview of the electronics scheme

The overall scheme for the calorimeter readout electronics is shown in Fig. 5.1. The front-end electronics (FEE) consists of two discrete and independent chips (Amp-HV), for each crystal, directly connected to the back of the SiPM pins. These provide the amplification and shaping stage, a local linear regulation of the bias voltage, the monitoring of current and temperature on the sensors and a pulse test. Each disk is subdivided into 34 similar azimuthal sectors of 20 crystals. Groups of 20 Amp-HV chips are controlled by a dedicated mezzanine board (MB), where an ARM controller distributes the LV and the HV reference values, while setting and reading back the locally regulated voltages. Groups of 20 signals are sent in differential way to a digitizer module (WD) where they are sampled and processed before being optically transferred to the DAQ system. The Detector Control System parameters read out/set by the MB are passed via I2C to the WD that then communicates them to the DCS system through an optical link.

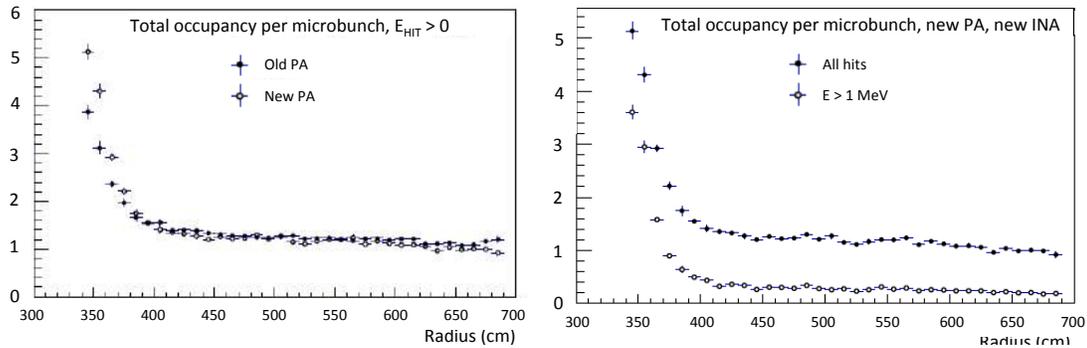

**Figure 5-2.** *Occupancy (hit crystals/microbunch) as a function of the calorimeter radius (left) and the dependence of the occupancy on the applied threshold (right). Data points are shown for different Proton Absorber (PA) and Inner Neutron Absorber (INA) configurations.*

The average input current driven by the SiPM and provided by the FEE depends on the distribution and on the average values of the background hits and on the instantaneous, as well as total, absorbed dose and neutron flux. In this section we summarize these 3 contributions:

1. **I(flash).** The calorimeter hits are due to two sources: (1.a) the flash of particles, called the "beam flash", produced within 200 ns of the interaction of the proton beam in the production target, and by the cascade of decays and interactions with the surrounding material, and (1.b) the background events generated by the muon beam interacting with the collimators, muon stopping target and the beam dump. From the simulation, the sum of all hits per channel corresponds to an equivalent energy deposition of ~ 130 MeV for each micro-bunch in the innermost ring of the first disk, 70 MeV of which are due to the beam-flash. The largest source of hits after the prompts dissolve comes from neutrons generated by muon capture, which produces an occupancy of more than 1 hit/channel. In Fig. 5.2.left, the calorimeter single channel ("cell") occupancy as a function of disk radius is shown. In Fig. 5.2.right, the occupancy is reduced to ~10% by applying a 1 MeV threshold. This environmental background will also create an average current from the SiPM, that is dominated by the beam flash and is equivalent to $I = N_{pe} \times MBR \times G_{SiPM} \times e$, where $N_{pe}$ is the average number of photoelectrons, MBR is the micro-bunch rate, $G_{SiPM}$ is the SiPM gain and e is the elementary electrical charge. In the CsI case, using $N_{pe}$= 70 MeV x 25 p.e./MeV = 1750 p.e., MBR = 600 x$10^3$ Hz and $G_{SiPM}$ = 0.6 x $10^6$, we obtain an I of around 100 μA. This "flash-current" is reasonably small compared to the other effects, **I(RIN) and I(neutrons),** and can also be compensated due to the fast recovery time of SiPM and FEE once providing the right capacitance values in the electronics.

2. **I(RIN).** The instantaneous dose seen by the crystals is instead the largest source of the RIN (sec. 3.6) that corresponds to an average "phosphorescence-like" current in the crystals and in the connected SiPM. Also this current reaches its highest value in the innermost ring of the first disk. Assuming to select, for this ring, the crystals with the smallest value of RIN, as measured with 2" PMTs, of $\sim (2 - 4) \times 10^9$ p.e./sec/rad/h, this corresponds to a SiPM current of $I_{SiPM}$ (RIN) = $N_{pe} \times I_{DOSE} \times G_{SiPM} \times \varepsilon_{SiPM}$= 0.6-1.2 mA, when using a Mu2e instantaneous dose, $I_{DOSE}$ of 3 rad/h and a product between collection and quantum efficiency $\varepsilon$(SiPM) = $0.2 \times 0.15 = 0.3$ and $G_{SiPM}$ of $0.6 \times 10^6$.

3. **I(neutrons).** While it is reasonable to assume that I(flash) and I(Rin) will remain stable along the whole data taking period, the most dangerous source of current increase along the time is the one related to the neutron flux irradiating the innermost crystal ring in the first disk. As shown in sec. 4.5, the dark current will quickly grow of large factor due to the irradiation and we will need to cool down the sensors around 0 C. Indeed we expect, with a factor of 3 safety, that we will handle SiPMs with current as large as 1-2 mA.

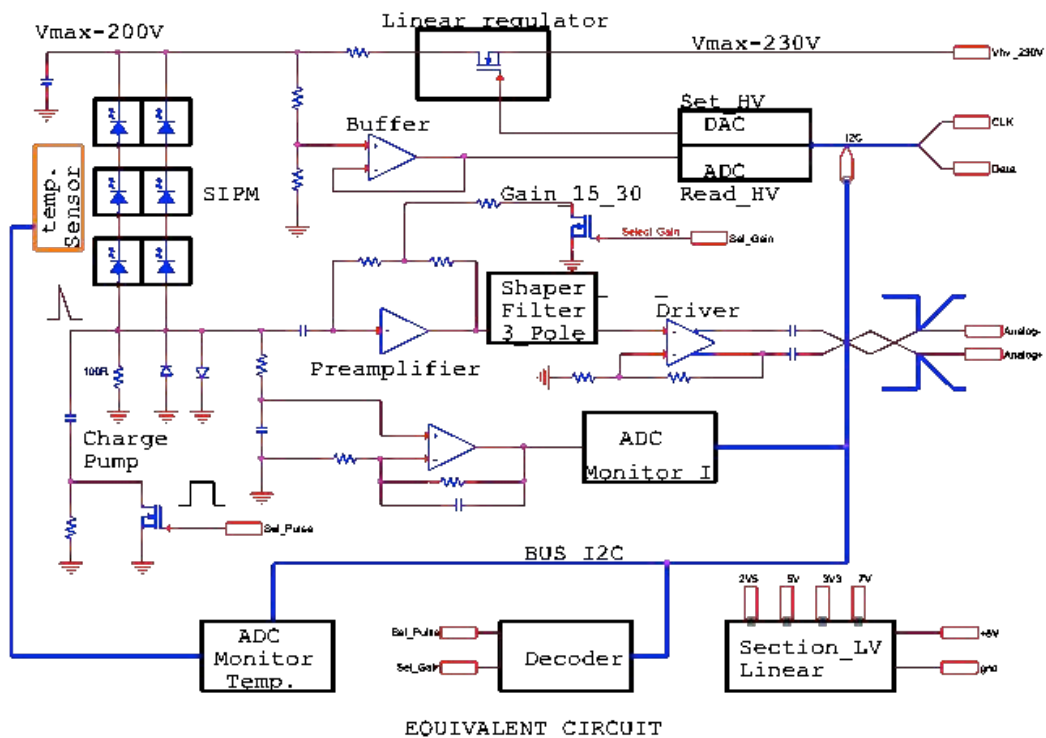

**Figure 5.3:** *Layout of the equivalent circuit for the AMP-HV chip.*

The rate capability and the pileup contribution are instead dominated by the source 1.b. As shown in Fig. 5.2.left, in the highest occupancy region we expect an average of 1 hit/MB in the 700 ns of "daq-acquisition" window. In the beam-ON period, we therefore have 550 kHz of beam-flashes and hits in pileup in the daq-window. This will be further discussed in the section of digitizer modules.

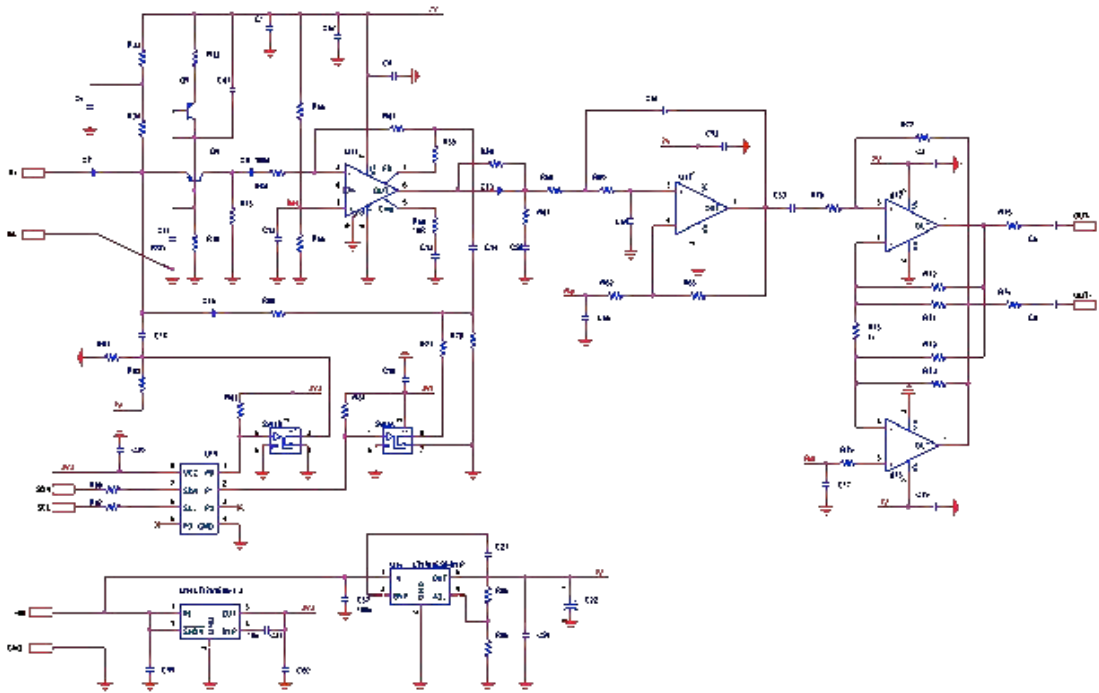

**Figure 5-4:** *Electrical scheme of the AMP-HV amplifier.*

## 5.2 FEE: The Amp-HV chip

The Amp-HV is a multi-layer double-sided discrete component chip that carries out the two tasks of amplifying the signal and providing a locally regulated bias voltage, thus significantly reducing the noise loop-area. The two functions are each independently executed in a single chip layer, named the Amp and HV sides, respectively. The AMP-HV chips are asked to provide: (i) two settable amplification values ($10 - 20$) with low noise, (ii) a signal rise time comparable with 3 to 5 times the WD sampling time for good time resolution and (iii) a short falling time to improve pileup rejection, (iv) a low detection threshold below the MeV level, (v) a high precision and stability in regulating and keeping the operation voltage (vi) a capability of operating in a rate environment of 500 kHz/channel without diminishing its gain, shape and pileup rejection capability, (vii) a stable output regardless to the average current due to Idark from the SiPM or radiation induced noise in the crystals assuming those to be contained below the mA, (viii) a stability to the response also in presence of 500 kHz beam flash signals and, finally, (ix) a low power consumption. Moreover, the AMP-HV chip has been added with three independent monitoring capabilities: (M1) the readout of the SiPM temperature, (M2) the readout of the current drawn in the two series and (M3) the generation of a pulse signal to test the amplifier gain. The three latter monitoring capabilities, as well as the readout and setting of the operation voltages, are controlled by means of I2C.

The development of the Amp-HV chip has been done by the Laboratori Nazionali di Frascati (LNF) Electrical Design Department. This final design has been built upon a long R&D phases carried out for the different calorimeter prototyping stages: forty prototypes were built during 2013 and have been used for testing a LYSO matrix readout with APD [5-1], 10 prototypes were build in 2015 for the test of BaF$_2$ readout with solar blind APDs and finally 150 prototypes (version-0) are now in preparation for the module-0 assembly with CsI and custom Mu2e SiPMs (c.f.r. sec.9). In Fig. 5.3, the architecture of the system is shown. The chip is constituted by a 8 layers PCB of compact dimension 30x60 mm$^2$. On the Top layer, there are located the preamplifier, the shaper filter and a differential buffer as well as the digital logic (DAC and ADC) to set and read back the bias voltage on the local regulator, read the local temperature and create a pulse signal. On the bottom layer, there is the local regulator for the bias voltage. The regulator is in linear technology, it can provide a maximum of 2 mA at 200 V and is fully programmable by remote.

### 5.2.1   The Amplification layer

The specifications for the amplification layer have been developed and tuned to work with the Mu2e SiPM connected to an undoped CsI crystal. Minor adjustments to the gain and power dissipation parameters will be implemented in the final production run.

The electronic scheme is that of a transimpedance preamplifier, with a final trans-impedance settable gain of 500 or 1k$\Omega$ (voltage equivalent, $V_{out}/V_{in}$ of 10/20). The basic characteristics are described in Tab. 5.1, while the schematic of the preamplifier circuit is shown in Fig. 5.3. To optimize the amplifier stability, an input cascade stage has been add. The output signal in U11 is shaped by a three poles + 1 zero Shaping Filter (U17) whose output has 18 ns rise time and a 100 ns falling time. This assumes in input a signal with a fast rise time and a falling time of 250 ns. The final output is provided by a buffer differential stage, DC balanced, (U12,U13) that drives the signal over a 100 Ohm differential cable. The power dissipation of the preamplifier is of 70 mW maximum.

### 5.2.1   The Linear Regulator layer

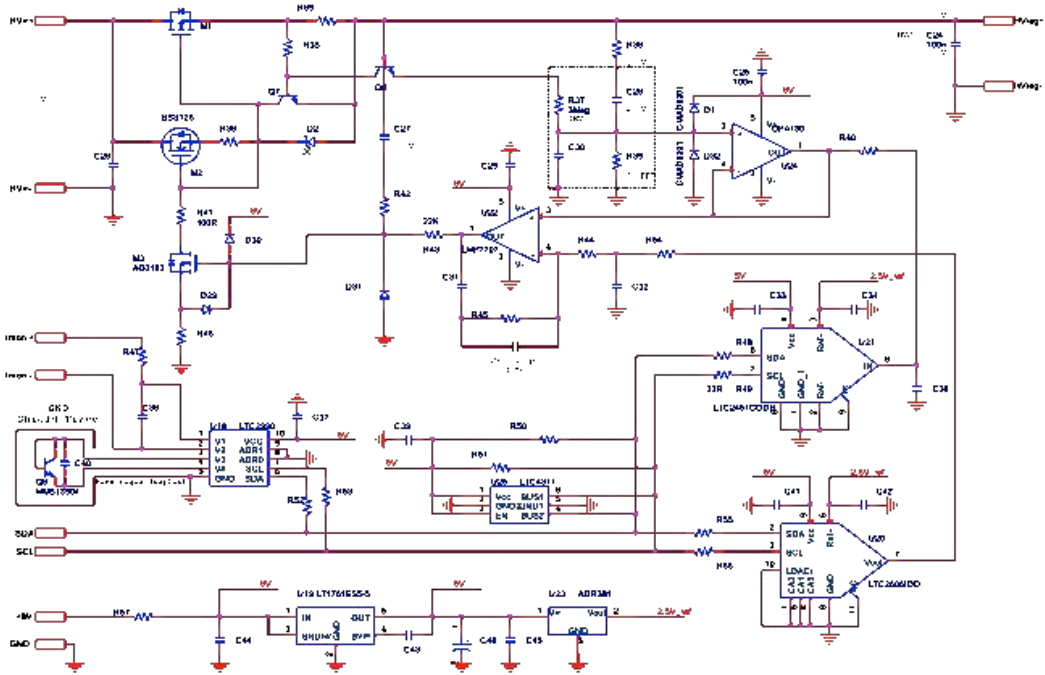

Figure 5.4 *The AMP_HV regulator layer: electrical scheme*

The linear regulator is required to provide extremely precise 16-bit voltage regulation and long-term stability of better than 100 ppm. The current limit of the SiPM is set to be at most 2 mA to match the situation where the sensors have been highly irradiated with neutrons. The list of characteristics measured on the first prototype are summarized in Tab. 5.1. The basic schematic of the linear regulator is shown in Fig. 5.4. The high voltage required for the Mu2e SiPM is produced by COTS primary generators residing outside the DS, providing 1 HV cable service for each 4 Mezzanine Boards. The local distribution of HV on the controller board is sufficient to power 20 channels in parallel. The highest requirement for the regulator is that of having an excellent stability also for high dynamic variation in the input current (from 100 uA to 2 mA). The serial regulator is constituted by a Mos M1, controlled by a depletion reference stage that modulates the voltage. A feedback circuit (differential U22) keeps the output voltage stable with respect to the reference voltage. A layer of passive elements (Q7) works as a current limiter to avoid supplying more than 2 mA. The output voltage is regulated by a DAC and is then read out again via an ADC, with 16-bit accuracy. An ADC with 14 bit and two differential channels are used for reading out the temperature and the current of the SiPM. The grounding scheme has been developed by the calorimeter group in accordance with the basic safety and ground rules decided with the Mu2e electronics integration groups but is not reported here. A detailed description can be found in ref. [5-2].

***Table.5.1. Characteristics of the Amp-HV chip: (left) for the amplification side and (right) for the linear regulator side.***

| • Dynamics Differential | ± 1 V | • Dynamic range Vout | 5-200V |
|---|---|---|---|
| • Bandwidth | 40 Mhz | • Reading and writing Vout | 16 bit |
| • Rise Time | 10 ns | • Current limit | 2 mA |
| • Output impedance | 100 Ω | • Stability DT < 10 C | 50 ppm |
| • Stability with source capacity - max | 1 nf | • Settling time | 100 μs |
| • Coupling output end source | AC | • Power dissipation max | 350 mW |
| • Noise, with source capacity of 1 pf | ---- enc | • Ramp-up programmable | 1-100 ms |
| • Power dissipation | 45 mW | • Ripple | 3mVpp |
| • Power supply | +8 V | • I current precision | 50 μA -2 mA |
| • Rise Time after 3 pole shaper | 16 ns | • Temperature range | -40, 50 °C |
| • Input protection over-voltage | 10 mJ | • Remote Control | I2C |
| • Injection pulse –fixed amplitude | 300 mV | • GND insulated to earth | Max 5pf |

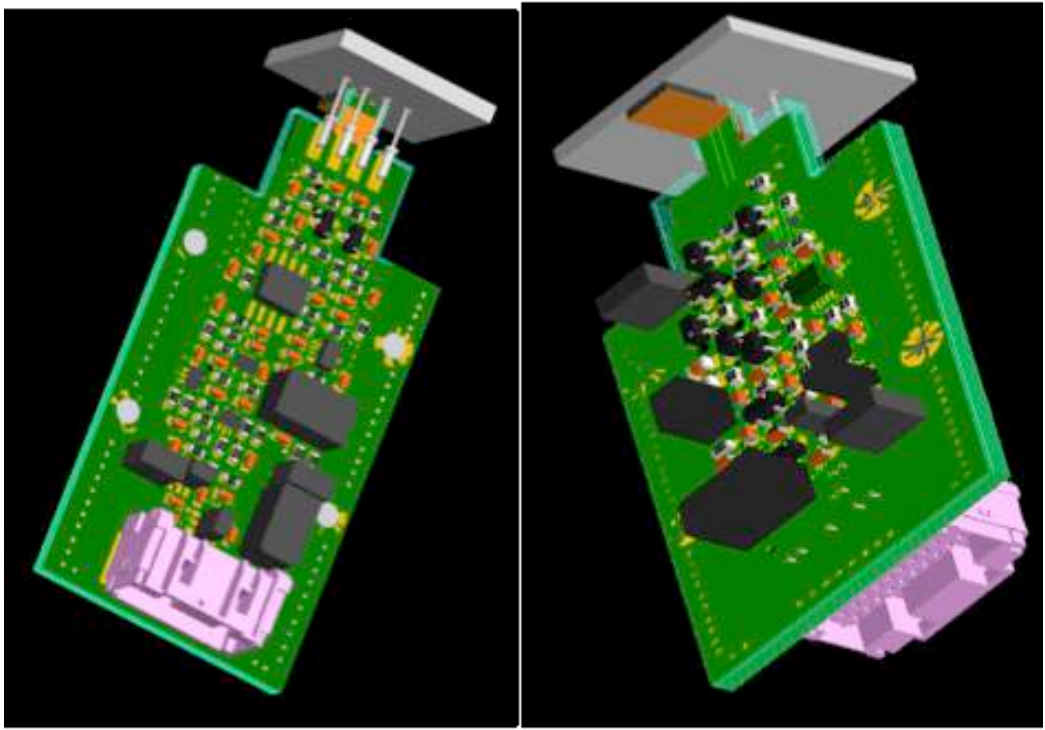

Figure 5.5 *CAD drawing of the two layers composing the AMP-HV chip*

### 5.2.2   FEE: the cooling of the Amp-HV chip

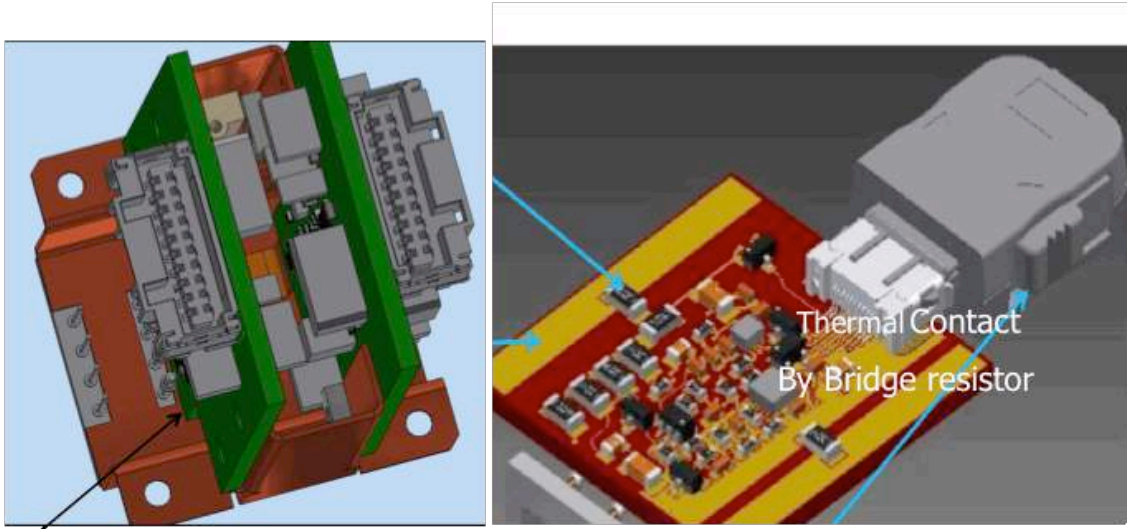

Figure 5.6 (*Left*) *Assembly of the two FEE boards inserted in the mechanical SiPM-FEE holder (right) example of thermal connection by Thermal Bridge Resistors.*

Integrating the cooling and the mechanics has been fully addressed for the Amp-HV chip, as shown in Fig.5.6-left.  Since the average power to be dissipated is ~350 mW per channel, the ground cannot be connected directly to the shielding surface or to the cold fingers.  Therefore the use of a bulk thermal bridge resistor [5-3], with a 1 pf capacitance, capable of transferring heat from the Amp-HV chip to the nearby mechanical structure, is used. The bridge resistor keeps the ground safely disconnected from the mechanical structure.  A CAD drawing of how this solution is implemented is shown in Fig.5.6-left where the insertion of the AMP-HV chip inside the cupper support for SiPMs and FEE is shown. The external shielding acting as Faraday Cage is not reported in this drawing. In Fig. 5.6-right the details of the Bridge Resistor connected to the cold finger are shown. A dedicated description of the way in which the cooling system works and the Ansys calculation of the heat removal are reported in the mechanics chapter.

### 5.3   FEE: The Mezzanine board

The Mezzanine board, MB, has many tasks to fulfill but in general is the interface between the FEE chips and the Digitizer board while performing all the needed Detector Control System steps. The MB should: (i) receive in input the 20 differential FEE signals and provide them to the digitizer board, (ii) set all the FEE parameters such as HV, preamplifier gain and pulse register, (iii) monitor HV and SiPM temperature and current, (iv) provide the low voltage for the FEE. All the DCS information are handled by a CPU system architecture, consisting of a series of ARM [5.4] processors, as shown in Fig. 5.7. The ARM processor controls each of the 20 connected Amp-HV cards.

In the NIM MB Prototype shown in Fig. 5.8, all these operations are done through firmware in the CPU via a standard Ethernet connection. Moreover, a local HV generator has been add for testing purposes. This generator will not work on B-field and, in the experiment, the primary HV source is coming through the IFB from external power supplies and is then distributed via the Mezzanine board. Multiple systems are connected via a hub. In the final board, whose first drawings are shown in Fig. 5.9, this will be done by I2C and then transferred to the WD. The WD will then communicates to the TDAQ by means of an optical link. Each FEE chip is connected to the ARM controller through a 50-200 cm cable. The cables are required for the transport of feeds for low and high voltage, as well as for the signals of the I2C control. A first version of the FEE cable [5.5] has been selected and it is under test right now.

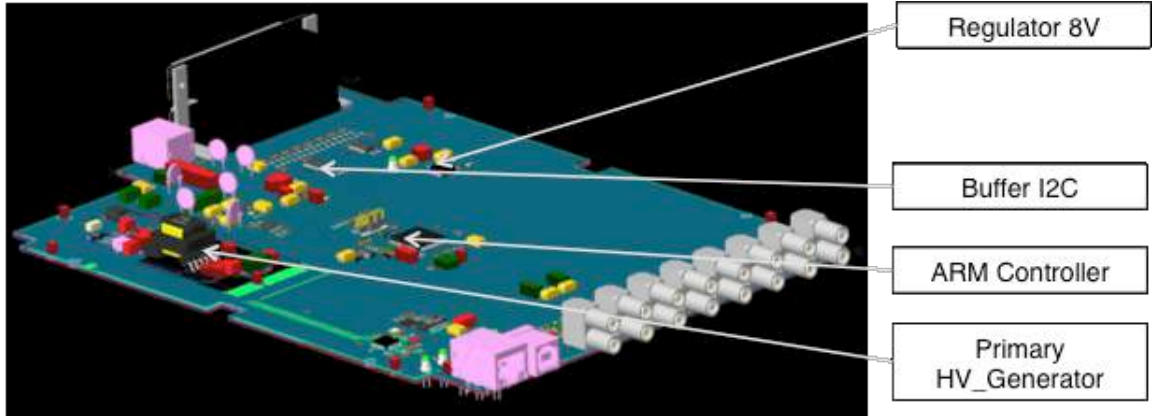

Figure 5.8: *Board schematic of the NIM MB prototype.*

Fig. 5.9 shows the first layout of the MB board to define the interface with the WD board.

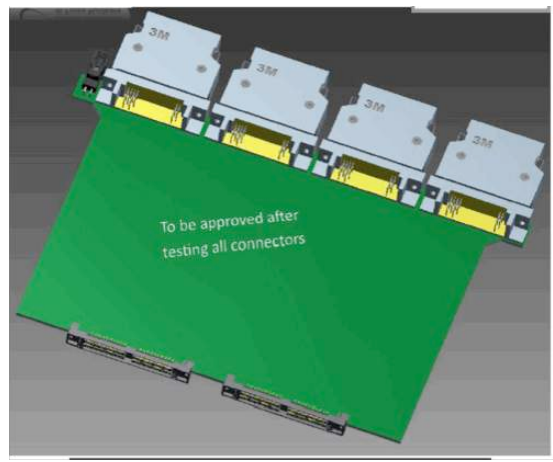

Figure 5.9: *first dimensioning of the MB board layout.*

## 5.4 FEE and Mu2e SiPM slice test

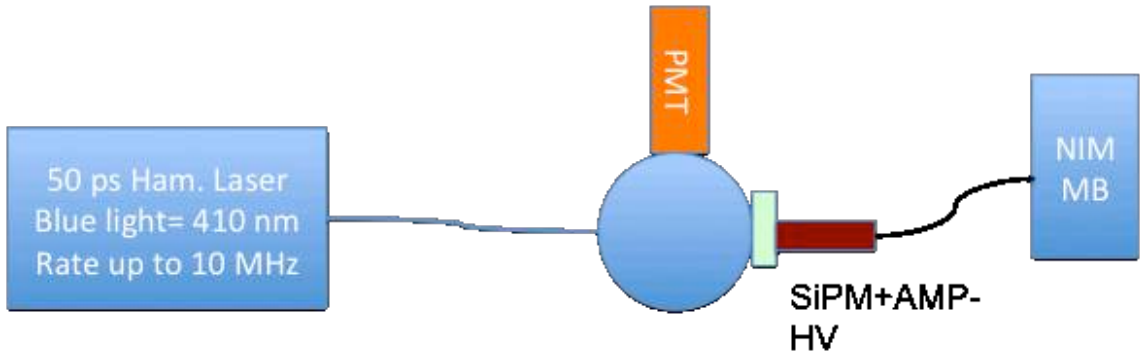

Figure 5.10: *FEE+SiPM slice test setup.*

To test the AMP-HV chip and the firmware of the MB NIM prototype, we have setup a small slice test by connecting the FEE to one Mu2e (Hamamatsu) SiPM that is illuminated by means of a light pulsing system. In Fig. 5.10 and Fig. 5.11 the schematic and a picture of the setup are shown. As light source, we have opted for a blue Hamamatsu laser that produces a very narrow light signal (FWHM=50 ps) and is easily adjustable in pulsing rate (from few Hz to 10 MHz) as well as in power output.

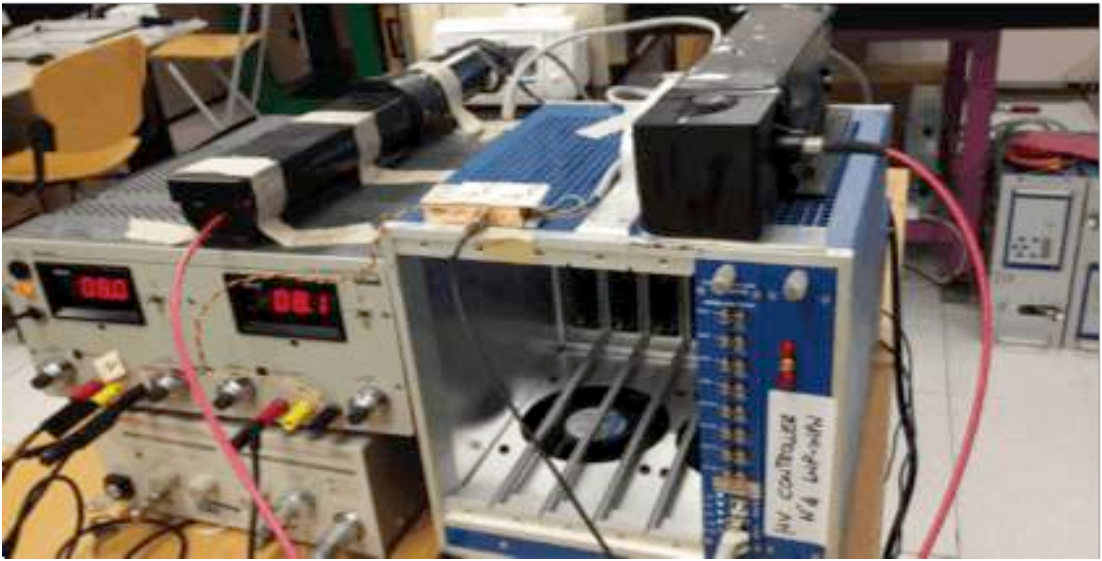

Figure 5.11: *Picture of the setup used for the FEE+SiPM slice test.*

The Laser light is brought by means of an optical fiber with SMA connector to the input of a 2" Thorlab diffusing sphere. Two output ports of the sphere are used to illuminate both the Mu2e SiPM under test and a reference 1" UV extended Hamamatsu PMT. The SiPM is connected to the FEE and inserted in a square black container in order to reside in a light tight environment. The FEE cables is inserted in the first input of the NIM MB prototype that is connected via Ethernet to a notebook. In Fig. 5.12 (left), an example of the signals observed at a digital scope are shown. In Fig. 5.12 (right) the first version of the DCS page for the setting and reading of the FEE information is also shown.

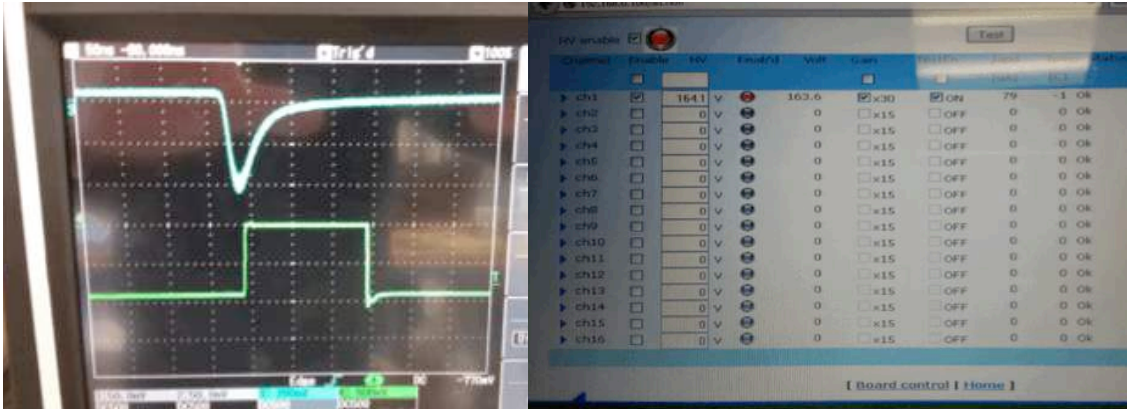

Figure 5.12: *(Left) SiPM+FEE signals at the digital scope and (right) Control web page for the NIM MB prototype.*

In the tests reported here, we have operated in a clean room at a temperature of 25 °C and RH of 45%. We have biased the SiPM to its operating voltage of 164.1 V and selected a preamplifier gain of 10. The dark current drawn by the SiPM was monitored in the MB web-page and, before pulsing, it reached a stable value of ~ 2 uA. The sensor signal after amplification and shaping was fed to a 1 Gsps CAEN desktop digitizer, after being properly adapted from differential to single ended input. Since the FEE dynamic range, in single ended connection, is 2 V while the CAEN desktop has only 1 V range, the signal under test was reduced by a factor of 2 by means of a passive divider. The PMT high voltage was set to 800 V following its specifications.

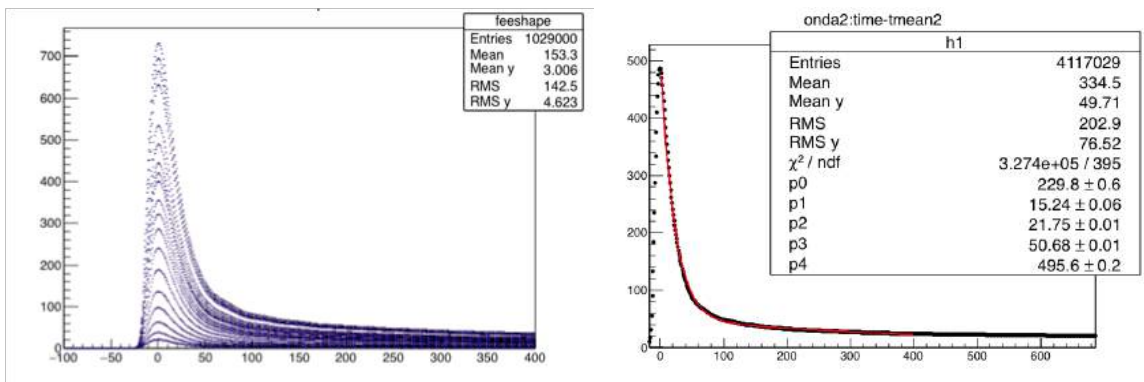

Figure 5.13 *(Left) distribution of signal shapes for SiPM+FEE for different values of the laser output, (right) signal shape for 1 V output. The fit with a sum of 2 exponentials is shown.*

We have divided the tests in 3 groups. In the first two groups, we have kept the same laser pulsing rate (few kHz) and made a scan in power output to study: (i) the signal shape and (ii) the linearity and resolution of the system. In the third group, we have instead kept stable the laser power output (with a signal in the middle of the dynamic range) and made a scan in pulsing frequency. A summary of the results found is shown in the next two sub-sections.

### Signal shapes

In Fig. 5.13, the distribution of the signal shapes, for the 20 points of the amplitude scan, is reported. Three clear points can be extracted from this plot:

1. The signal shapes for different amplitudes are similar and do not show any deterioration or saturation. Additional tests demonstrated that a first hint of saturation is observed around 2 V;
2. The rise time is around 18 ns as in the specifications of Tab. 5.1;
3. The falling time is composed of two slopes: the fast component presents a $\tau$ of 22 ns, while the long component has a $\tau$ of 500 ns.

While the shape stability in amplitude is excellent, the rise time and the falling times are not optimal for the matching with the digitizer and for pileup rejection. The long tail is related to the residual component of a long tail present in the SiPM array. Additional shaping cannot be done at the level of the FEE chips due to the small space available for the circuit. To improve on this, we have designed an additional receiver section in the digitization board that will increase the rise-time to > 25 ns and reduce the positive tail.

### Linearity and resolution tests

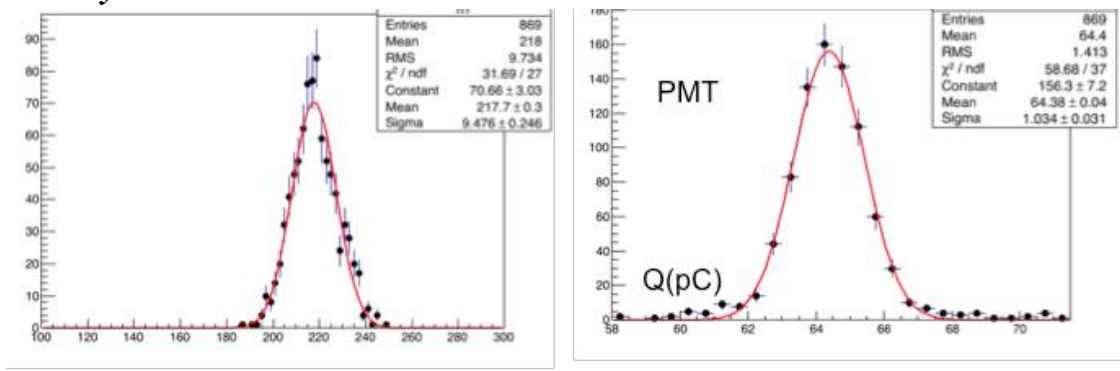

Figure 5.14: *Distributions of the integrated charge for (Left) SiPM and (right) PMT .*

For each laser beam amplitude, the SiPM+FEE and PMT signals have been integrated to extract their charge and measure linearity and resolution. In Fig. 5.14, an example of the SiPM (left) and PMT (right) charge distributions is shown. Each distribution has been fit with a Gaussian shape that well follows the experimental data. For the provided example, the resolution of the SiPM charge is around 5% while the one of the PMT is of around 1.5%, as related to the different quantity of light arriving to the sensor. The distribution of the ratio of these two signals, event by event, allows to evaluate the resolution and the differential linearity assuming granted linearity and negligible contribution to the resolution for the PMT. While the amplitude in pulse height are similar for the two sensors, the ratio between the charge distribution is around 3.5 due to the SiPM larger time width. Also the distribution of the ratio SiPM/PMT has been fit with a Gaussian function. In Fig. 5.15.left the distribution of the peak value for the SIPM/PMT ratio is reported as a function of the SiPM peak. A good differential linearity (at few permil level) is observed. This linearity holds up to around 300 pC (300 mV) output while a linear dependence appears above this value as due to an evident saturation of the PMT. In Fig. 5.15.right the distribution of the sigma for the SiPM/PMT ratio is reported as a function of the SiPM peak. The resolution dependence is well described by $1/\sqrt{Npe}$ behavior demonstrating that the excess noise is negligible and that the maximum charge collected (1000 pC) corresponds to around 2000 pe. Taking into account the factor of 2 reduction in pulse height due to the passive divider, the response of the system is well characterized by 1 pe/pC or, equivalently, 1 pe/mV.

***Stability test as a function of rate***

In order to control the stability of the response as a function of rate, we have kept stable the laser output and measured the response at different rates. We have taken 8 points at different rates (500, 1 kHz, 5 kHz, 10 kHz, 50 kHz, 100 kHz, 500 kHz, 1 MHz). Both the SiPM shape and charge looks practically not disturbed (< 1.5 %) by the rate variation as shown in Fig. 5.16 whee the charge distributions for 4 of these cases is reported.

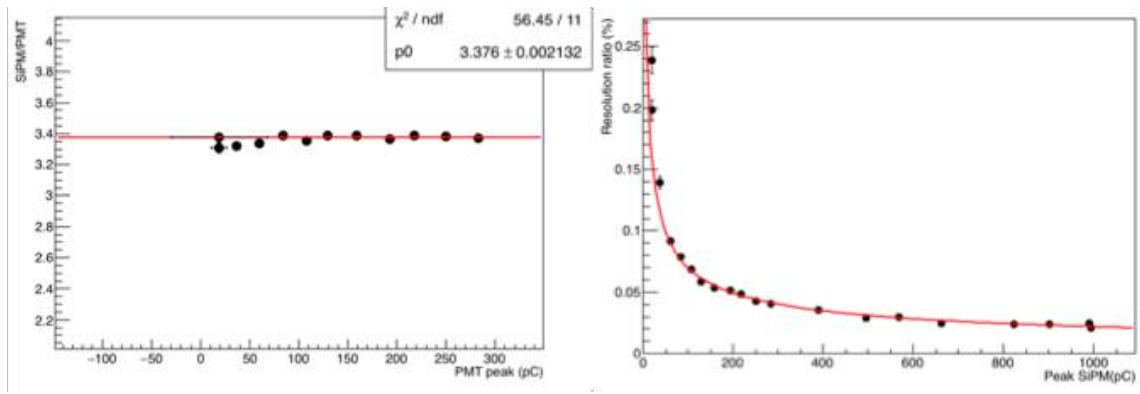

Figure 5.15: (left) *Differential linearity SiPM/PMT as a function of the PMT charge and (right) energy resolution as a function of the SiPM charge.*

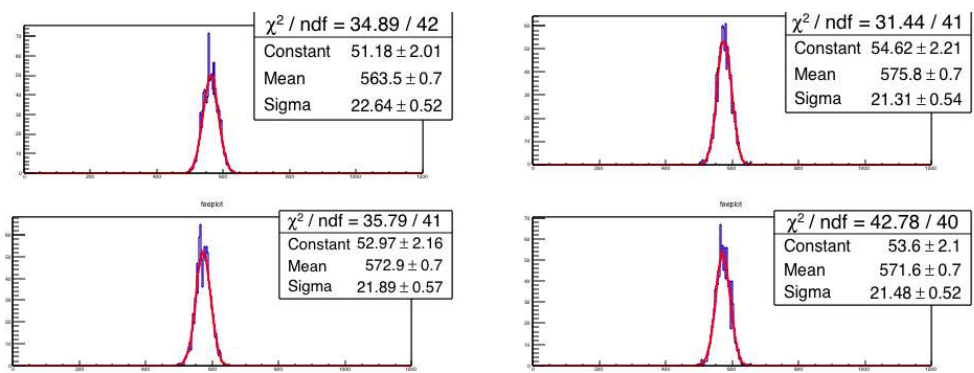

Figure 5.16: *Distribution of the SiPM+FEE output for 4 different laser pulsing rates.*

## 5.5  The Digitization system

### 5.5.1  Introduction

The calorimeter is composed of 1348 crystals, each equipped with 2 arrays of SiPMs, for a total of 2696 fast analog signals that must be digitized after being amplified and shaped by the FEE. The shape of the signal, see Fig. 7, is a function of the CsI emission time, the SiPM quenching time and the FEE amplification and shaper parameters.

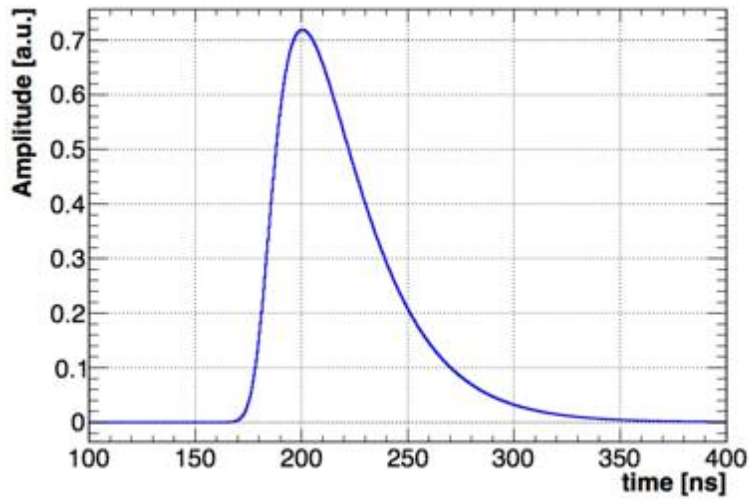

Figure 5.17: *Monte Carlo simulation of the CsI + FEE waveform*

The average pulses of 150 ns maximum width with a rise time of 20 ns is expected as input to the digitization state with a maximum pulse height of 2 V dynamic range.

The Mu2e Calorimeter Waveform Digitizer subsystem (WD) is an electronic printed circuit board that digitizes analog data, serializes it and sends it upstream to the DAQ via a fiber optic transceiver. The WD must also perform some digital signal processing (DSP) operations, removing data below threshold (zero suppression) as well as providing the mean charge and time for each channel by means of running averages.

Simulations from the Mu2e offline software were used to optimize the digitization frequency and the bit resolution required to satisfy the Mu2e requirements. Table 5.2 and 5.3 show the expected calorimeter time and energy resolution, respectively, for $\mu \rightarrow e$ conversion electrons as a function of the digitization frequency and the bit resolution. 200 Msps and 12 bits of resolution are a good compromise between performance, power dissipation and costs.

|         | 150 MHz | 200 MHz | 250 MHz |
|---------|---------|---------|---------|
| 8 bits  | 470 ps  | 440 ps  | 440 ps  |
| 10 bits | 370 ps  | 250 ps  | 250 ps  |
| 12 bits | 300 ps  | 170 ps  | 170 ps  |

Table 5.1: *Expected time resolution for* μ → e *conversion electrons as a function of the digitization frequency and the bit resolution.*

|         | 150 MHz | 200 MHz | 250 MHz |
|---------|---------|---------|---------|
| 8 bits  | 9.8 MeV | 8.0 MeV | 7.8 MeV |
| 10 bits | 6.5 MeV | 5.5 MeV | 5.5 MeV |
| 12 bits | 6.2 MeV | 5.5 MeV | 5.5 MeV |

Table 5.2: *Expected energy resolution (FWHM/2.35) for* μ → e *conversion electrons as a function of the digitization frequency and the bit resolution*

## 5.5.2  Specifications of Digitizer modules (WD)

To limit the number of pass-through connections the WD boards will be hosted inside the cryostat. This is a very harsh environment and requires to follow special design rules. In particular:

- The boards will be operated in vacuum ($10^{-4}$ torr ) and they will be quite difficult to be serviced (the DS will be opened no more than once per year). This imply to follow in the design high reliability rules, commonly used in space applications.

- The thermal dissipation will be mainly for conduction, so the mechanics of the board (and of the supporting brackets or crates) and the PCB will have to designed to keep account of this. The design will follow extremely low power requirements but nevertheless an appropriate cooling system has been designed (see sect. 6.2).

- The environment will be radioactive, both due to neutrons and ionizing particles. The boards are specified to sustain a TID of 15 krads and a NIEL 1 (Mev eq) of around $2x10^{11}$ n/cm$^2$ over the data taking period (safety factors not included) [5.6].

- Inside the cryostat, a strong magnetic field is present. The boards are specified to sustain a B field of 1 Tesla. This implies that at the component level only air wounded inductors will be allowed and a special care will be needed in the choice of the DC-DC converters.

### 5.5.3   WD Technical design

A block diagram of the calorimeter WD is shown in Fig. 5.18.

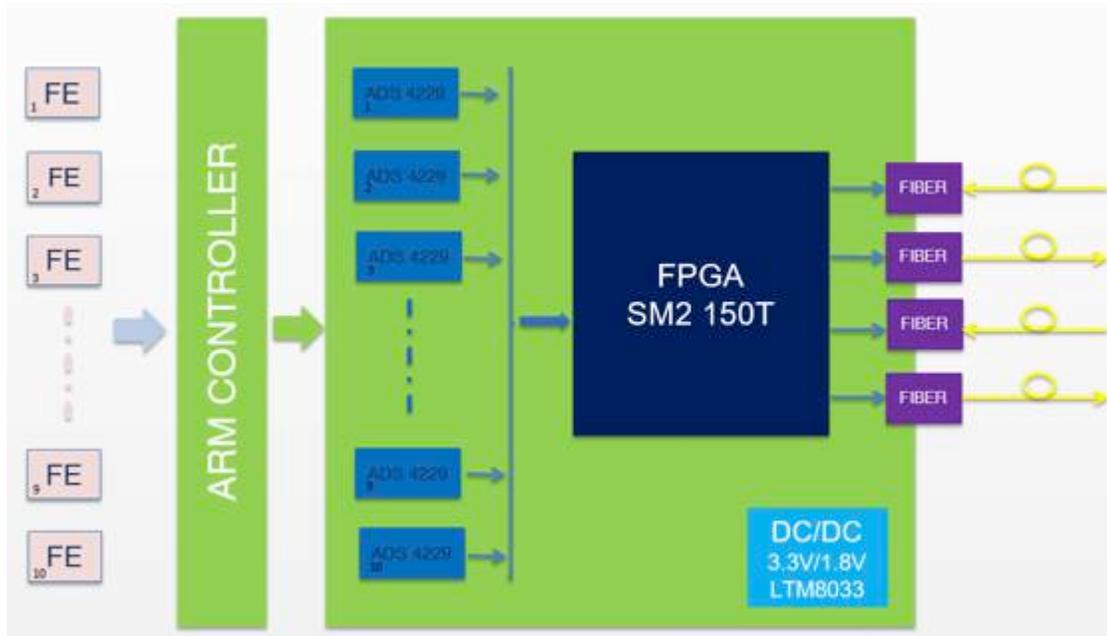

Figure 5.18: *Block diagram for the Calorimeter Waveform Digitizer.*

The design is based on a SoC component (FPGA + CPU integrated in the same package) belonging to the Microsemi Smartfusion 2 family, model SM2150T1152F. It combines a 150K logic elements FPGA with a 166 MHz ARM Cortex M3 processor, plus 5 Mbits of static RAMS, 4 Mbits of flash memory  and 16 fast serializers (5Gbits/sec). By using a preliminary firmware, these resources have been shown to be adequate in handling the data flow related to 20 ADC channels sampled at 200 MHz.The smartFusion2 devices are guaranteed to be SEU immune at the configuration flash level and to be reliable at the foreseen radiation levels [5.7].

The ADC has been selected to be the ADS4229 from Texas Instruments. It converts with a maximum sample rate up to 250 MHz with 12 bits of resolution. The guiding parameters were the relative low cost, in comparison to the competitors, and the extreme low power. The choice will be confirmed when the first prototype will be available for testing. Each ADS4229 includes 2 converters, so a total of 10 parts are needed for each board.

The WD board will receive a single 28 volts power line (plus return) from the outside, through a terminal block, and will downscale it internally to form all the needed voltages (+3.3, +2.5, +1.8, +1.2). The conversions will be made through 5 DCDC converters.

Several studies has been done in the past 2 years to identify DCDC converters that could survive at high radiation and magnetic field levels [5.8]. At the end we have chosen the Linear Technologies LTM8033 as a good candidate.

The clock signal will come from outside through a 125 um fiber to a custom clock fanout board, placed inside the crate. From there it will be converted to electrical (LVDS) and distributed to the WD boards. A very accurate jitter cleaner (LMK04828) will reduce the incoming clock jitter to less then 100 ps and distribute it to the ADCs and to the SoC.

### 5.5.4   Qualification tests of the WD components

Several test campaigns has been settled down to qualify for radiation and magnetic field survival the main WD components. At the moment of writing, we have carried out a complete test for the DCDC converter and the ADC. Both of them were irradiated with gamma and neutrons. The DCDC was also tested in a high magnetic field. For the SoC we are trusting the producer internal tests. All these tests will be repeated at board level .

***DCDC qualification tests***

The LTM8033 parameters measured during the tests were the output voltages and the conversion efficiency. In Fig. 5.19, the result of irradiating the LTM8033 up to a TID of 20 krad is shown. A slight, still tolerable, increase of the output voltage is observed, while the efficiency remained practically constant.

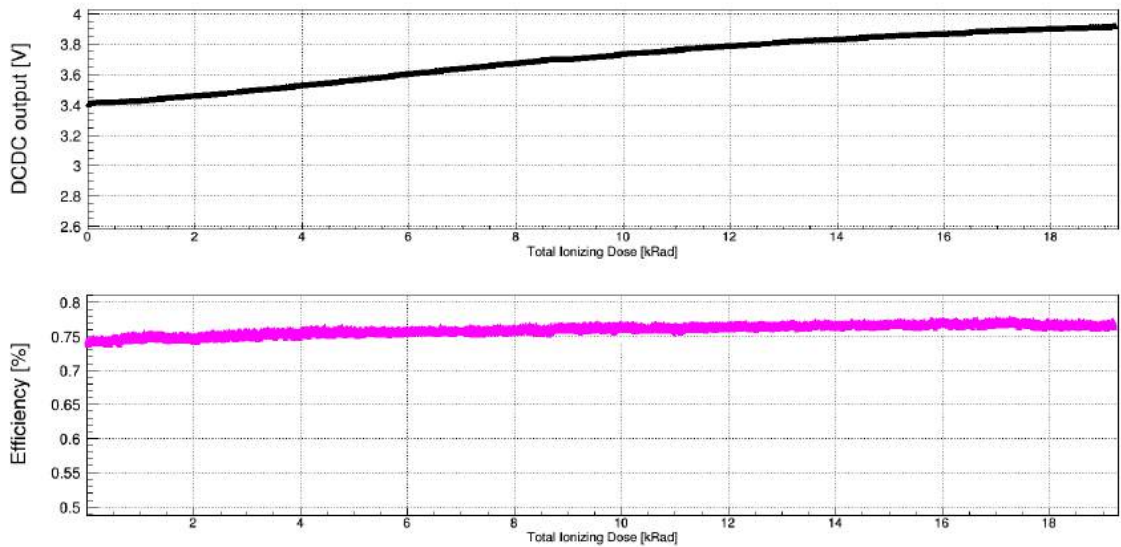

Figure  5.19 *Dependence of LTM8033 parameters as a function of the integrated dose: (top) output voltage and (bottom) efficiency.*

In Fig. 5.20, the results of irradiating the LTM8033 with a fluence of 5 x $10^{12}$ 1 MeV (Si) n/cm$^2$ are shown. No relevant changes are  observed during and after the irradiation.

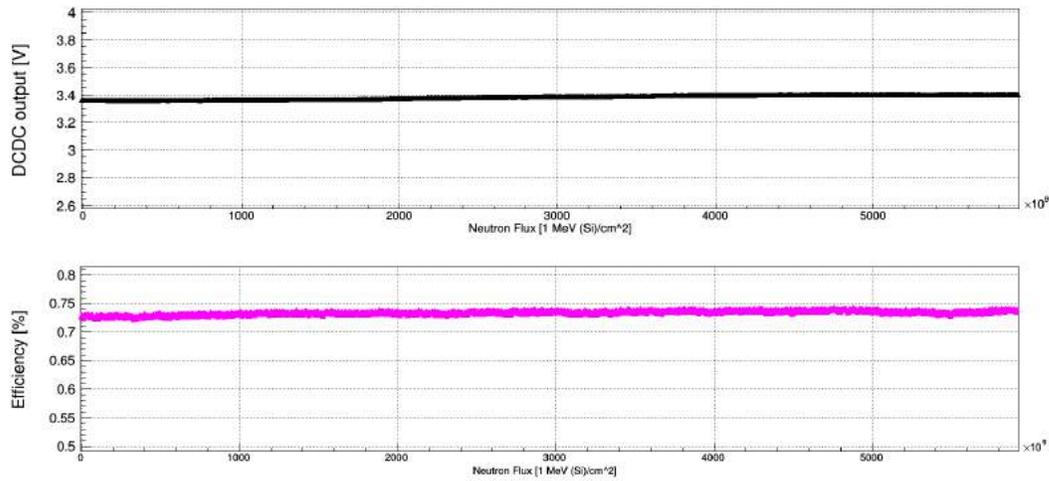

Figure 5.20 *Dependence of LTM8033 parameters as a function of the neutron fluence: (top) output voltage and (bottom) efficiency.*

The DCDC converter module has been exposed to B-field in three different orientations. Fig. 5.21 shows the measurement of the efficiency as a function of  the field for the x-axis  component.

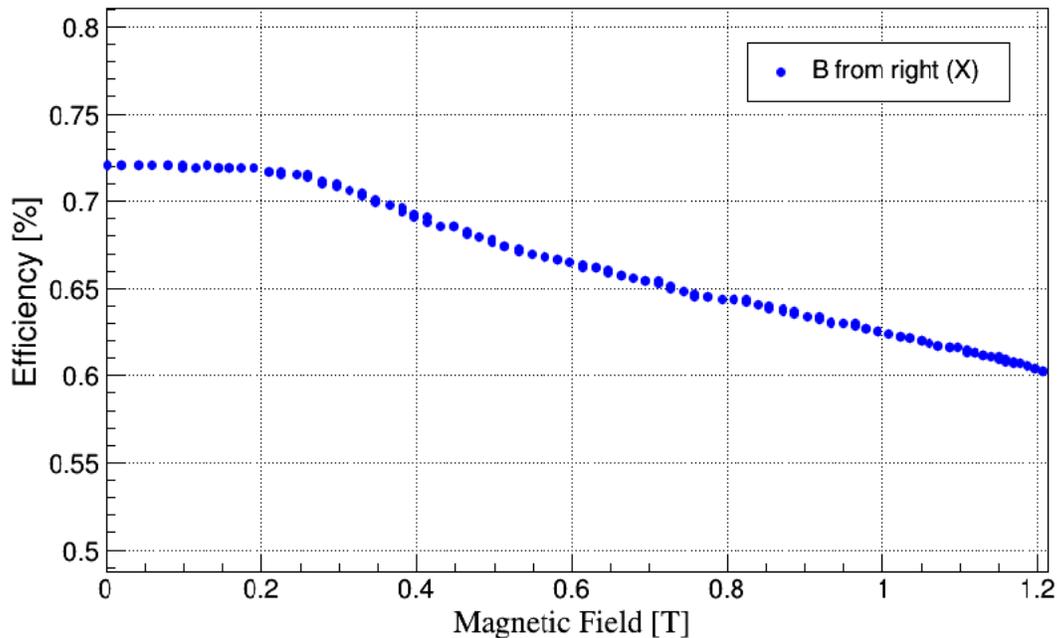

Figure 5.21 *Dependence of the LTM8033 efficiency as a function of the X-component of B field.*

The input current tends to increase slightly with the increase of the applied magnetic field, with a consequent decrease in the conversion efficiency starting from 0.2 T. The device still provides the 3.3 V output at 1 T B-field, despite a total drop in efficiency of around 10%.

***ADC  qualification tests***

The ADS4229 was irradiated both with gamma and neutrons radiation.
The ADC performances have been monitored using a custom setup. A 200 kHz standard sinusoidal, tuned to fulfill the ADC dynamic scale, has been used as input for both the ADC channels. A custom DAC board, located far away from the radiation zone, read the ADC channels. The DAC output waveforms has been acquired through a digital oscilloscope, while irradiating up to 20 krad and 5 x $10^{12}$ 1 MeV (Si) n/cm$^2$. Gathering more than 300 GB of data from both tests, no significant changes in the analog output or bit flips have been observed.

### 5.5.5   WD system organization

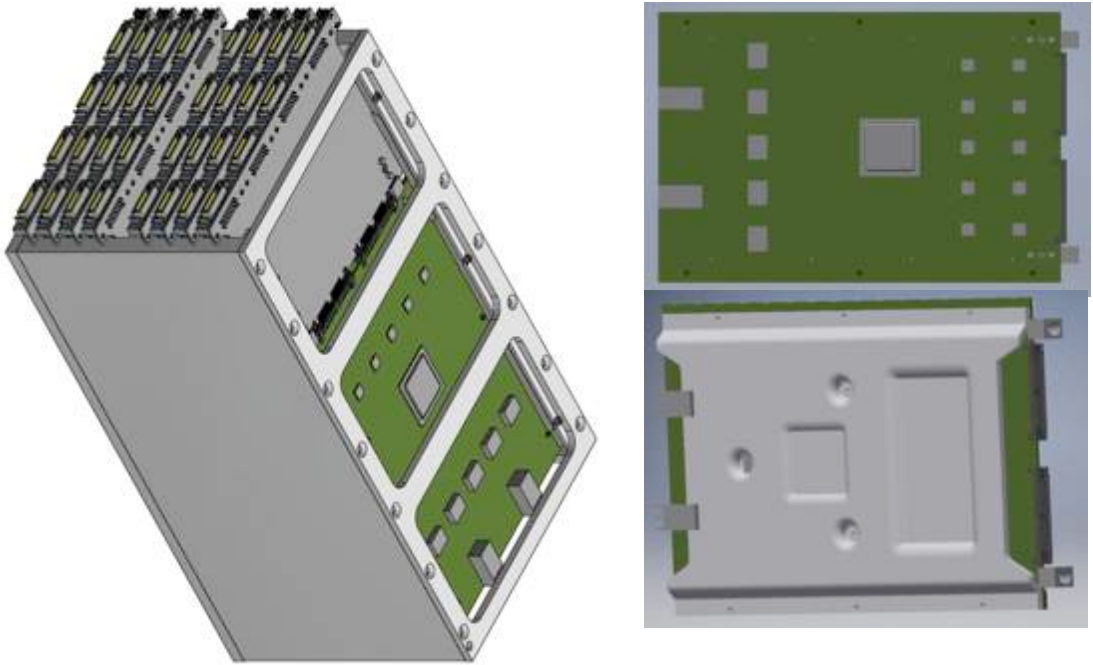

Figure 5.22 *(Left) CAD drawing of the crate. (Right) naked and  cold plate equipped WD.*

The 2696 SiPMs will be readout through 136 WD boards, each supporting 20 channels. Each card is 233 x 160 mm long, with the FEE connectors located on the short edge. Half of the cards reads one of the disk. The cards will be hosted on 10 crates for each disk, each comprising up to 9 slots, 6 or 8 MB+WD boards, while the 9$^{th}$ slot is reserved for the clock distributor. A description of the crate distribution on the disks can be found in the description of the cooling system in sec. 6.2. Here we show only a 3D drawing of the crate Fig. 5.22 (left) and explain the basic concepts behind the WD cooling. Several simulations have been performed to understand the thermal requirements of the system (see assuming an approximate value of 15 W of power dissipation for each WD and around 140 W for a full crate. Each WD will be cooled through a custom designed aluminum cold plate, kept in strict contact with the more dissipating components (SoC and DCDC converters). The cold plate will be thermally connected to the crate structure through card locks. Fig. 5.22 (right) shown pictures of a naked WD (top) and one (bottom) equipped with the cold plate.

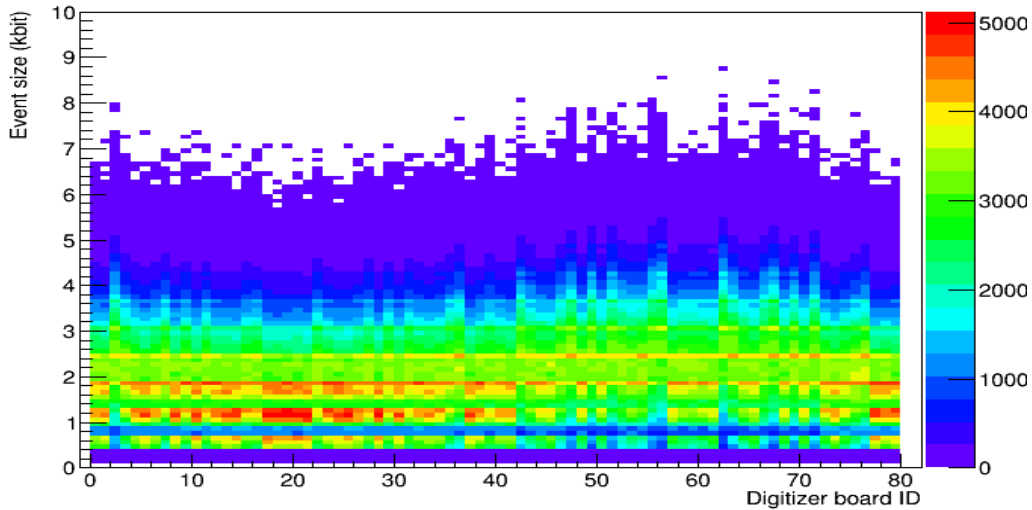

Figure 5.23: *Buffer size for a chain of 2 boards. A safety factor of 2 is included in the data rate estimate.*

### 5.5.6 Ring organization

Simulation studies were performed to optimize the arrangement of the channels within the digitizer boards. Given the cylindrical symmetry of the calorimeter, grouping radially the channels provides a homogeneous load among the digitizer boards, as shown in Fig. 5.23. This result was combined with the expected Mu2e beam structure in order to evaluate how many digitizer boars can be arranged in a daisy chain for streaming out the data. Fig. 5.24 shows that the configuration with two boards in a daisy chain (ring) allows to empty the buffer.

The redundancy of the calorimeter DAQ system is ensured by the following specifications:

- L and R channels belonging to the same crystal are plugged in different WD boards;
- Each ring is equipped with 2 bi-directional optical fibers.

This daisy chain system avoids data losses also in case the link between the boards is lost or damaged, because in that case we simply need to plug in both fibers.

For a total of 136 WD boards reading out the full calorimeter, 68 bi-directional optical fibers, plus other 68 for redundancy, are needed to link the WD boards with the 12 calorimeter DAQ servers.

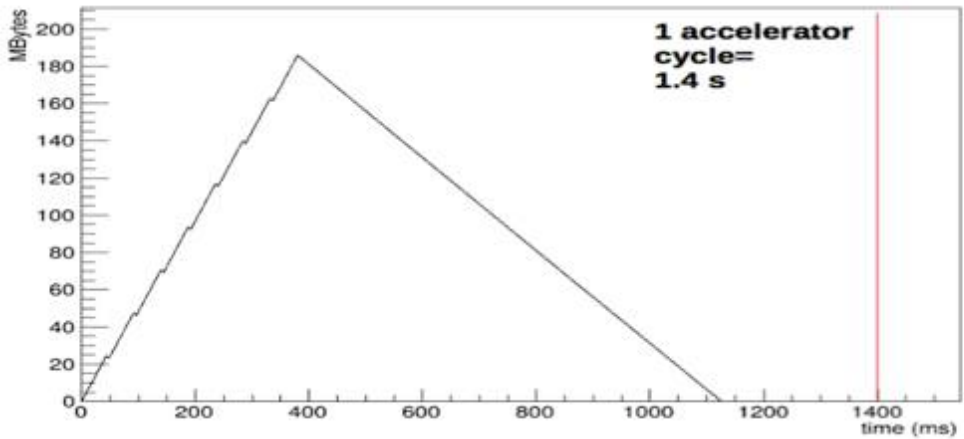

Figure 5.24: *Timing for empting the buffers as simulated with 2 WDs per ring.*

### 5.5.7   Initial firmware studies

The Smartfusion2 FPGA on the WD board has to perform several functions:

- Provide control signals to the ADCs;
- Zero Suppression and packing of data;
- Timestamp to the data;
- Write the data to memory. Data will be set to DAQ in the period between spills.
- Read data from memory and serialize.
- Serialized data, 8b/10b encoded, are sent to the DAQ through an electrical-optical converter.
- To limit the number of fibers reaching the DAQ, the WD will be daisy-chained in rings. Each FPGA will receive data from the previous WD, add its own data and pass data onto the next board.
- Slow control commands will be received and acted on by the FPGA through the same fibers used for data transmission.

An intermediate version of the firmware that partially carried out these functions has been developed in order to test the FPGA performances and the full compatibility with the selected ADC at 200 MHz sampling frequency. A schematic view of the data pipeline is shown in Figure 5.25 and 5.26.

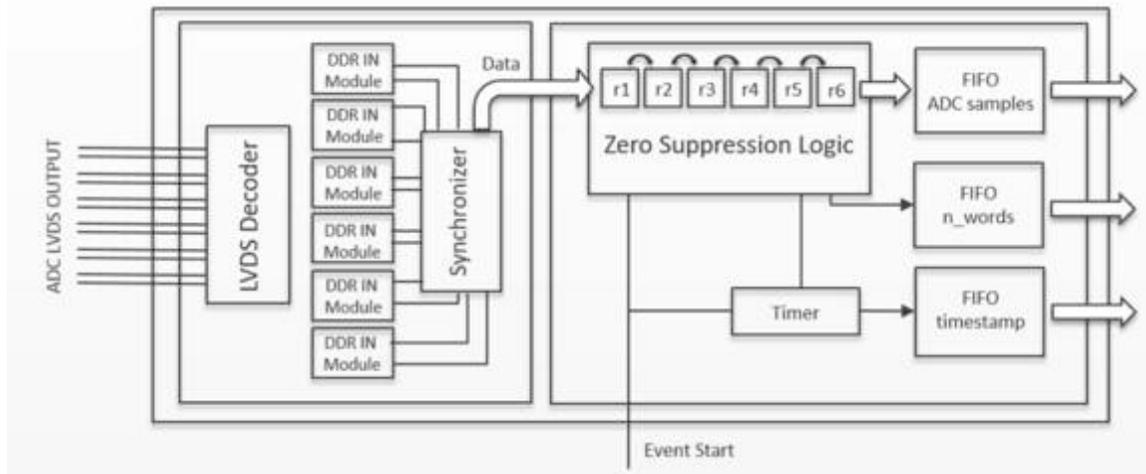

Figure 5.25: *FPGA firmware block diagram of the one ADC channel interface.*

The first block transforms the parallel LVDS DDR ADC input signals in a 12 bits bus. Data are zero suppressed, packed in '*hits*' and stored in three FIFOs. One contains the ADC samples of the hits, one the numbers of samples in each hit and one the hits timestamp. When a new hit is stored in a channel, a flag is sent to a control block so as to be able to be sequentially read back the information.

The data in the FIFO channels are then ordinated and merged into another single deep FIFO, ready to be serialized and sent to the DAQ through the optical link.

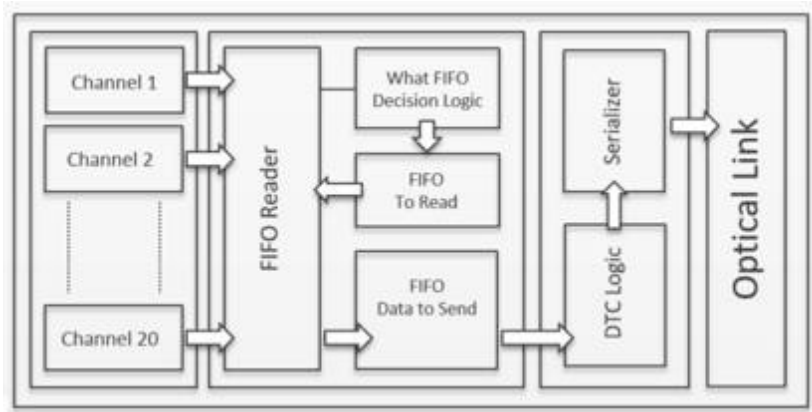

Figure 5.26: *FPGA firmware block diagram of the WD digitizer.*

# Chapter 5 References

# 6 Calorimeter Mechanics

We tailored the calorimeter mechanical structure around the best ideal layout for the crystals obtained by simulation to maximize acceptance. Given the chosen pattern, the best solution is to pile up the crystals in a self-standing array organized in consecutive staggered rows.

The active area of the Mu2e calorimeter consists of two annuli with an Inner radius of 375 mm and an Outer radius of 657 mm containing a number of 674 staggered scintillating crystals with a square prism shape (34 x 34 x 200 mm), each wrapped with a 150 μm thickness reflective sheet (Tyvek®) for a total weight of 700 Kg.

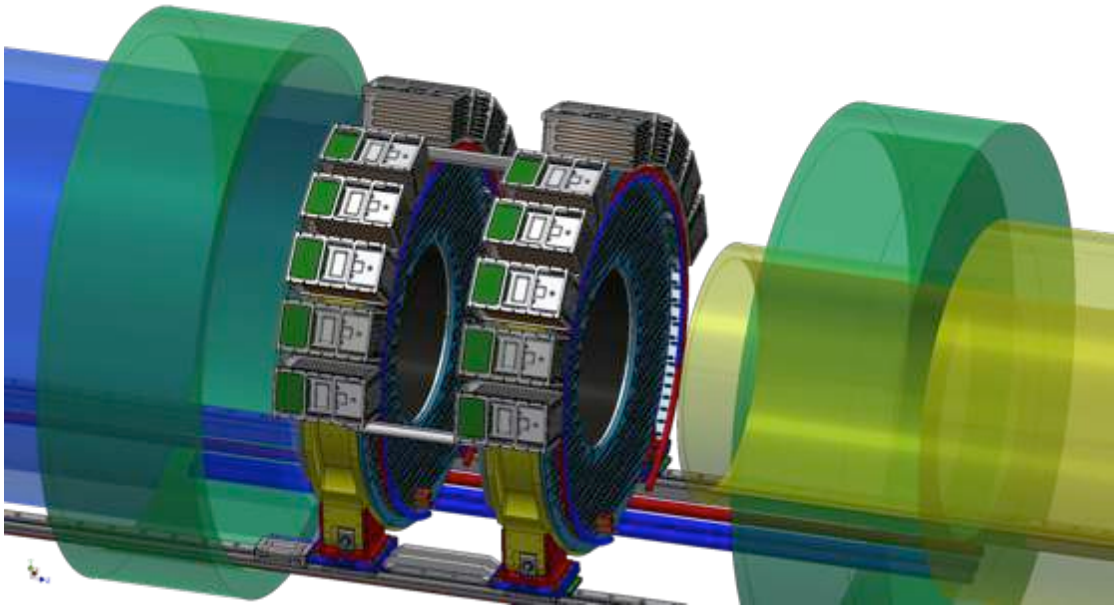

Figure 6.1 *The Mu2e Calorimeter annular disks.*

The arrangement of the two calorimeter annuli inside the detector solenoid is shown in Fig. 6.1. Each annulus has an inner radius of 336 mm, an outer radius of 910 mm including all the electronics crates.

Fig. 6.2 is an exploded view of all the elements composing each annulus. As shown, each crystals' array is supported by two coaxial cylinders. The inner cylinder must be as thin and light as possible in order to minimize the passive material in the region where spiraling background electrons are concentrated. The outer cylinder is as robust as required to support the load of the crystals. Each disk has two cover plates. The plate facing the beam is made of Carbon Fiber to minimize the degradation of the electron energy, while the back plate can be very robust. The back plate will also support the SiPMs, the front-end electronics and the cooling.

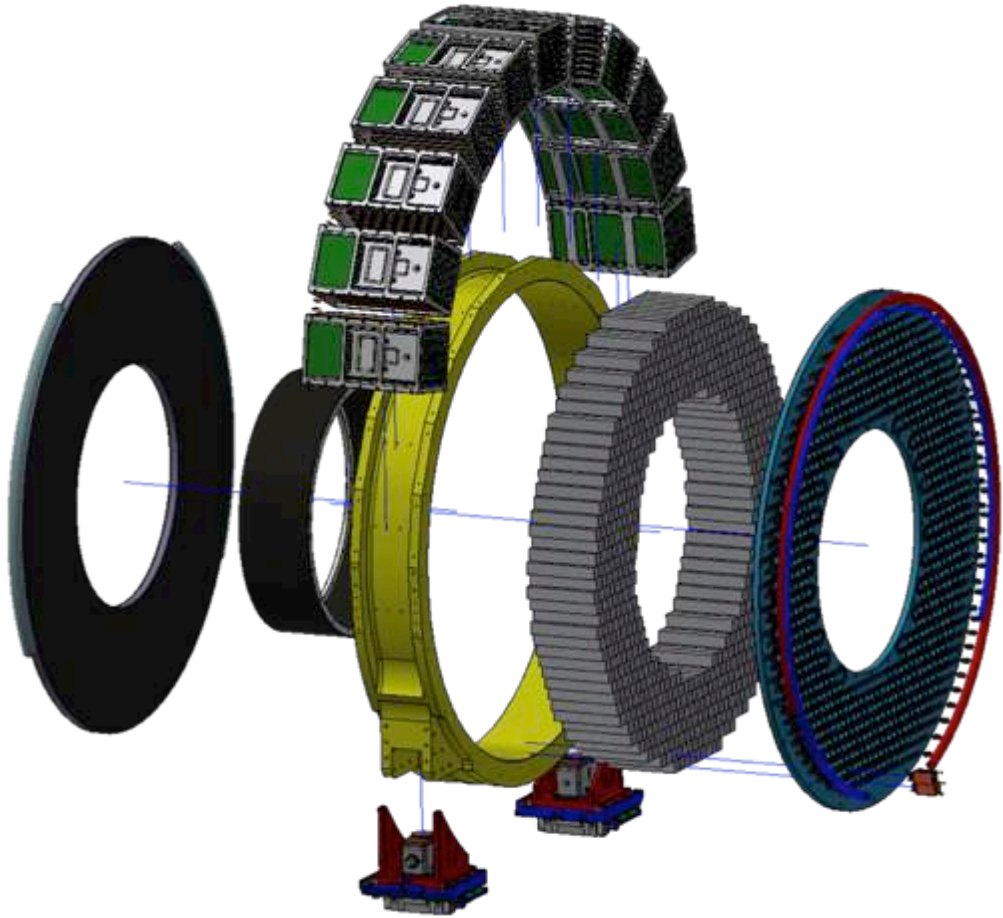

Figure 6.2 *Exploded view of the components of the Mu2e calorimeter annular disks.*

The crystal arrangement is self-supporting, with the load carried primarily by the outer ring. A catenary structure resembling a Roman arch is constructed to reduce the overall load on the inner cylinder. The mechanical properties of the crystals are critical for this type of configuration. These include the Young's modulus, tensile modulus, Poisson ratio (or torsional modulus of elasticity), yield strength and ultimate strength. A Finite Element Model, using the crystal properties as input, has been constructed to optimize the design of the structural components and to verify that the crystals could be piled up and sustain their own weight. The boundary conditions of this layout are fixed and the structural analysis has been used to verify displacements and deformations of the various components. We have built a first prototype of the outer cylinder and the supporting feet that integrate X-Y adjustment mechanism. The mock up prototype is shown in Figure 6.3.

The back plane will be built of plastic material with good outgassing properties, either PEEK or Zedex. It provides support for the FEE electronics and SiPM holders and hosts the cooling pipes to dissipate the power of the electronics and cool down the sensors. A readout unit is composed of a crystal, two SiPMs and two AMP-HV chips. The back plate will provide visual access to each crystal. An example of the concept is shown in Fig. 6.4. The Front Plate will be made of Carbon Fiber and will embed the piping for running the source calibration fluid.

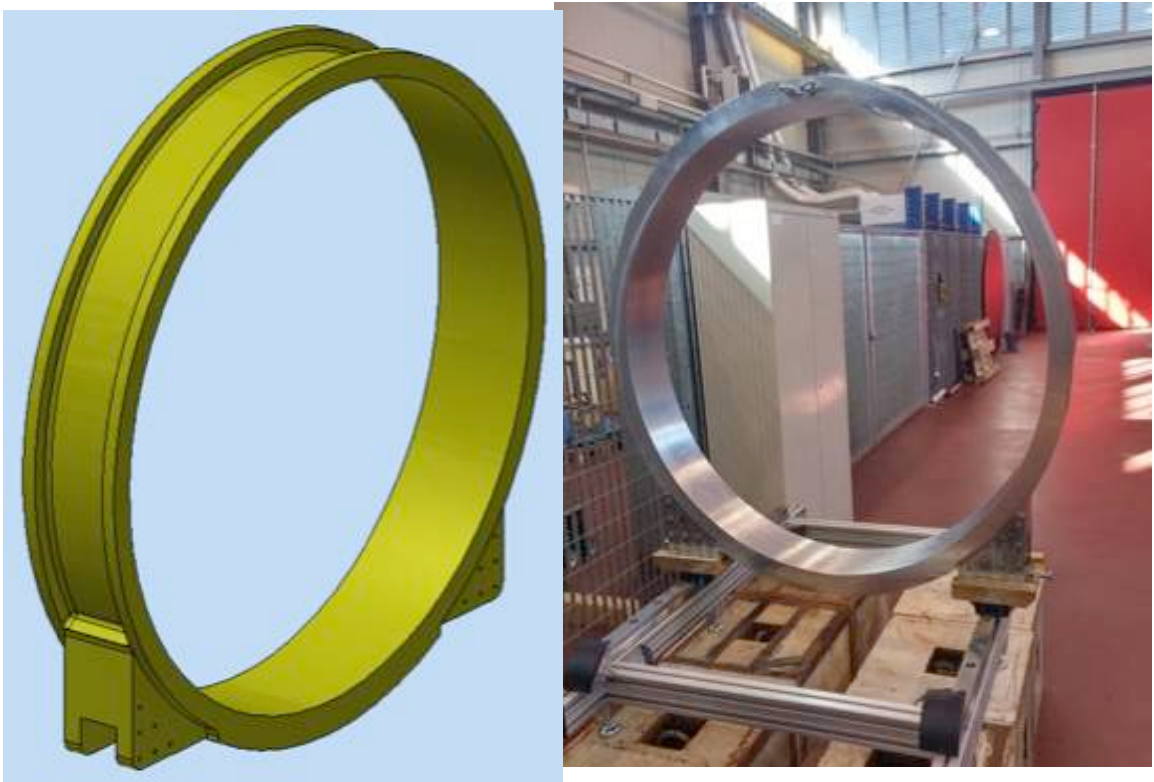

Figure 6.3 *CAD drawing of the outer support ring (left) and full size prototype in aluminum (right).*

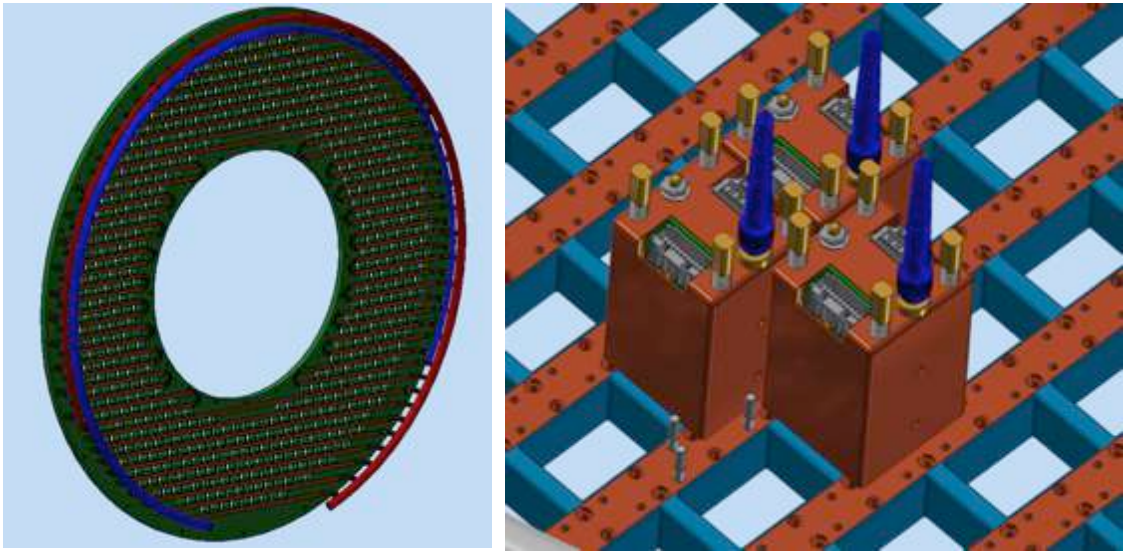

Figure 6.4  *Back plate layout with a detail of FEE holder and cooling pipes*

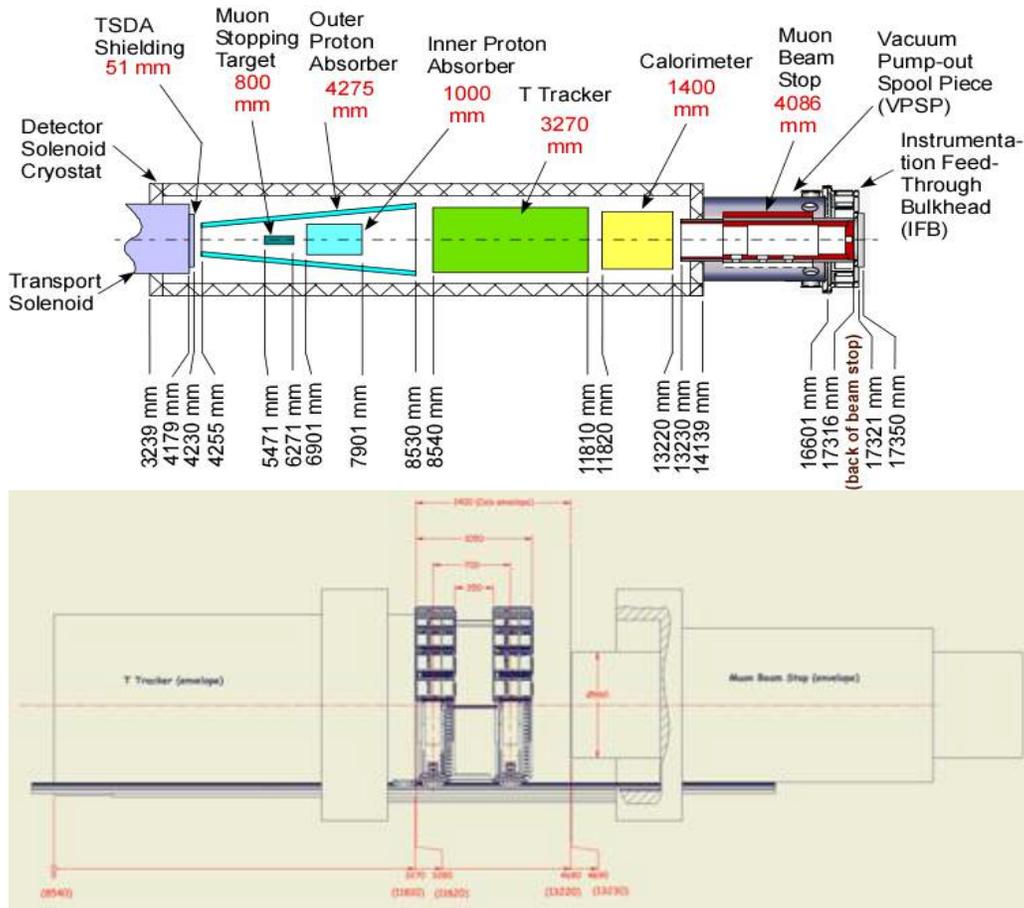

Figure 6.5 *Longitudinal space of the calorimeter inside DS: (left) all DS area, (right) Zoomed around the calorimeter region.*

## 6.1 Envelope and overview of the Mechanical structure

### Envelope and integration

The Calorimeter mechanics design had to fulfill the requirements to be integrated in the Detector Solenoid respecting the available designated volume and all the interferences with other detector and components. In Fig. 6.5, the longitudinal space available for the calorimeter (top) and its actual occupancy in the current design (bottom) are shown.

The calorimeter is 1060 mm long. Fig. 6.6 shows the transverse occupancy of the calorimeter. Particular care has been put in trying not to fill the transverse area and leave empty volumes to facilitate the evacuation of the area. In the transverse plane a dedicated area for the routing of the Mu2e Tracker cabling and cooling pipes has been reserved.

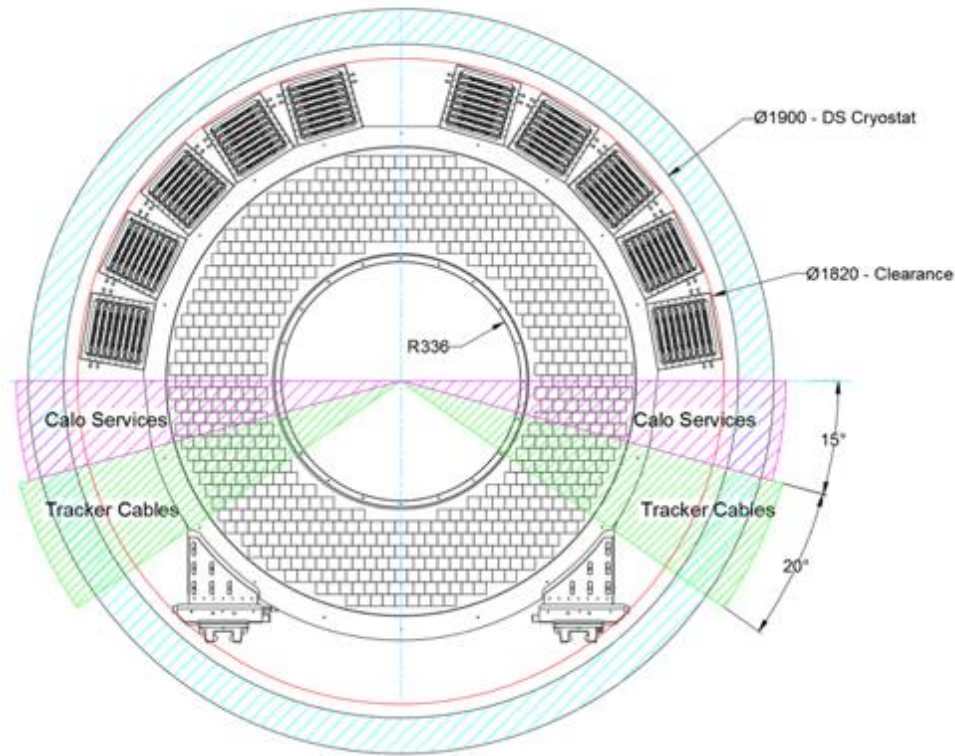

Figure  6.6  *Transverse view of the Calorimeter. Calorimeter services area is up to 12°
below the horizontal axis. The area reserved is much smaller than the one drawn.*

### Description of the components

As shown in Fig. 6.2, each annulus disk is composed by:

- Two support feet and X-Y adjustment mechanisms;
- An outer Al supporting ring with integrated stands;
- An inner carbon fiber ring;
- Carbon fiber front plate integrating the source calibration system;
- PEEK back plate, housing photo sensors and FEE electronics and embedding the cooling copper pipes;
- Array of 674 crystals;
- 674 SiPM holders;
- 10 Read out/service electronics crates (9 boards/20ch each).

The Outer ring is a monolithic Aluminum component; it is machined from a whole Aluminum block as shown in Figure 6.7.left. This to maximize its support properties and minimize stresses and displacements on the crystals. The geometry has been verified via a Finite Element Method analysis as described later in this section.

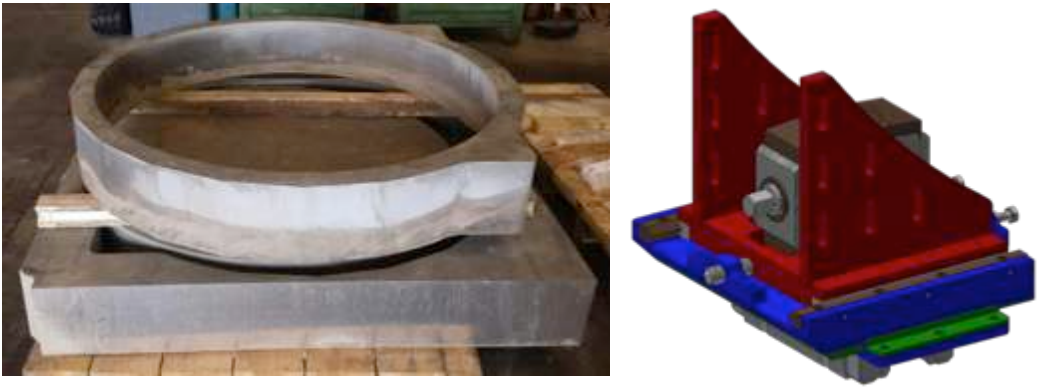

Figure 6.7 *Outer Al ring initial machining. Detail of the Calorimeter Feet and adjusting mechanism.*

The stands integrate a X-Y adjustments mechanism (Fig.6.7 right) consisting in a wedge clamp system for the X regulation and a slide that provides Y regulation, both with locking screws. The Feet are connected to the bearing block inserted in the Mu2e train rails.

Figure 6.8 shows the Inner Carbon Fiber cylinder and its connection rings to the Front Plate and Back plate. It also shows the detail of the Carbon Fiber front plate composed by 2 Carbon Fiber layers sandwiching the source pipes embedded in Al honeycomb. While for the prototypes on the mockup a simple Epoxy Carbon Fiber has been used, the final version will be in a Cyanate Easter resin or similar material. This is due to the outgassing properties in vacuum that are much better than the standard Epoxy CRFP.  Moreover, this resin offers also a good radiation resistance.

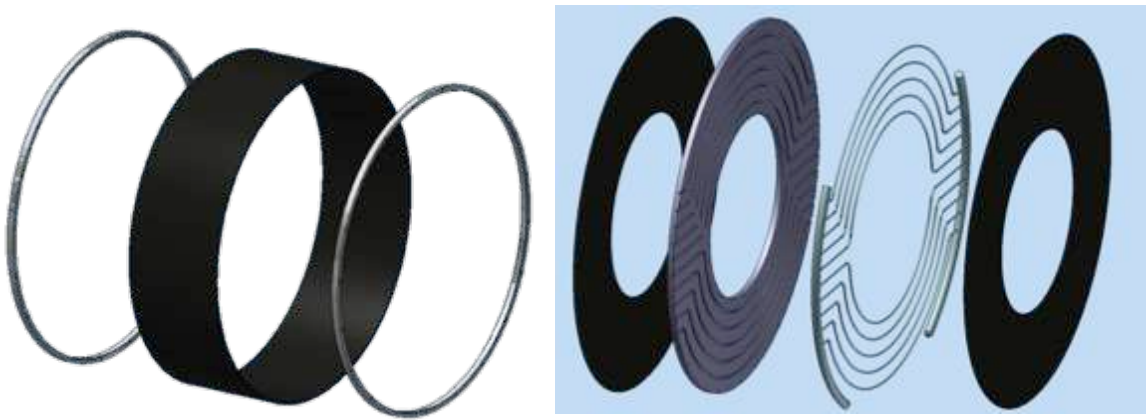

Figure 6.8  *Details of the Inner cylinder and front plate Carbon Fiber components.*

The Back plate is machined to obtain the windows facing the front face of the crystals where the SiPM copper holder will be inserted. The plate is also machined to insert the copper cooling pipes to dissipate the power generated by the FEE electronics and thermalize the photosensors. All the cooling pipes are connected to the IN/OUT manifold. The Back plate is thermally isolated from the Outer ring and from the crystals. Figure 6.9 shows the details of the Back plate design. A dedicated discussion of the cooling system can be found in sec. 6.2.

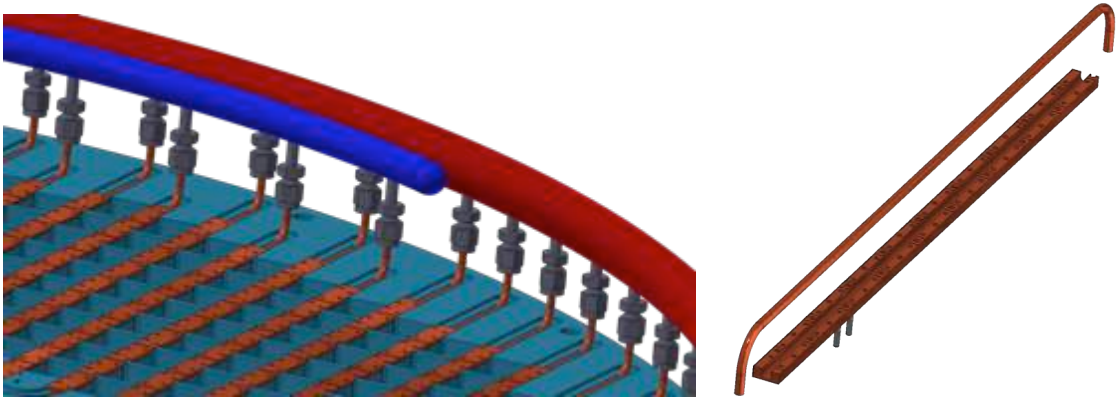

Figure 6.9 *Details of the Back plate and cooling pipes.*

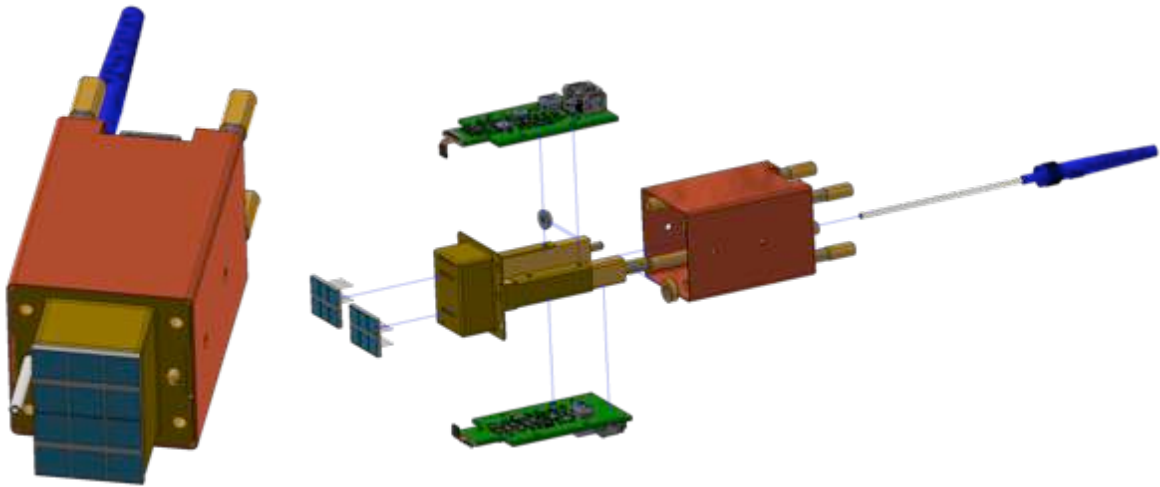

Figure 6.10  *(Left) Assembly of the FEE holder, (right) explosion view of the FEE holder with all components: 2 SiPMs glued on the inner cupper support, FEE AMP-HV chips, Faraday cage and needle for insertion of the optical fiber.*

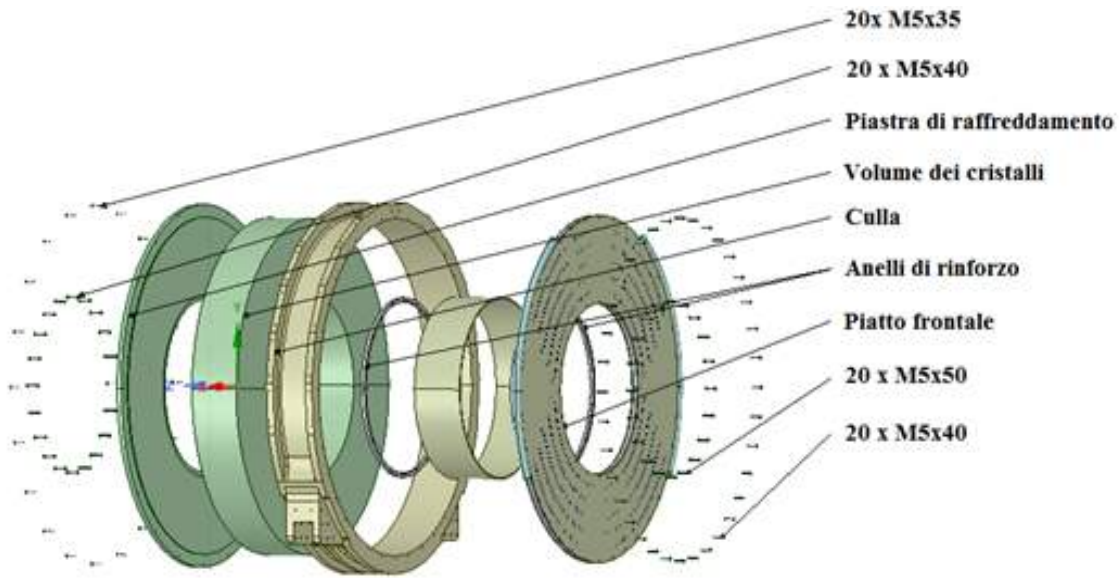

20x M5x35
20 x M5x40
Piastra di raffreddamento
Volume dei cristalli
Culla
Anelli di rinforzo
Piatto frontale
20 x M5x50
20 x M5x40

Figure  6.11 *Exploded view of the components used in the FEM analysis.*

The CAD drawing of the FEE and SiPM holder is shown in Figure 6.10. It is composed of a copper mask where the 2 SiPMs of each crystal are going to be glued to and that holds 2 FEE electronic boards in thermal contact by means of bridge resistors. The FEE boards are surrounded for shielding by a copper Faraday cage. The last component of the holders is the optical fiber needle that will bring the light from the laser calibration system inside each crystal. The shape of the FEE holder ensures the needed light tightness.

On top of the Outer cylinder there are 10 electronic crates with 9 slots; each slot hosts a Digitizer board and mezzanine HV/LV controller board. Each board has 20 channels granularity. To dissipate the power of the crate electronics, the crate mechanics is in contact to cooling copper coils mounted on the lateral sides and connected to the cooling main pipes. On sec. 6.2, the cooling of the crates and boards will be discussed.

### FEM analysis of the overall structure

All the components of the supporting structure have been designed and verified through extensive Finite Element Method analyses. Crystals have been considered as a monolithic load. Figure 6.11 shows all the parts included in the simulation. The outer ring is the most important component. High number of rules have been applied on a tetrahedrons meshing method that ensures regularity and refinement of the elements. The main load condition is represented by the gravity. Besides the structure weight, a further load is due to 10 crates with a weight equal to ~10 kg each.

The structural analysis, carried out considering several conservative hypothesis, allows to verify the structure robustness and the magnitude of the local stresses.

The maximum total deformation is due to the preload given on the bolts (100 N). This deformation is equal to 51 μm and is localized in a small area around the holes of the front plate. The maximum vertical deformation of the cradle is due to standard Earth gravity and is equal to 17 μm. The absolute values obtained through the FEM analysis on the whole system can be considered negligible. The results of the analysis are shown in Figure 6.12.

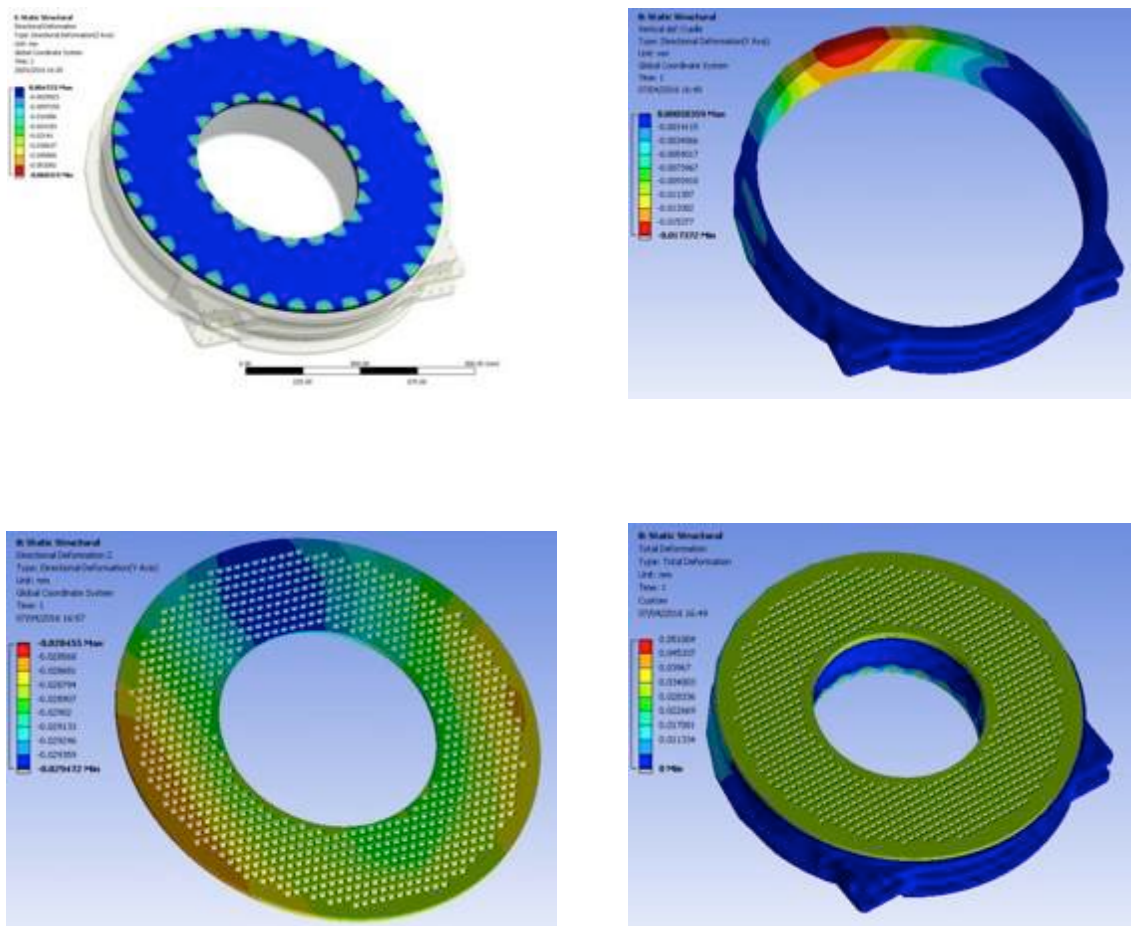

Figure 6.12 *Maximum deformation and stresses on the calorimeter structure.*

## 6.2 The calorimeter cooling system

The cooling system is a crucial element of the Mu2e calorimeter for several reasons. The power generated by the SiPMs, front end and read out electronics must be removed with temperature values acceptable.  The maximum operating temperature of components is set to guarantee a high level of reliability in vacuum and high radiation environments. The complexity of the accessibility of components and the high apparatus cost requires a cooling system free from fault and maintenance for a period of time at least one year. A leak or an operating temperature above the limit will compromise the whole experiment. For all these reasons the cooling system will be made with very high standard components and technologies compatible with vacuum and radiation environments. .

The main cooling requirements are:

1) The SiPMs' temperature must be maintained below 0 °C;
2) The electronic components on the WD read-out normally operate in air at 120 °C. We have set the operating temperature to 60 °C to improve the reliability considering the hostile environments in which they will operate;
3) The maximum operating temperatures of the DAQ and service card have been set to the same value of 60 °C;
4) The uniformity of the temperature is considered important to guarantee the mechanical detector precision. So we decided to set the max temperature raise of the cooling fluid to 2 °C.

### 6.2.1  Choice of the Cooling fluid

In order to operate the SiPMs at 0 °C is necessary to use a fluid below the water freezing temperature. We choose a mixture of 35% mono propylene and water that has the freezing temperature of -17 °C for several reasons. The proposed fluid has good thermal properties without having a high viscosity, it is not flammable, not corrosive and not expensive.  It is a proved fluid in several industrial applications and it is easy to procure.

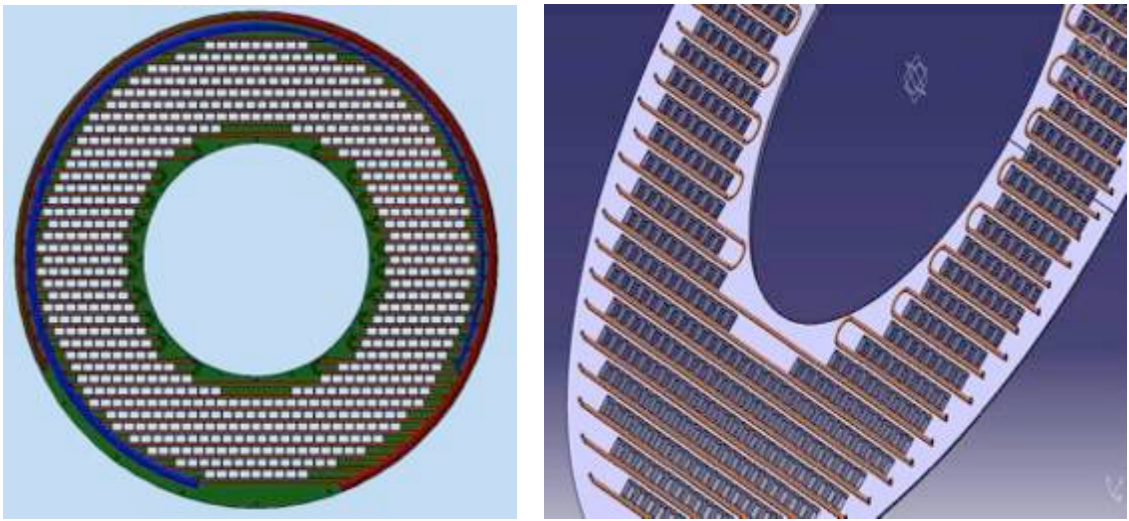

Figure  6.13:  *Views  of  the  cooling  back  plate:  (left)  transversal  view  with  Inner  and  Outer cooling pipes, (right) with zoom of the small cooling lines passing in the back plate.*

***Cooling flow and heat transfer coefficients.***
The process to calculate the flow is driven by two main parameters: (i) the raise of temperature in the cooling lines and (ii) the heat transfer coefficient. This has been determined knowing the flow distribution and the sizes of the cooling tubes. This process has been carried-out for a long period of time considering several aspects. Fig. 6.13 and Fig. 6.14 show, respectively, the lines distribution on the back plane of the calorimeter and on the DAQ-services. The FEE cooling lines are made around a cooper tube of inner and outer diameter respectively of 3 and 4 mm. The DAQ-services uses cooper tubes a little be larger, with inner and outer diameter respectively of 4 and 5 mm.

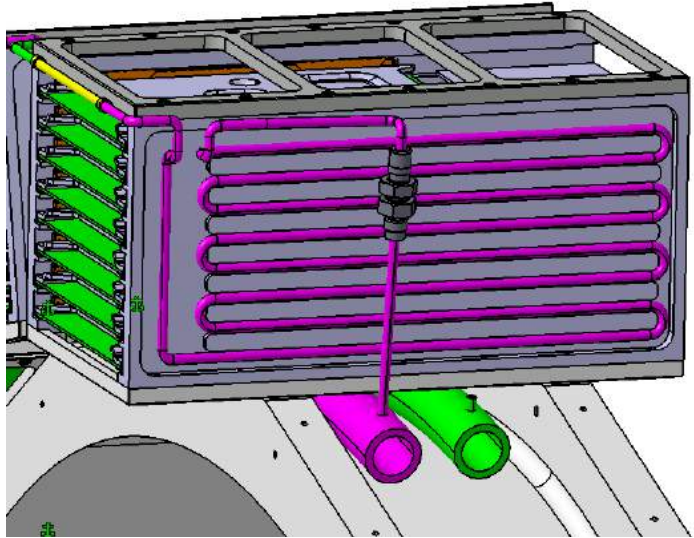

Figure 6.14: *View of the Inner and Outer cooling pipes*

The average speed inside the channel must be maintained above to 3.8 m/s in order to have a good heat transfer coefficient and minimize the temperature gradients. In this condition the film coefficient is ranging 4500-5000 watt/m$^2$ C in the considered tubes.

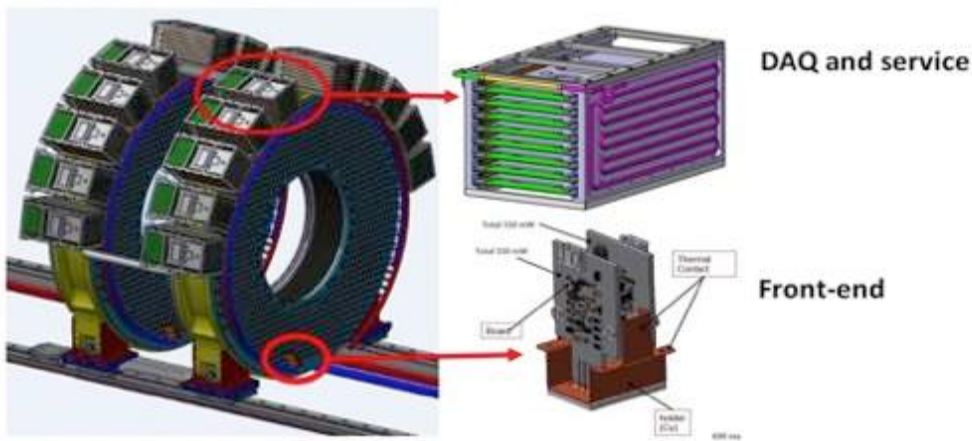

Figure 6.15 *Distribution of the 3 regions where power is generated: (top) The calorimeter WD crates and (bottom) the SiPM+FEE holders.*

## 6.2.2   Power generated by each calorimeter component.

The power is generated in three main area of the calorimeter as shown in Fig. 6.16: in the SIPM, in the FEE and in  the read-out electronics.

Each SIPM+FEE unit has a max power of 2 Watt distributed as shown in Fig. 6.17.

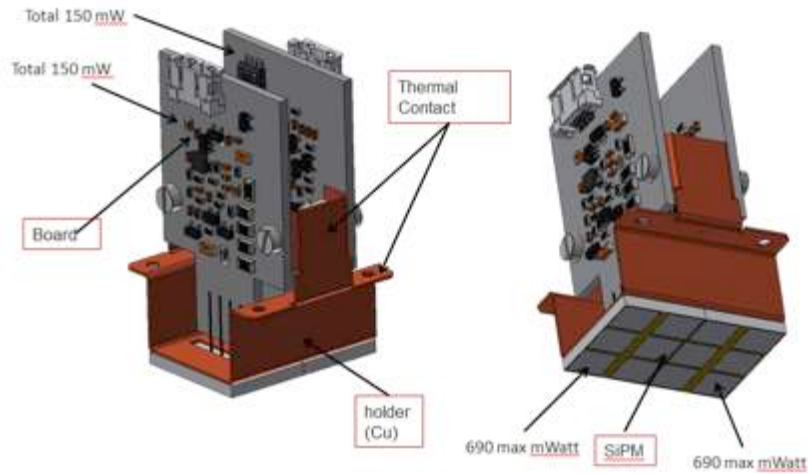

Figure 6.17: *distribution of power sources in the SiPM+FEE holders..*

Each crate unit can contain up to 9 cards (see Fig.6.18) and the power consumption of digitizers is shown in Tab. 6.1.

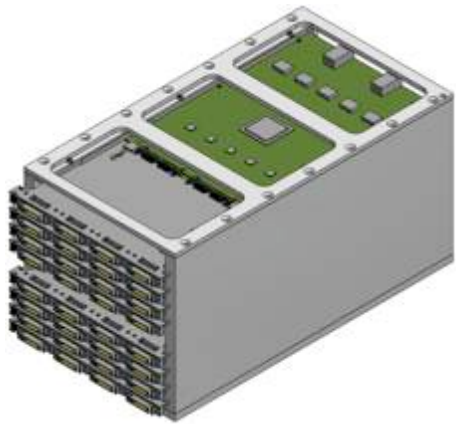

Figure  6.18: *CAD drawing of a read out crate with all boards mounted.*

### 6.2.3  Thermal analysis:

***SIPM/FRONT END Electronics:***

An adequate film coefficient and heat transfer area are necessary to assure a moderate gradient with respect to the power generated by the SiPM and the front end board. The equivalent thermal resistance of the circuit is shown in Fig. 6.19.

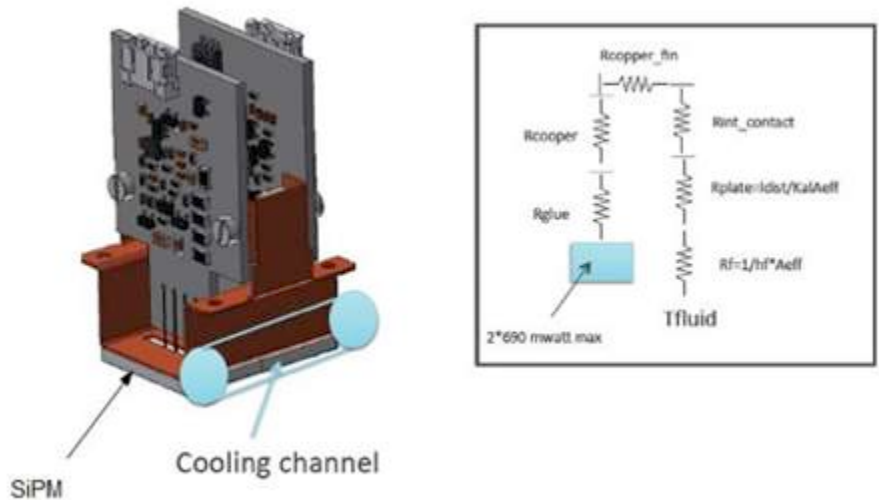

Figure 6.19: *Equivalent thermal resistance of the cooling circuit on the SiPM/FEE holder*

The thermal analyses have been carried out using a finite element modelling. All SiPM end front end electronic details are shown in Fig. 6.20. The thermal gradient has been optimized using the FEA results. The plot in Fig. 6.21 shows the temperature distribution from the fluid down to the SiPM temperature. As mentioned in the specifications, the SiPM must be maintained below 0 $^o$C at the maximum expected power.  Fig. 6.21 shows the results of the Front-End thermal analysis that confirms how the current design satisfies requirements.

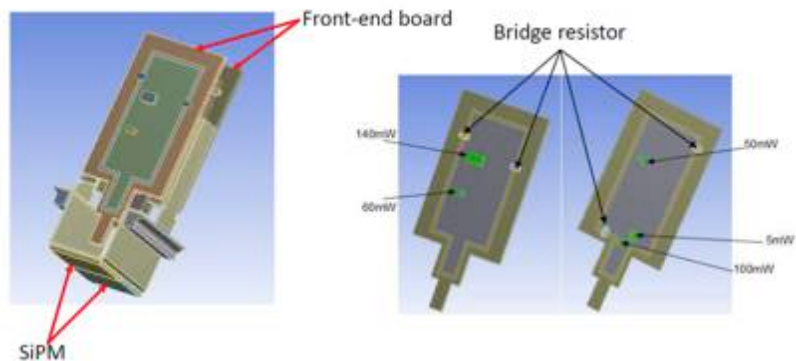

Figure 6.20: *SiPM and FEE electronics details.*

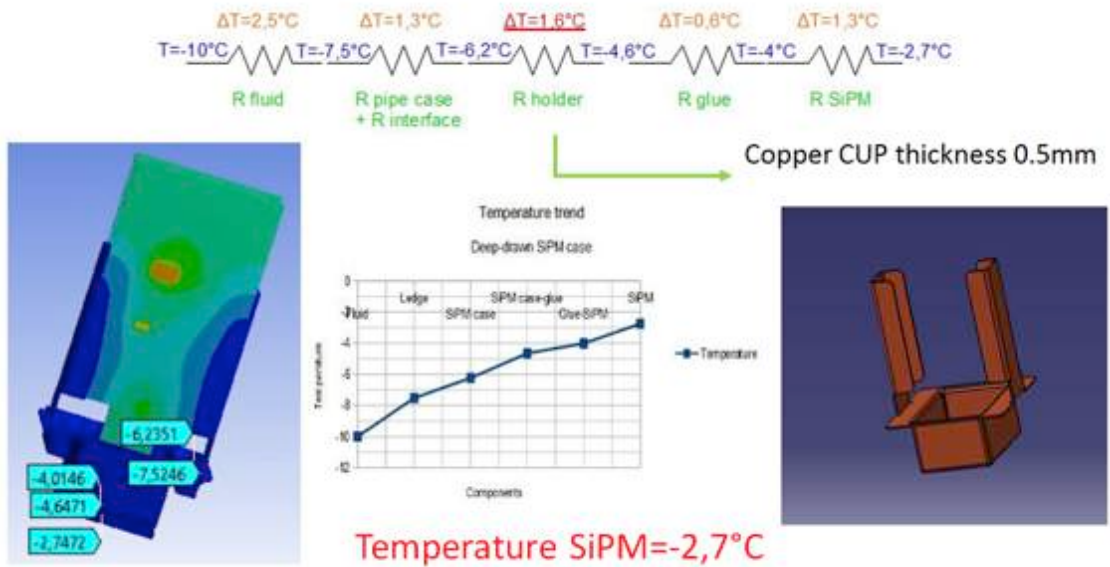

Figure  6.21: *Results of the SiPM/FEE thermal analysis.*

### DAQ AND SERVICES:

The crate must cool down the boards contained.  Each unit board is made by two cards the MB and WD as shown in Fig. 6.22. The powers generated are reported in Tab. 6.1.  A figure 9 shows the board and the cooling arrangement. Two crate plates are maintained at lower temperature by a brazed cooling coil. Two card locks assures the thermal contact with the crate plates. We need to use an additional aluminum plate to remove efficient the power generate by the electronic components. The thermal analyses circuit and the temperature distribution of the cooling plate are shown in figure 6.23

| Board | Waveform Digitizer | Waveform Digitizer | Waveform Digitizer | Service | Service |
|---|---|---|---|---|---|
| Components | FPGA | 5DC DC converted | 10 ADC | Voltage regulator | ARM controller |
| Power | 4 watt | 5X3W=15W | 10X0.5 W=5W | 10 W | 0.2 W |

Table 6.1: *Power consumption in the digitizer board.*

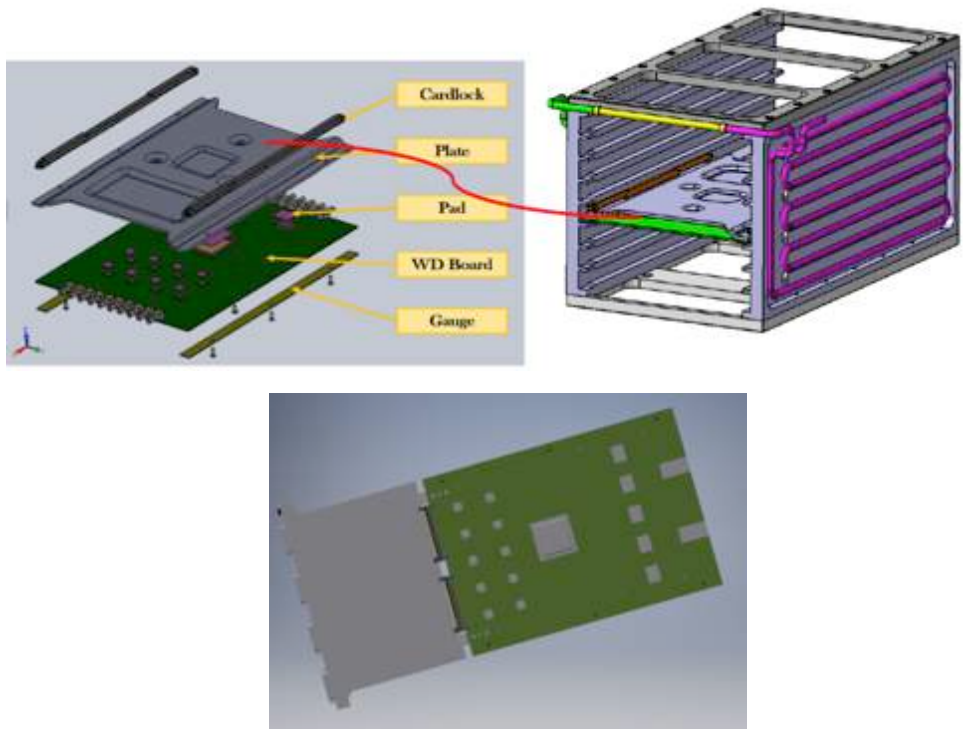

Figure 6.22: *Details of the crate cooling: (top) layout of the WD boards and (bottom) CAD drawings of MB+WD board.*

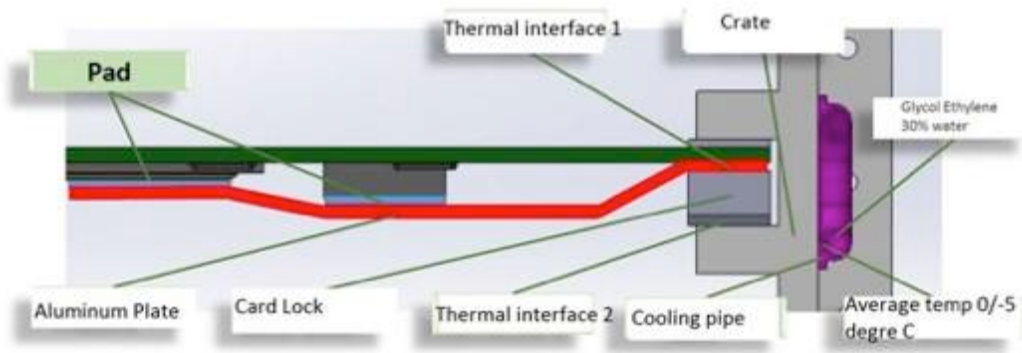

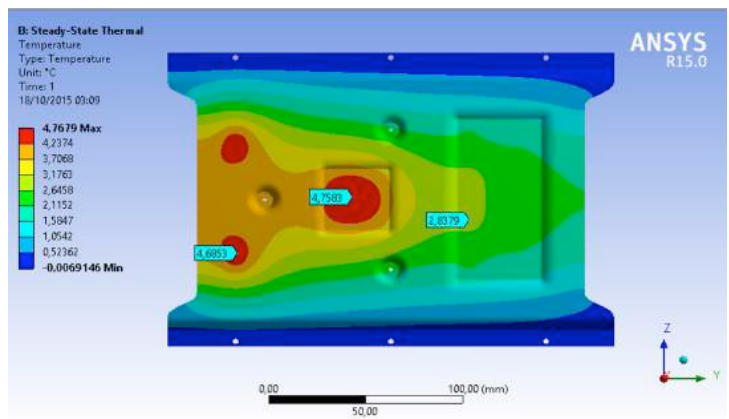

Figure 6.23: *Details of the card cooling: (top) layout of the plates and card locks, (bottom) results of the thermal analysis on the WD board.*

## 6.3   Back plate design

The back plate has to position the SiPM/FEE holders. Each SiPM holder is connected directly to a cooling line see Fig. 6.23. The cooling tubes are vacuum brazed to a C profile. The choice of cooper type, the brazing material and the join technology is compatible with high vacuum application. The SIPM holder is restrained with four M2.5 screw to the copper profile. This adopted solution has several advantages. It assures a good thermal performance because the cooling tube is embedded in the copper and has a good thermal contact with SIPM holder. Furthermore, it is a simple part that can be tested and mounted screwing it on the calorimeter back plate. It has been made such that the thermal contact with the plate is minimized and it is optimized to be evacuated from the air (see Fig. 6.23).

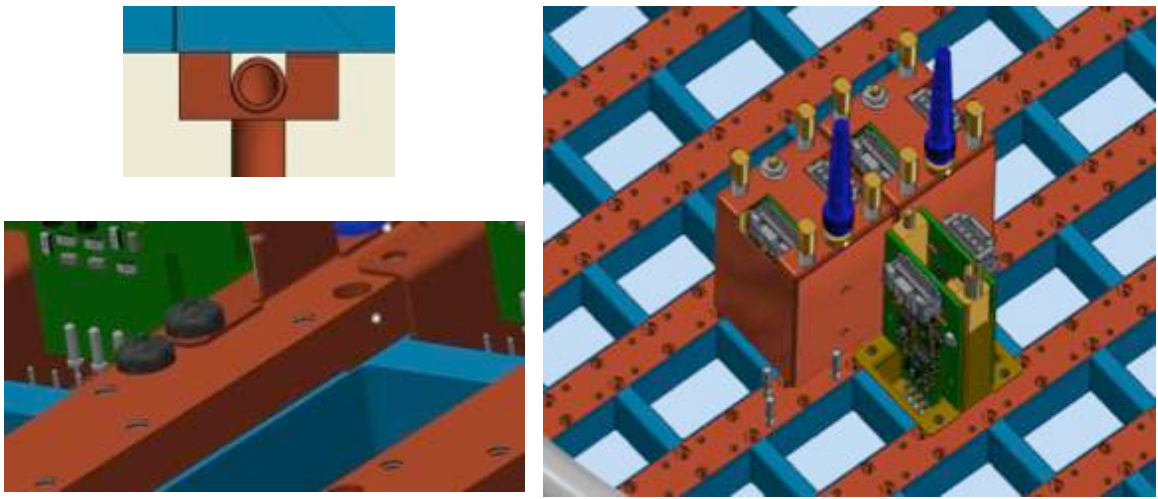

Figure 6.23: *Details of the cooling lines in the FEE back plate.*

The back plane material is special plastic compatible with vacuum. Two valid candidates, PEEK or ZEDEX, are under evaluation. This material choice has some advantages like the low thermal conductive and weight. These plastic candidates have a good machinability. They can be threaded in order to obtain good tolerances.

### *Crate design and board cooling*

The crate design is still in progress and it is following the board design update. The design is challenging because the space available is very limited. We have to respect an envelope, the number of boards and their complexity require detailed topological studies. It is a cooled box that must restrain and cool the electrical component mounted on the cards. Fig. 6.24 (left) shows the body of the crate.

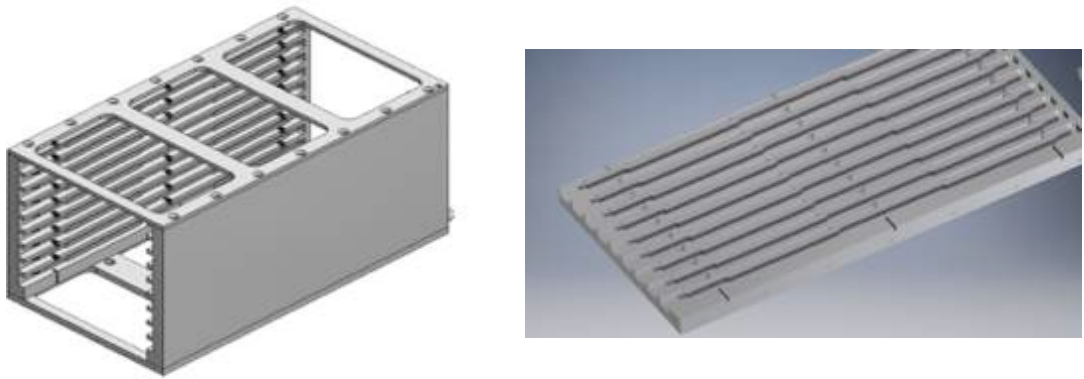

Figure 6.24: *(left) body of the WD crates, (right) lateral plates.*

It is an aluminum box made from four machined plates. The lateral plate (Fig. 6.24 (right)) integrates the cooling channel and the restrain system for the electronic cards. The design is to connect a card lock to the crate, as dictated by the fact that we have two cards (MB+WD) connected together that cannot be mounted as a single piece. Indeed, the space available in front of the crate does not allow mounting the two cards as a unit and the standard use of a card lock implies to have it connected to the card. So we end up having two card locks mounted on a single card. The operation of restrain the two cards and connect them to the connectors is difficult. So we are testing the solution in which we start inserting the WD in the crate but not fully and at a certain moment we connect the MB and we finish pushing both of them in position. Then we can tight the card lock. Fig. 6.25 shows the crate with MB/WD cards mounted.

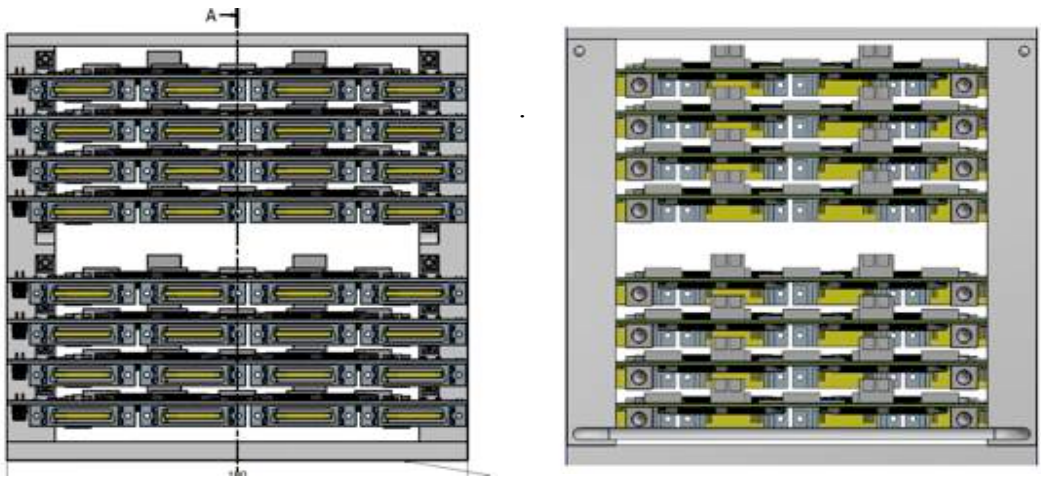

Figure 6.25: *(Left) Rear, (right) front view of a readout crate with MB/WD boards mounted.*

*Calorimeter* – 1860 crystals in 2 disks. There are 240 Readout Controllers located inside the cryostat. Each crystal is connected to two avalanche photodiodes (APDs). The readout produces approximately 25 ADC values (12 bits each) per hit.

# 7 The Calibration systems

*Cosmic Ray Veto system* – 10,304 scintillating fibers connected to 18,944 Silicon Photomultipliers (SiPMs). There are 296 front-end boards (64 channels each), and 15 Readout Controllers. The readout generates approximately 8 values for each hit. CRV data is used in the offline reconstruction, so readout is only necessary for timestamps that have passed the tracker and calorimeter filters. The average rate depends on threshold settings.

*Extinction monitor* – ... will be implemented as a standalone ... does not provide ... information will be forwarded to the DAQ for inclusion in ... database and optionally in the event stream.

*Tracker* – 23,040 straw tubes, with 96 tubes per "panel", 12 panels per "station" and 18 stations total. There are 240 Readout Controllers (one for each panel) located inside the cryostat. Straw tubes are read from both ends to determine hit location along the wire. The readout produces two TDC values (16 bits each) and typically six ADC values (10 bits each) per hit. The ADC values are the analog sum of the two ends of the straw.

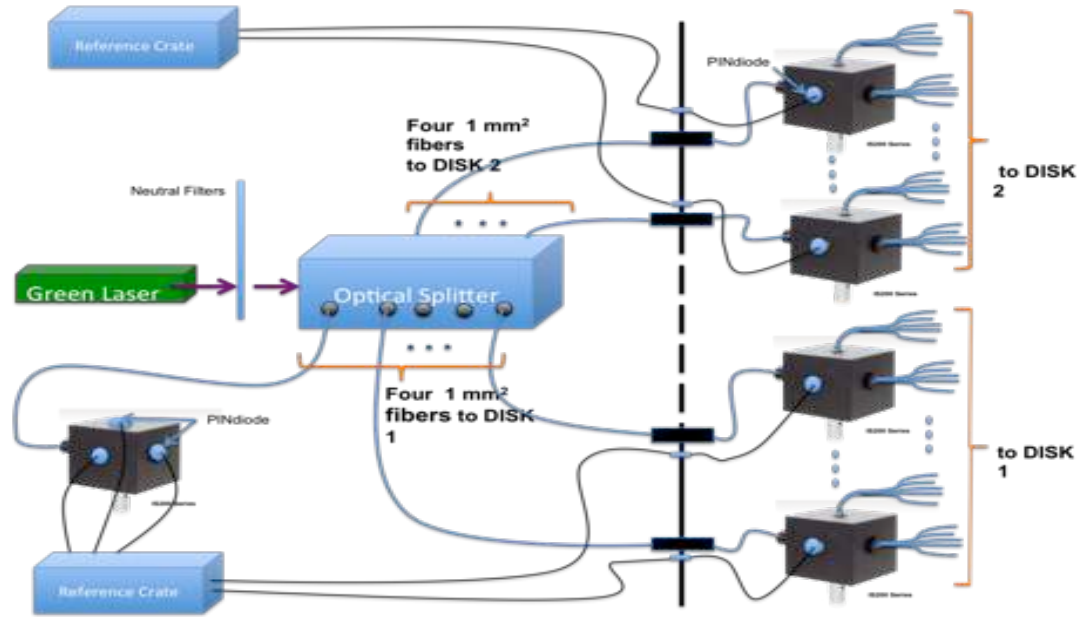

Figure 7-1 *Schematics of the Laser Monitoring System.*

## 7.1 Laser Monitoring System

In order to continuously monitor variations of the SiPM gains and of its charge and timing resolutions, as well as performing a fast equalization of the timing offsets, a laser system has been designed similar to the one used for the CMS calorimeter [7-1]. The UV extended SiPMs require a laser with a wavelength where the sensor has reasonable quantum efficiency. However, in order to disentangle the contribution due to variation of the transmittance in the crystal, we have opted for a green light laser. The contemporary usage of the source system and of cosmic rays samples will allow us to disentangle the contribution due to the crystal changes. The laser light is transmitted by means of an optical distribution system and bundles of optical fibers on the readout side of the detector. The end of each optical fiber is inserted in a long needle that is positioned inside the FEE/SiPM holders in a reproducible way, see Fig. 6.10 (left). The light is then transmitted through the crystal and then diffused by the crystal and the wrapping material to illuminate the active area of the photosensor.

A schematic of the overall system is shown in Fig. 7.1. A high-precision, high-power, pulsed laser sends light through standard collimation optics to an optical splitting system, done with mirrors, to subdivide the beam into 8 equal parts. By means of eight 1-mm diameter, ~ 60 m long quartz fibers, the light is brought to the Detector Solenoid

Instrumented bulkhead where, through a vacuum feed-through, continues its travel to the back face of the calorimeter disks. On each disk, there are four, 2" diameter, integrating spheres with one input for the incoming fiber and three outputs for the bundles, 1 output for pin diode. Running from three of the outputs is a bundle of 70 200-μm diameter fused silica fibers, for a total of 840 fibers/disk. Of the 840 fibers/disk, 674 are used for gain calibration, the remaining 166 are replacements in case some fibers are broken during handling or installation. The 4 hole of the sphere has a pin-diode for reference.

The light output from the laser system is monitored with pin-diodes that measure the output light from the laser and the light diffused from the integration spheres. A total of 8+4 pin diodes are needed. The pin diode monitors are required to track pulse amplitude variations larger than 1%. The laser and the monitor boxes will be temperature controlled to reduce the variation of the laser to a few percent and to minimize the pin-diode temperature correction. In order to monitor calorimeter response linearity, a neutral filter wheel with gradually changing absorption values is inserted between the primary beam and the light distribution system.

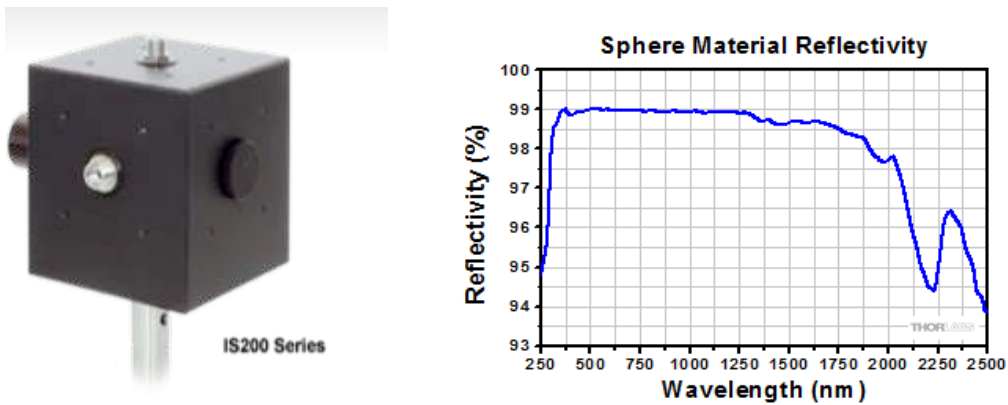

Figure 7.2 *Picture of the ThorLab IS-200 integrating sphere (left); and the sphere's reflectivity dependence on wavelength.*

There is not a stringent requirement on the laser pulse width, since the SiPM readout electronics has a rise time of 25 ns, thus setting an upper limit on the width of 10 ns. Similarly, the pulse frequency is not strongly constrained since, as shown in the prototype test, running at 1 Hz provides better than per-mil statistical precision in one hour of data-taking. It is instead mandatory to synchronize the laser pulse with an external trigger to allow the light to reach the detector at the correct time with respect to the proton beam pulse. Plan is that laser data can be acquired during the gaps between beam when the calorimeter is quiet. Special running with laser data acquired simultaneously with the calorimeter acquiring physics data is foreseen. The laser pulse energy is strongly attenuated by the distribution system. However, the laser signal is required to simulate a 50 MeV energy deposition. For CsI  this corresponds to ~ 1,500 p.e. in each photosensor.

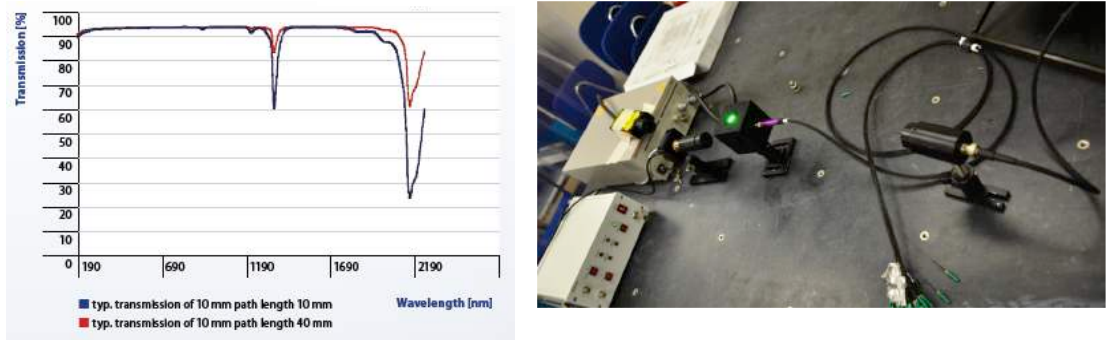

Figure 7.3 *Transmission as a function of wavelength for fused silica fibers (left) and a picture of the light distribution system prototype (right).*

This roughly translates to few nJ energy source. A safety factor of 20 is designed into the system to account for the eventual degradation of the signal transmission with time, resulting in an energy pulse requirement of ~ *0.1* μJ. There is a stringent requirement on the fibers. They should have high transmission at 540 nm, a small attenuation coefficient and they must be radiation hard up to O(100 krad). The best choice is fused silica fibers, both for their transmission properties (see Fig. 7.3-left), a nearly flat wavelength dependence, a long attenuation length and high radiation tolerance.

### Laser monitor prototype for the LYSO crystals

The setup used for the transmission test and for the calibration of the LYSO calorimeter prototype is shown in Fig. 7.3-right. The light source was an STA-01 solid-state pulsed laser emitting at 532 nm with a pulse energy of 0.5 μJ, a pulse width < 1 ns, good pulse-to-pulse stability (3%), and synchronization to an external trigger for frequencies up to 100 kHz. Tab. 7.1 summarizes the performance of equivalent STA-01 lasers emitting in the UV region that are being evaluated for the final implementation. The prototype distribution system uses a 2" integrating sphere, the ThorLab-IS200, with one input port and 3 output ports. Each of the output ports has a 0.5" diameter. Pictures of the sphere and of its reflectivity diagrams are shown in Fig. 7.2 (left). A Hamamatsu Pin-Diode S1722-02 is mounted in one sphere port to monitor the laser pulse variation, while a bundle of fifty 2 m long, *Leoni* fused silica fibers of 200 (400) μm diameter core (core plus cladding) is inserted with an SMA connector to another port.

The number of photoelectrons, Npe, observed at the end of the transmission line has been determined by a direct measurement of the APD charge seen in the calorimeter. The input laser source was first reduced by a factor $T_{filter}$ = 200 by means of a neutral density filter, in order to avoid signal saturation. The average APD charge, with the APD gain set to 50, was around 120 pC, with a channel-by-channel spread of ± 10%. This corresponds to Npe = 33,600, a factor of 20 more than required in the CsI case. However, this determination does not take into account the reduction factor of 14 that results from the initial optical splitting system and for the factor of 2 in the energy ratio between UV and green light.

Table 7.1 Main properties of STANDA Lasers operating in the UV region

| Models | STA-01-TH | STA-01-FH |
|---|---|---|
| Wavelength, nm | 354 | 266 |
| Average output power (max), mW | 15 | 20 |
| Pulse energy, μJ | > 1.5 | 20 |
| Pulse duration, ns | < 0.5 | < 0.5 |
| Repetition rate (max), Hz | 10,000 | 0.1 - 1000 |
| Beam profile | | $M^2 < 1.2$ |
| Pulse spectral structure | | Single longitudinal mode |
| Polarization ratio | | > 100:1 |
| Beam waist diameter inside the laser head 1/e$^2$, μm | | 25 - 200 |
| | | < 5 (near transform limited) |
| Pulse spectrum FWHM, pm | | < 0.6 |
| Pulse-to-pulse energy stability rms | | < ± 1.5% |
| Power stability over 6 hours | | 100 - 240 |
| External power supply voltage, VAC | | 15 - 40 |
| Operating temperature °C | | USB, External trigger (TTL rising edge) 1 Hz max repetition rate |

The measured Npe is consistent with the pulse energy and distribution losses. One photon at 520 nm corresponds to $4 \times 10^{-19}$ J, so that in a single laser pulse $\sim 10^{12}$ photons are produced. Using the measured $T_{fiber}$ and $T_{filter}$, the light transmitted at the end of the chain is estimated to be $N_{photon} = 10^{12} \times (7 \times 10^{-5}) \times 0.005 = 3.5 \times 10^5$. Correcting this estimate for the PD quantum efficiency of 70% and for the APD/crystal area ratio of 1/9, 27,000 detected photoelectrons are expected, in reasonable agreement with the measurement.

The prototype calibration system has been tested by measuring the transmission at one of the output ports, $T_{port}$, and by measuring the transmission at the end of the fiber bundle, $T_{fiber}$. The transmission in one port can be written as $T_{port} = (S_{port}/S_{sphere}) \cdot M$, where S represents the surfaces and $M = R/(1-R \cdot (1-f))$ is the sphere multiplication factor. R is the sphere reflectivity and f is the ratio between the ports and the sphere surfaces. At $\lambda > 400$ nm, R is 98%, f is ~5% and M is ~16, so that the transmission factor is ~0.012×16=0.192. The first measurement was performed by calculating the ratio between the light emitted by the laser and the light exiting from the sphere port. A calibrated photocell of 13 mm diameter has been used, positioned at zero distance from the hole. The transmission measured is ~0.12, in reasonable agreement with our simplified model. Similarly, the transmission at the end of the fiber bundle has been measured, resulting in an average factor of $T_{fiber}=7 \times 10^{-5}$ that, as expected, is much better than the product of the simple geometrical ratio, $10^{-5}$, and the fiber numerical aperture. The spread of transmission values for the best 43 fibers in the bundle has a $\sigma = 8\%$. The seven remaining fibers were accidentally cracked before the test, showing deviations worse than a factor of two.

Finally, in Fig. 7.4 (left) the variation of the observed laser pulse as a function of the running time is shown. The average laser fluctuation observed in 12 hours of running has a $\sigma$ ~5% and is mainly due to the variation of the APD response. This is shown by comparison with the reference pin-diode (green circles), which is much flatter than the calorimeter response. The residual fluctuation of the calorimeter to pin-diode ratio is 3.5%. This is much worse than the ratio between two calorimeter channels (red points) that is at a level of 0.4% and of the PIN-diode, which is 1.6%. To confirm this, the dependence of the calorimeter response on the temperature has been studied by measuring the temperature in the APD region with a PT-100 probe (Fig. 4.right). The APD gain dependence on temperature is consistent with the observed residual calorimeter/pin fluctuation as shown by the anti-correlation between the APD temperature and the calorimeter response in Fig. 4 (right). The gain variation of the APD corresponds to ~ -4 %/C°.

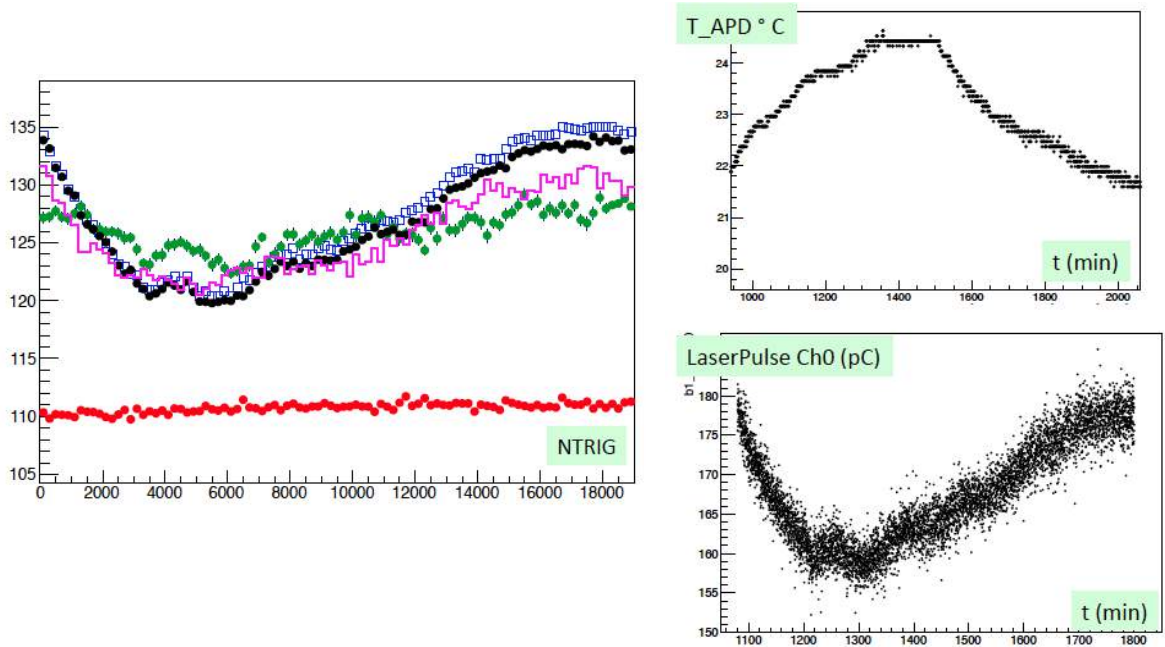

Figure 7.4 . *The left plot shows the distribution of the average laser pulse energy as seen by two calorimeter channels as a function of the trigger number (black and blue points), of the average pin diode response (green points) and of the ratio between the calorimeter channels (red points). The purple histogram shows the ratio between calorimeter channels and the PIN-diode, isolating the residual fluctuations of the APD gain. Also shown is the distribution of the temperature as a function of running time in minutes (top right), and the variation of the laser pulse for channel 1 during the same period (bottom right).*

## 7.2 Source Calibration System

Calibration and monitoring while physics data is being accumulated is an important ingredient if the best possible performance of the calorimeter is to be realized. A suitable system must provide precise, independent crystal-by-crystal calibration. The use of radioactive sources is a proven technique for accomplishing such a calibration. However, most long-lived sources are limited to an energy around 1 MeV, which makes it difficult to secure a signal that is significantly above electronic noise, and sources that must be deployed individually are not practical with a system of ~2000 crystals. Mu2e has adopted an approach formerly devised for the *BABAR* electromagnetic calorimeter [7-2]. In this system, a 6.13 MeV photon line is obtained from a short-lived $^{16}$O transition that can be switched on and off as desired. This system was successfully used for routine weekly calibrations of the *BABAR* calorimeter. It is an ideal match to the Mu2e requirements, and we have salvaged components from the *BABAR* system that may be used in Mu2e, and for use in the prototype that has been constructed.

The decay chain producing the calibration photon line is:

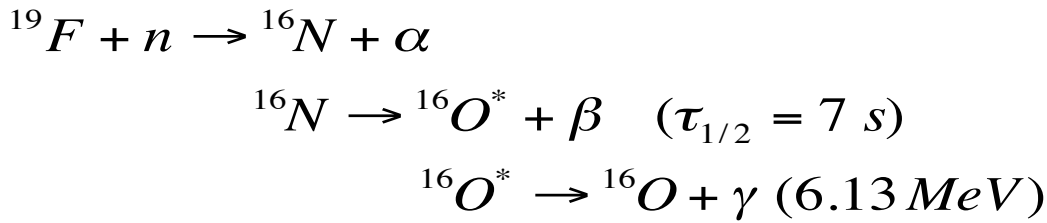

$$^{19}F + n \rightarrow {}^{16}N + \alpha$$
$$^{16}N \rightarrow {}^{16}O^* + \beta \quad (\tau_{1/2} = 7\,s)$$
$$^{16}O^* \rightarrow {}^{16}O + \gamma\ (6.13\,MeV)$$

The fluorine, a component of Fluorinert™ coolant liquid, is activated with a fast neutron source, producing the $^{16}$N isotope. This isotope then $\beta$-decays with a half-life of seven seconds to an excited state $^{16}$O*, which in turn emits a 6.13 MeV photon as it cascades to its ground state. A source spectrum collected with the Mu2e prototype is shown in Figure 7.5. There are three principal contributions to the overall energy distribution: one peak at 6.13 MeV, another at 5.62 MeV and a third at 5.11 MeV, the latter two representing $e^+e^-$ annihilation photon escape peaks. Since all three peaks have well-defined energies, they simultaneously provide both an absolute calibration and a measure of the linearity of response at the low end of the calorimeter energy scale.

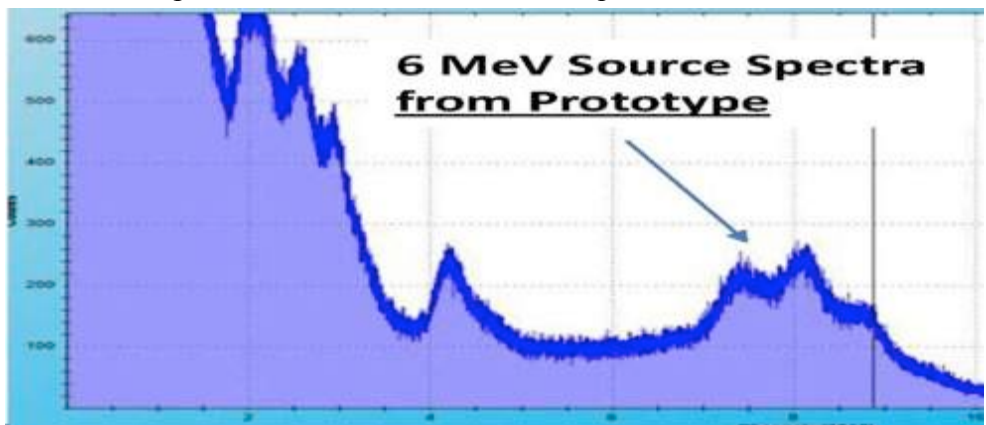

Figure 7.5. *Spectrum using the Mu2e 6 MeV prototype calibration system. The crystal is BaF$_2$. Three peaks from the source are indicated on the right. The lower peaks are due to radium contamination in the crystal.*

The fluorine is activated using neutrons provided by a commercial deuterium-tritium (DT) generator producing 14.2 MeV neutrons, at a rate of about $10^9$ neutrons/second, by accelerating deuterons onto a tritium target. The DT generator is surrounded with a bath of the fluorine-containing liquid Fluorinert™, which is then circulated through a system of manifolds and pipes to the calorimeter crystals. Many suitable fluorine-containing liquids are commercially available; Fluorinert™ "FC-77" was used in *BABAR*. In Mu2e, FC-770 will be used, since FC-77 is no longer considered environmentally acceptable. We have analyzed samples of FC-770 and find that it contains a good concentration of fluorine. The FC-770 fluid is stored in a reservoir near the D-T generator. When a calibration run is started, the generator and a circulating pump are turned on. Fluid is pumped from the reservoir through the DT activation bath and then to the calorimeter. The system is closed, with fluid returning from the calorimeter to the reservoir.

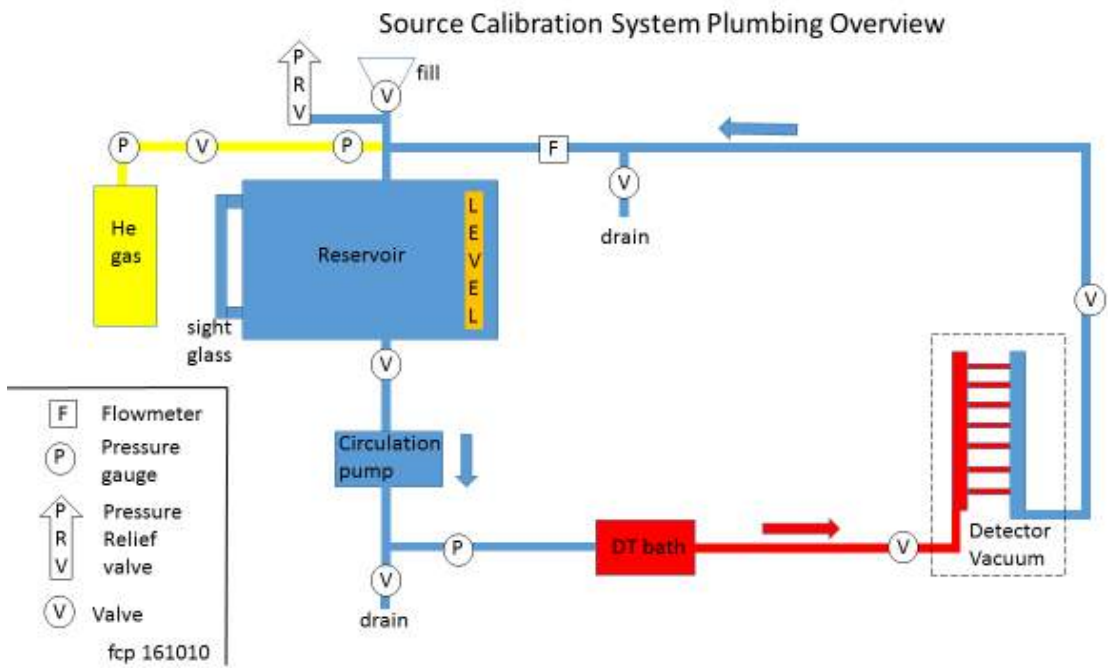

Figure 7.6 *Schematic of the pumping system for the source.*

A schematic of the Mu2e system, based on the *BABAR* system, is shown in Figure 7.6. In operation, the circulation pump pumps the fluid at about 3.5 l/s to the ``DT bath'', a reservoir containing the DT generator. With the DT generator on, the $^{16}$N is produced in the bath. The activated fluid exits the bath and travels through piping to the detector, where it is distributed to 20 (10 per disk) thin wall aluminum tubes arranged near the front faces of the crystals. After passing in front of the crystals, the fluid is returned to the reservoir, where it is available for recirculation through the pump.

The arrangement of the thin wall tubing that irradiates the crystals is shown in Figure 7.7. The tubing is 0.5 mm aluminum, with 3/8 inch outer diameter. The tube centerlines are held in a milled Rohacell foam structure at a distance of 20 mm from the front face of the crystals. The arrangement is designed to provide nearly uniform irradiation of the crystals with minimal material.

When the calibration is completed, the FC-770 is drained by gravity from the thin wall tubing on the detector to the reservoir. While draining is not strictly necessary, this has three benefits: (i) The material in front of the calorimeter is kept at a minimum (Fluorinert has a radiation length of 20 cm, presenting less than 0.2% of a radiation length at the maximum distance); (ii) Potential background in the calorimeter from fluorinert activation by background neutrons is eliminated; (iii) The potential for leakage into the vacuum chamber is minimized. Because the reservoir must be located not far (about 5 feet) above the pump to accommodate draining, a small gas pressure (several PSI) is maintained in order to provide sufficient pressure head during operation.  This is represented with the ``He gas'' bottle in Figure 6.  Other gases could potentially be used, as long as they are not susceptible to activation or highly soluble in Fluorinert.

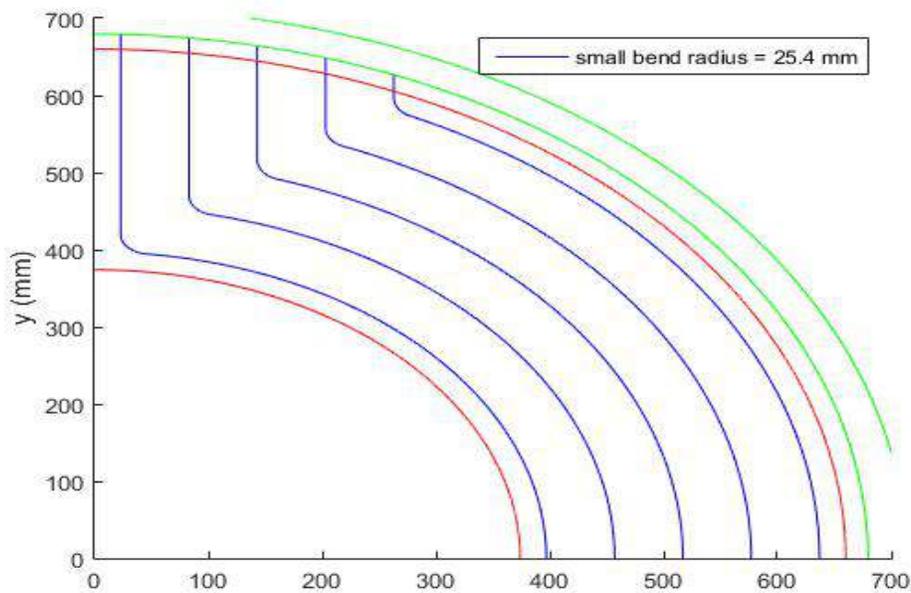

Figure 7.7 *Distribution of source system piping in front of the calorimeter surface.*

The DT neutron generator is a small accelerator. Radiation safety protocols factor into the design of the calibration system, and operation of the source will be done remotely. The half-life of the activated liquid is 7 seconds; residual radioactivity is thus not a substantial concern when the DT generator is not operating. The DT generator will be shielded according to FNAL safety regulations. The fluid reservoir is capable of holding the entire volume of Fluorinert™ fluid required for operation of the system. In the event of a fluid leak, the maximum exposure for the *BABAR* system was calculated to result in a maximum integrated dose of less than 1 mrem. For Mu2e, a detailed hazard analysis will be performed in collaboration with Fermilab radiation safety experts. Operation of the system is anticipated to be approximately weekly during Mu2e running.

### *Salvage of System Components from SLAC*

Several of the components of the source calibration system used at *BABAR* have been salvaged for use by Mu2e, and are presently used in the prototype at Caltech. These items include:

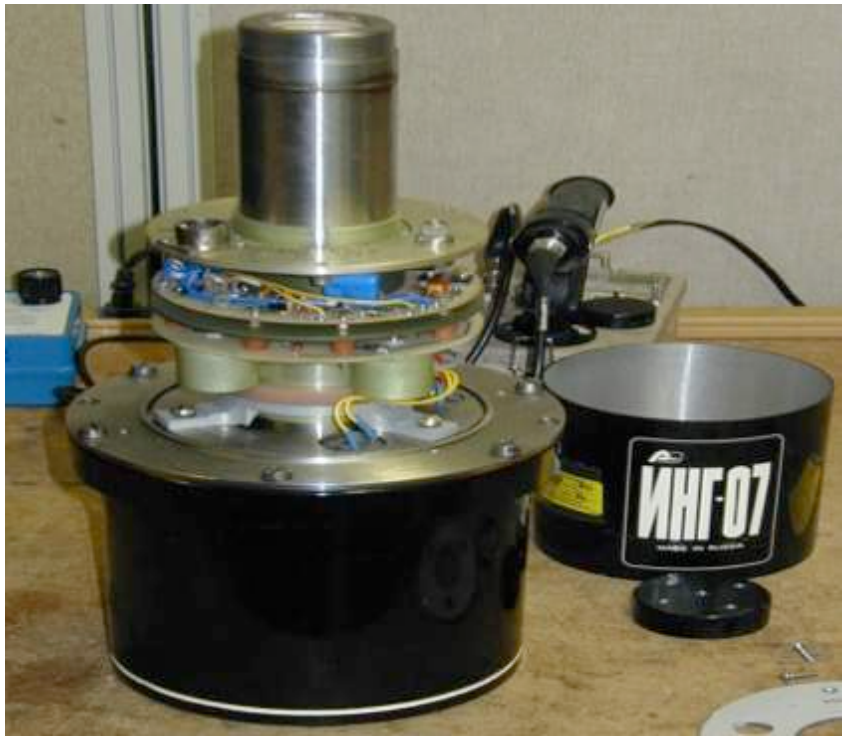

Figure 7.8 *The model ING-07 DT generator salvaged from BaBar and in use in the Mu2e prototype source calibration system.*

- the *BABAR* DT generator, model ING-07, manufactured by the All-Russia Institute of Automatics (shown partially disassembled prior to installation at *BABAR* in Figure 7.8), including HV power supply, PC-interface controller card and cabling;

- elements of the fluid distribution system, including the primary outgoing and incoming manifolds, valves and pressure gauges, and the main distribution panel on which many of these items are mounted;
- the circulation pump for the fluid activation loops.

The circulation pump has been refurbished and is in use in the prototype system. It will also be used in the final system at Fermilab. The DT generator from BaBar is being used in the prototype system. However, because of its age and the 12.3 year half-life of tritium, the plan is to purchase a new generator for the final Mu2e system.

### The Prototype Source System

A prototype for the source calibration system has been constructed and operated in the underground isotope laboratory at Caltech. Figure 7.9 is a photograph of the completed system. The bunker containing the DT generator and surrounding Fluorinert bath is constructed from layers of lead, paraffin, and concrete. The reservoir is salvaged from BaBar. A new reservoir will be constructed for Mu2e to meet ASME specifications.

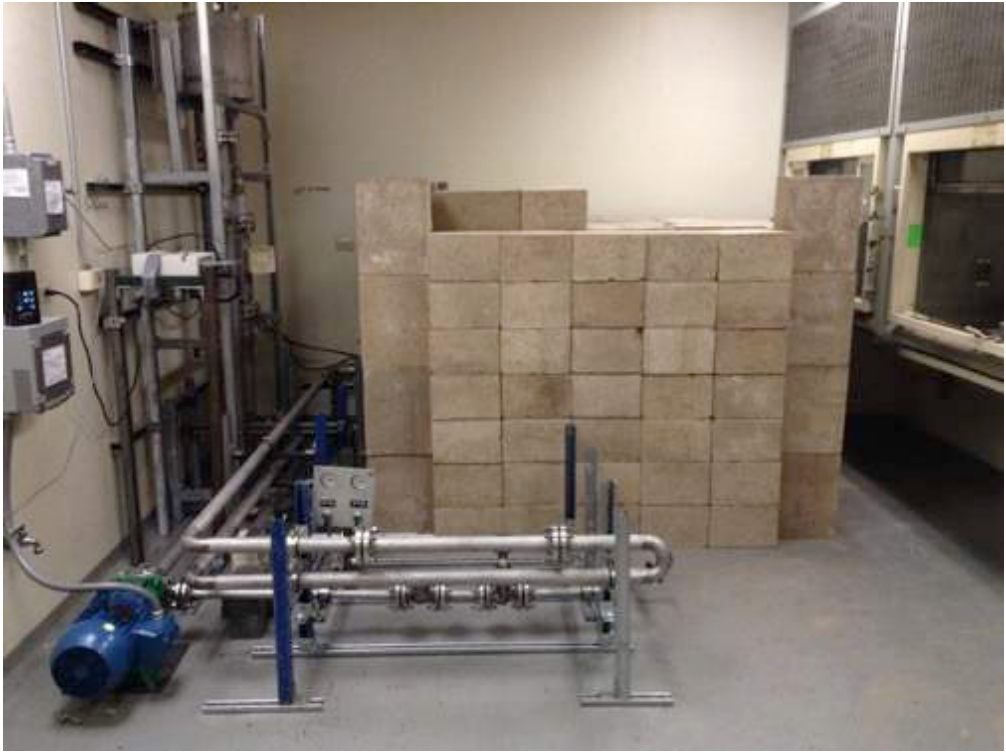

Figure 7.9. *The prototype source calibration system at Caltech. The DT generator and Fluorinert bath are hidden in the bunker. The refurbished circulation pump is in the lower left with its associated controls and power source above. The reservoir is visible in the upper left of the picture. The "test" section, where crystals may be irradiated, is the section of plumbing between the two flanges in the front.*

The DT generator and surrounding Fluorinert bath are shown in the partially constructed bunker in Figure 7.10. The Fluorinert bath is a new one meeting ASME standards, and

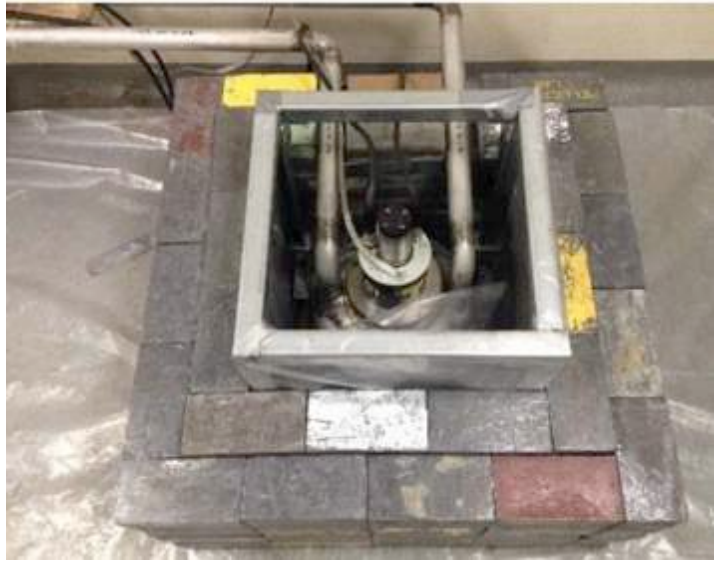

Figure 7.10. *The DT generator in place in the Fluorinert bath in the partially constructed prototype source bunker.*

will be used in the final Mu2e system. The DT generator has been re-trained, with electrical performance consistent with that achieved prior to BaBar decommissioning, Figure 7.11. Figure 7.5 illustrates a spectrum that has been obtained with this prototype

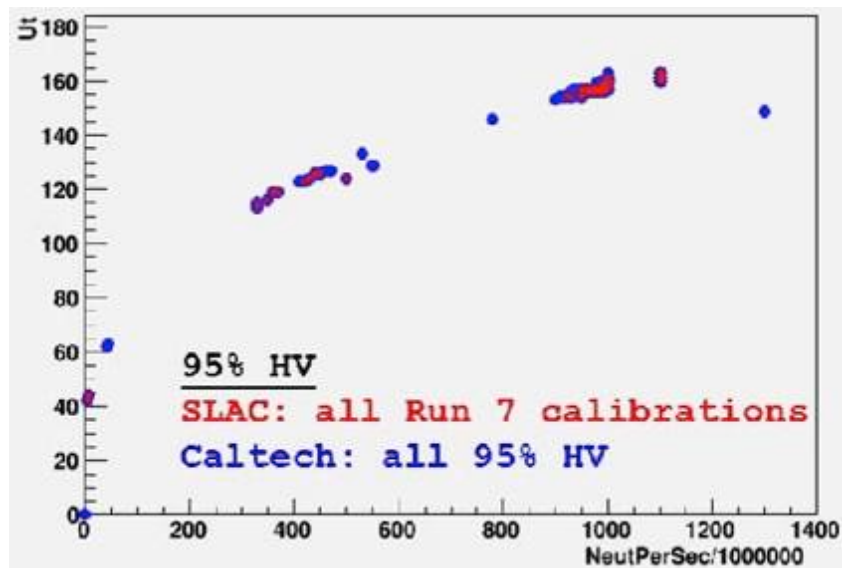

Figure 7.11. *The DT generator performance in the prototype at Caltech (blue), compared with before decommissioning of BaBar (red). Note that the horizontal axis corresponds to neutrons/second for a fresh charge of tritium; it is not an actual measurement of neutrons.*

source.

### Implementation for Mu2e

The source calibration system for Mu2e is designed to provide a weekly calibration of the entire calorimeter [7-3] in about 10 minutes of data acquisition. The design precision requirement for the CsI calorimeter is better than 0.14 MeV at the 6.13 MeV line, or better than 2.3%. This yields a negligible contribution to the overall resolution of the calorimeter.

The number density of fluorine in Fluorinert™ FC-77 is approximately $4 \times 10^{28}$ m$^{-3}$, essentially all in the desired $^{19}$F isotope.  There is some uncertainty in this number density as the proprietary formulation is not precisely known; we work with a worst-case assumption. We have verified by chemical analysis of samples that the number density of fluorine in FC-770 is approximately the same as in FC-77. The viscosity, at 0.8 centiStokes, is similar to that of water. The radiation length of FC-770 is approximately 20 cm. The density of FC-770 is 1.8 g/cm$^3$ (3M MSDS, accessed 161201)

The relevant $^{19}$F(n,alpha)$^{16}$N cross section is about 24 mb  [7-4]. The total inelastic cross section is around 80 mb, dominated by $^{19}$F(n,2n)$^{18}$F. The elastic cross section is much larger, at about a barn.

The bath irradiated by the DT generator has a volume of about 20 liters, with the fluid pumped at a rate of 3.5 l/s; for a neutron rate of $10^9$ n/s, the density of $^{16}$N at the bath exit is about $3 \times 10^9$ m$^{-3}$. With decays, this is attenuated by a factor of 0.4 by the time the fluid reaches the furthest crystals in the calorimeter.

The presence of the DT generator was designed into the Mu2e experimental hall, such that it sits in a below-ground bunker, shown in Figure 7.12. The bunker is covered with layers of lead, polyethylene, and concrete so that the operating radiation level produced by the generator in the hall is < 5 mrem/hour, Figure 7.13. The DT generator (ING-07) is sensitive to magnetic fields above 7 gauss. The field (from the detector solenoid) in the pit is likely to be of this order, perhaps as much as tens of gauss, so we plan to incorporate magnetic shielding when the DT generator is installed.

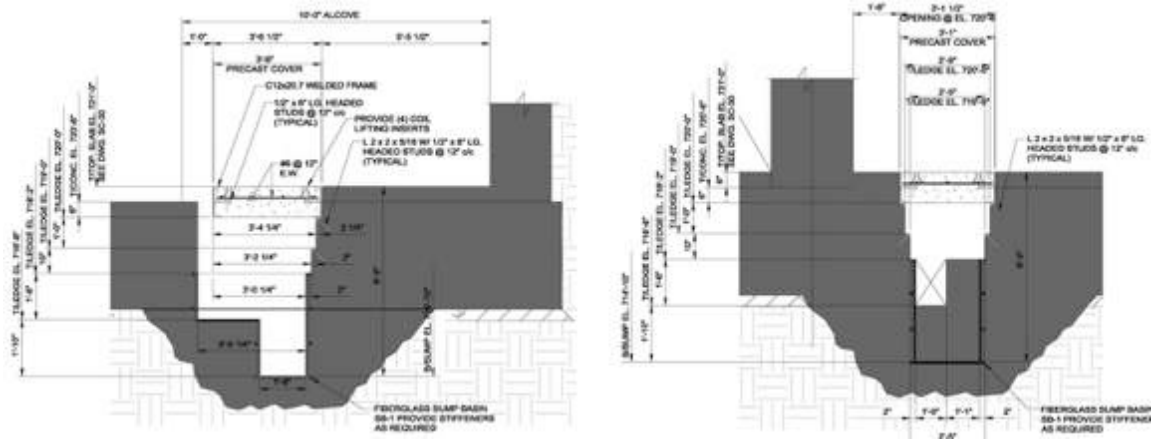

Figure 7.12. *Two elevation views of the DT generator bunker poured below floor level in the Mu2e experimental hall. The stepped structure in the top portion provides support for the lead, polyethylene and concrete layers of shielding covers.*

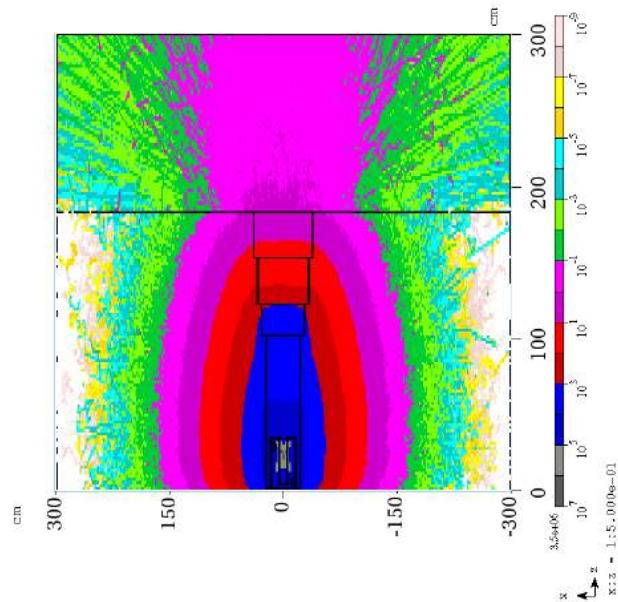

Figure 7.13. *MARS calculation (A. Leveling) of the radiation does from the DT generator in its bunker at Mu2e. The colored scaled units are mrem/hr.*

The source calibration bunker, reservoir, circulation pump, and associated instrumentation is located in "equipment alcove A", near the detector feed-through bulkhead. The basic layout may be seen in Figure 7.14, showing the Mu2e detector hall during construction. From the DT generator bunker, a 1.5" NPS, schedule 10S stainless steel pipe passes through a small trench connecting to a larger longitudinal trench carrying detector cables and services. The plumbing then follows this larger trench from which it emerges at the detector feedthrough bulkhead. Provision is made for operation with the detector in both the commissioning position and in the installed position in the detector vacuum chamber, by rerouting the plumbing in the longitudinal trench. The return line from the detector passes through the same trenches up to the alcove, where it bends to follow the alcove wall trench. It emerges from the trench at the back of the alcove to rise to the reservoir. From the reservoir a supply line drops to the circulation pump, which pumps the fluid back down into the bunker to the DT generator bath. It is possible that the return lines from the detector will be 2 inch tubing instead of 1.5 inch NPS. Welded connections will be used wherever possible, with Swagelok™ fittings used where disconnects are needed.

Inside the detector vacuum the main supply and return lines will be 1.5 inch 316L

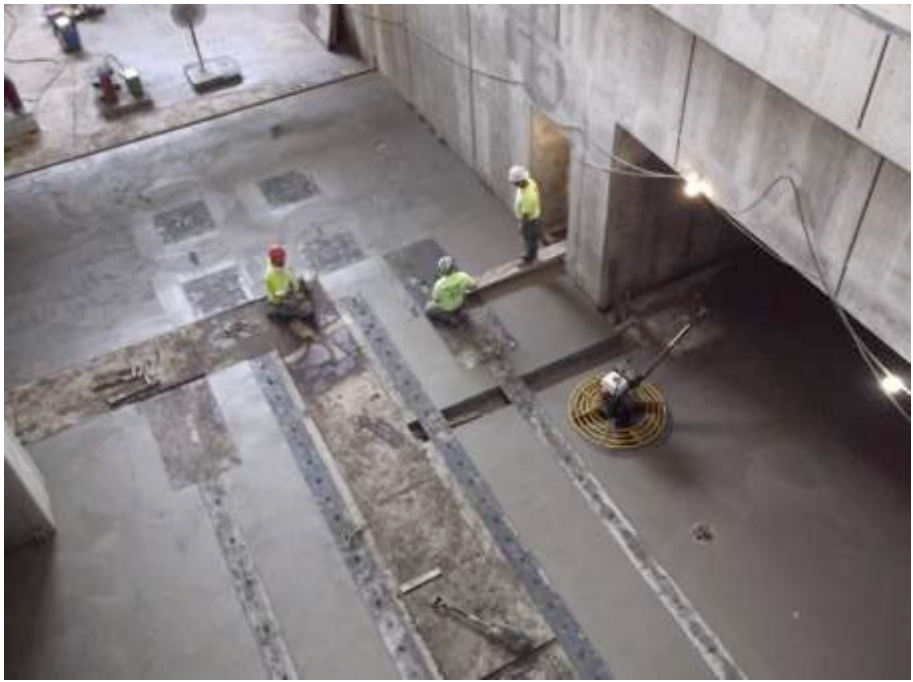

Figure 7.14. *Looking down on the Mu2e detector hall. Equipment alcove A is to the right. Towards the top of the picture within the alcove, a square outline of the DT generator bunker is visible, with a trench extending out to the hall and another trench following the top wall of the alcove. There is a larger covered longitudinal trench connected where the source trench terminates near the center of the hall.*

stainless tubing. This alloy has low magnetic permeability, and is suitable for use in the magnetic field. The tubing wall thickness is 0.095 inches, which is thick enough for Swagelok™ connections, but otherwise chosen to be thin to maximize the internal diameter. These lines must traverse the distance past the muon beam stop before reaching the calorimeter. At the calorimeter, a tee connection for the two disks is made. The thin-wall tubing in front of the disks is connected to aluminum tubing manifolds, with 33 mm OD and 1.5 mm wall thickness. The aluminum manifolds are connected to the stainless tubing with a transition piece probably formed with explosion bonding, so that the aluminum is welded to the aluminum side of the transition piece and the stainless is welded to the stainless side.

### Simulation study

The source calibration has been simulated using the Mu2e Monte Carlo. Figure 7.15 shows the overall spectrum from the calibration source. The largest contribution is from Compton scattering and leakage. The calorimeter resolution is not sufficient to resolve the individual calibration lines, but a broad peak is produced that may be fit to the sum of three lines.

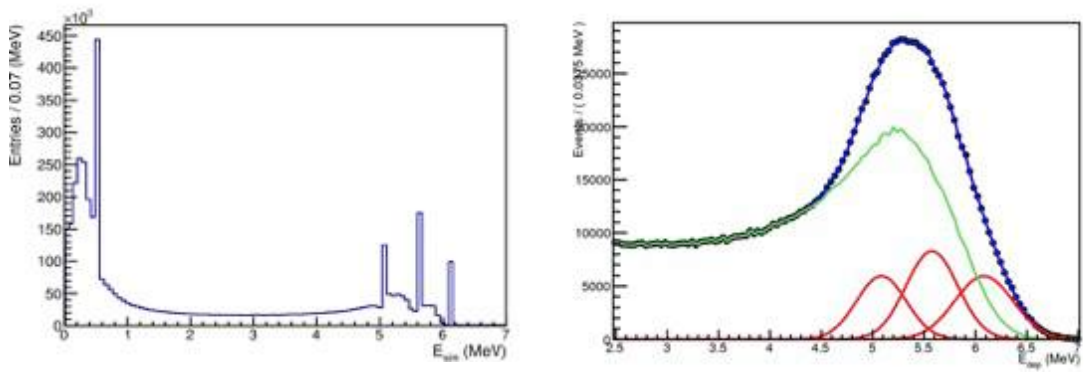

Figure 7.15. *The simulated spectrum in a CsI crystal from the source calibration. Left: Generator-level spectrum. Right: Spectrum as measured. The overall spectrum is shown in blue, the contribution from Compton scattering in green, and the three source lines (including the two escape peaks) in red.*

A routine weekly calibration is planned to take approximately ten minutes, for between 10-20k photons per crystal. Figure 7.16 shows what a 10k photon calibration spectrum would look like. This represents a typical spectrum that may then be fitted to determine the calibration constants for a given crystal. Figure 7.17 shows the distribution of fit results for a calibration with 10k photons in each crystal, according to the simulation.

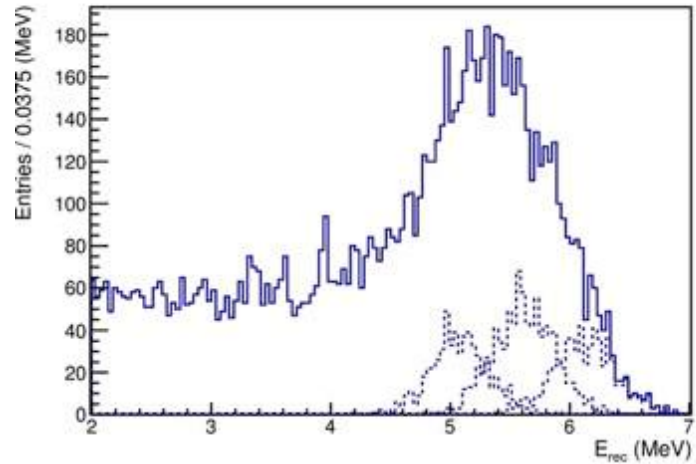

Figure 7.16. *A source calibration spectrum with 10k calibration photons. The dashed spectra show the contributions from the three peaks.*

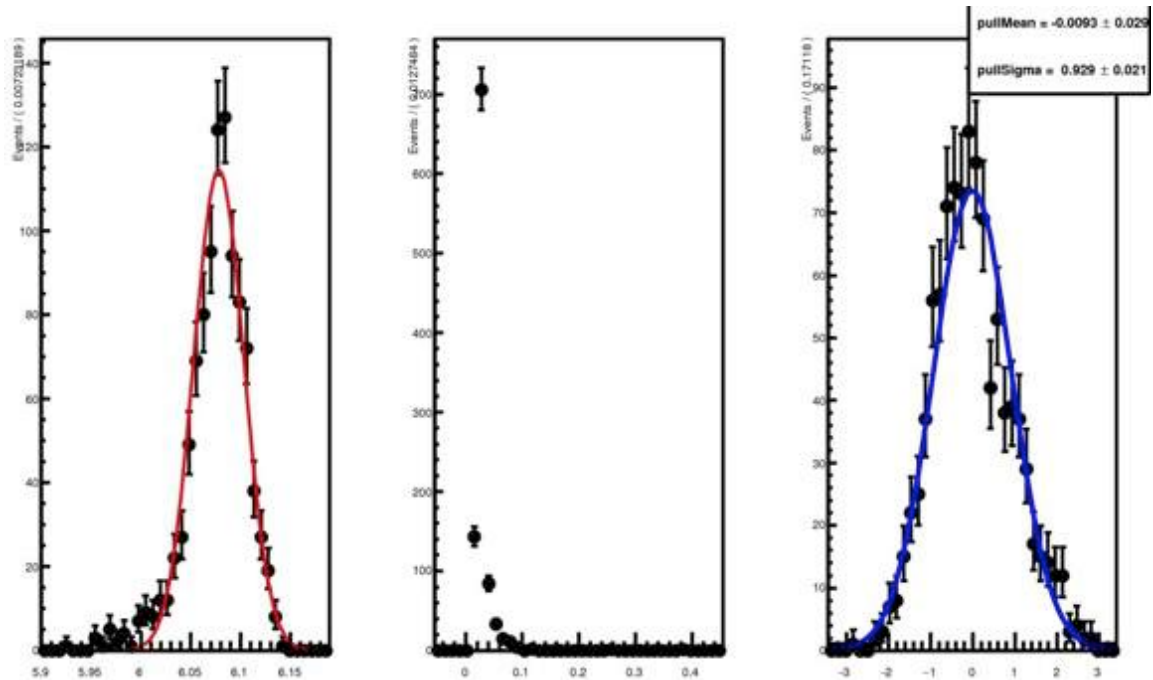

Figure 7.17. *The distribution of fit results in a simulated calibration run with 10k photons in each crystal. Left: The fit result for the mean of the 6.13 MeV peak (note that the calorimeter has not been corrected for the overall offset yet). Middle: The distribution of estimated uncertainties in the mean from the fit. The units are MeV. Right: The normalized error distribution for the fits.*

Figure 7.18 shows how the fit precision depends on the number of photons in a crystal. It is observed that the precision is substantially improved if the relative strengths of the three peaks is known and constrained, although the required precision is achieved even without doing this. We plan to do an occasional long calibration to determine these relative strengths well, on a crystal-by-crystal basis, and then use this constraint in the weekly calibrations.

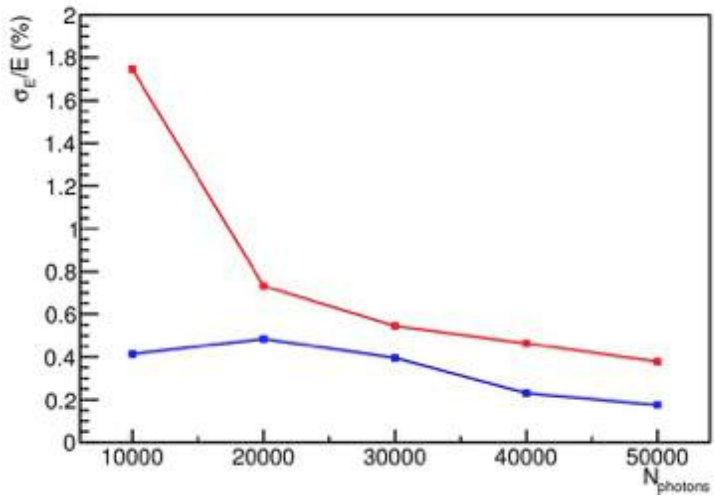

Figure 7.18. *The dependence of the average calibration uncertainty on the number of calibration photons in a crystal. The red curve is for fits in which the relative strengths of the three peaks is unconstrained, and the blue curve is for fits in which the relative strengths of the three peaks is fixed at a known value.*

## 7.3 Cosmic calibration

Cosmic muons represent an important calibration source for the Mu2e calorimeter. They have some unique characteristics which make this calibration complementary to the other calibration techniques described above:

• they can be acquired during normal run operations, in the same experimental conditions of the physics data sample;

• their flux is high enough to collect a large amount of calibration data in a relatively short time, allowing a continuous monitoring of the detector response;

• since they are minimum ionizing particles (MIPs) their energy loss is practically independent of their initial energy;

• they are relativistic particles and, thanks to their negligible energy loss, their speed is practically always equal to the speed of the light c; the time they take to travel through the calorimeter can be used to align the time offsets of all the channels without any external time reference.

Cosmic muons crossing the calorimeter are selected by the dedicated trigger described in section 2.4. Additional cuts on the cluster energy and on the path length inside the crystal can be imposed to purify the calibration sample [7-6].

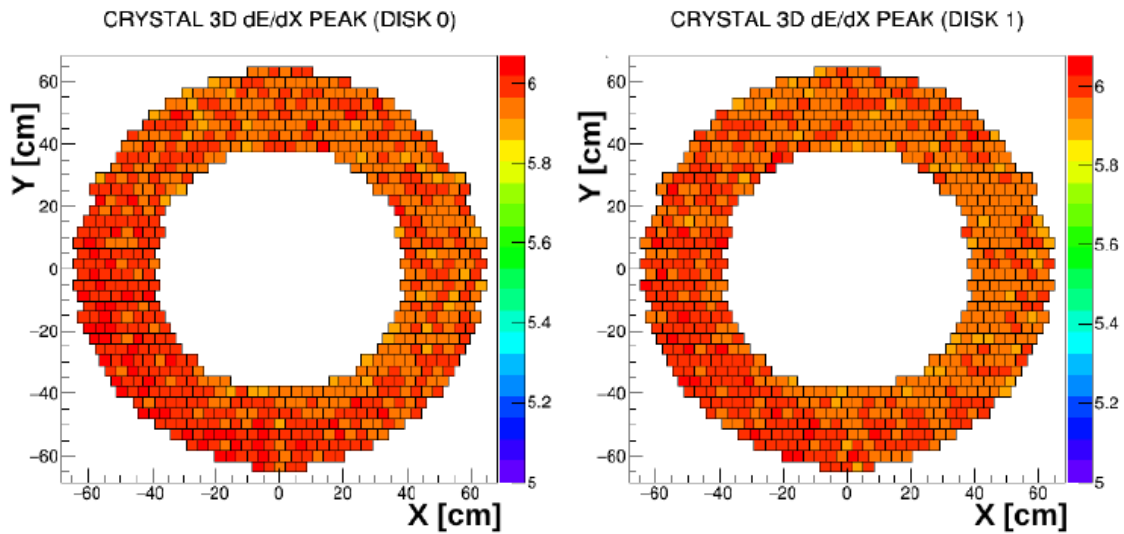

Fig.7.19  *Specific energy loss of selected cosmic muons crossing the calorimeter.*

The specific energy loss, dE/dx, of the selected muons is uniform along the calorimeter (Fig. 7.19). Since the Mu2e calorimeter does not provide a measurement of the z coordinate, the total three-dimensional path lenght cannot be directly obtained by the crystal energy deposits. Nonetheless, if the two-dimensional path length in the transverse plane, that is independent of the z coordinate, is used instead of the total path length, the 2D specific energy loss has still a distribution with a good homogeneity along all the calorimeter. An equalization at the level of 1% can be easily achieved (Fig. 7.20.left).

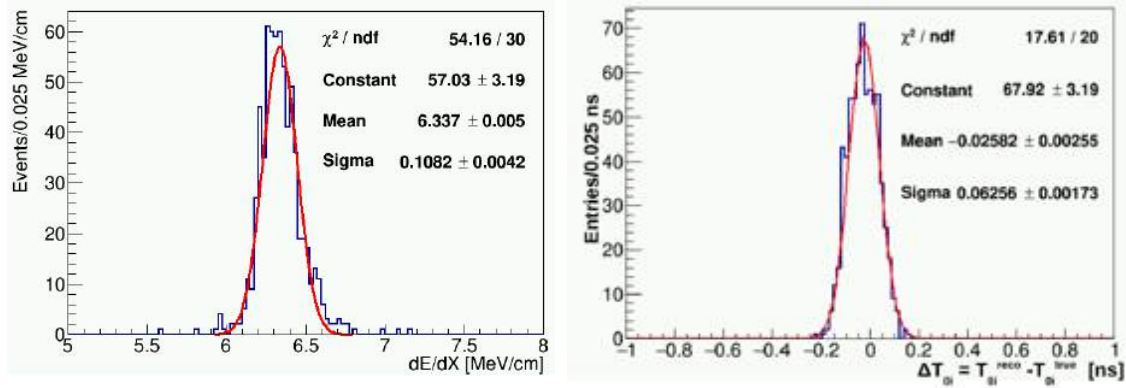

Fig. 7.20  *Specific energy loss after equalization using cosmic muons.*

Cosmic muons time of flight can be used to align the time offsets of the different channels. Time offsets can be aligned using the following procedure:
• the 2D fit of the muon trajectory in the transverse plane is performed. The slope obtained from the fit gives the angle - in the x-y plane;
• only the crystals which have a center located at a distance lower than ½ of the crystal size (17 mm) with respect to the fitted trajectory and a path length in the transverse plane longer than the crystal thickness (34 mm) are considered;
• the crystal center position is converted in the distance travelled in the x-y plane:

$$\Delta y' = \frac{\Delta y}{\sin \phi} = \frac{y - y_0}{\sin \phi}$$

where $y_0$ is the y coordinate of the cosmic muon starting point;
• the measured time in the crystal as a function of its distance $y_0$ is fitted;
• considering all the calibration sample, the average value of the fit residuals is used to estimate $T_0$ for each crystal;
• the values of the measured time are corrected in all the crystals by subtracting their estimated $T_0$.
The procedure can be iterated 5 times to converge to stable results [7-7].
After 5 iterations, an alignment at the level of 60 ps is obtained as shown in Fig. 7.20.right.

# Chapter 7 References

# 8. Risks, EVMS and QA

## 8.1 Risks

There are several risks that could jeopardize the success of the calorimeter subproject. The risks as well as potential mitigation strategies are described below.

The large total ionization dose, TID, and the neutron flux in Mu2e pose several risks to the calorimeter performance. The highest risk is associated to the radiation hardness of the SiPMs. While the effects related to the TID look small, the neutron flux generates a large increase in Idark (c.f.r. 4.5) and thus deteriorates the calorimeter performances. The highest irradiation problem is located in the innermost ring of the first disk where, according to the simulation, the neutron flux is at least three times larger than in the outer regions of the disk. Since the Mu2e front-end has been designed to provide up to 2 mA current to the sensors, we have planned to mitigate the current drawn by the irradiated sensor by cooling them down at temperature of around 0 °C. All mechanics supporting the sensors, the cooling FEE disk and a *tracking independent* cooling station have been designed on purpose for fulfilling this task, while providing a large safety margin in operation. As shown in the cooling section, while running the cooling fluid at -10 °C, the temperature reached by the SiPM is of -4 °C. The cooling fluid can be cool down of additional 5-6 °C without problems providing a margin of -10 °C in operation. We are now completing the prototypes of the cooling disk and we will use this test to certify the cooling studies done with Ansys. At the same time, we will complete a new campaign of sensor irradiation to test their behavior at higher doses and fluencies to incorporate larger safety factors.

A second large risk related to radiation is the possible degradation of the light yield of the crystals and of the related value of the radiation induced noise, RIN. To mitigate this, our QA programs will take into consideration the survey of the produced crystals and the selection of the ones with smaller RIN for positioning them in the innermost radius. Moreover, for large neutron fluxes the pileup in and around the reconstructed cluster could become very important, depending upon the timing characteristics of the selected crystals. Pulse shape analysis can mitigate this risk to a large degree. Minimization of the signal width is underway in the shaping sections of FEE and WD.

The radiation hardness of the electronics, both for FEE, MD and WD, have to be studied further. Dedicated irradiation tests will be planned in order to certify the safety factors assumed and clarify if there is the need of: (i) produce a given amount of spares for replacement during running or (ii) define a more radiation hard piece of electronics for given hot-spot locations.

Finally, the risk that INFN might not be able to commit to the calorimeter construction has been highly mitigated since INFN has signed a Statement of Work with FNAL regarding the participation and the commitment to the experiment. The INFN in-kind contributions have been specified as well as the sharing of work load for assembly and installation. The number of INFN physicists participating in Mu2e can also limit the funds that INFN is willing to commit or limit the funds for travelling, but this last point has been mitigated with the participation to the European Network MUSE [8-1] that provide a large support for INFN travelling.

## 8.2 Value management

A careful value management has been carried out after CD-2 through final design and will be kept also during construction. In particular, a careful examination and validation of detector requirements coupled with evaluation of alternative engineering and design choices has continued, with special attention to balancing performance with schedule and cost.

The option of using large-area SiPMs instead of APDs and pure CsI crystals instead of BaF$_2$ has been considered and selected thus reducing the cost of the basic material. The inherent high gain and lower noise of SiPMs is allowing for a simpler design of the front-end electronics, reducing the HV needs and simplifying the amplifier design. The basic layout of the FEE chain has been kept unchanged but there would be no need to have a DC-DC converter working inside the magnetic field.

## 8.3 Quality Assurance

For a calorimeter of this complexity, Quality Assurance is a fundamental component of the procurement, fabrication and assembly phases. Quality Assurance will be applied to all components and subsystems, building on the relevant experience from the *BABAR*, CMS and PANDA calorimeters and from the Mu2e group itself. Indeed, within the Mu2e collaboration, the expertise that already exists in the construction of the KLOE-2 calorimeter upgrade, as well as the *BABAR* and Super*B* calorimeters have been useful for carrying out the calorimeter R&D program where the QA procedures have been developed.

### 8.3.1 QA for Crystals

In order to construct a high-performance calorimeter that satisfies the Mu2e physics requirements, strict requirements are imposed on various crystal parameters that must be controlled both at the production sites, and upon receipt by Mu2e. The calorimeter is composed of ~1350 un-doped CsI crystals of parallelepiped shape. Each crystal has to satisfy three different QA tests in order to be accepted by Mu2e. These include

1. an optical and dimensional inspection;
2. a test of light yield and longitudinal uniformity of response, LRU;

3.   a test of the RIN.

To ensure timely feedback on crystal production (~100 /month), the use of automated stations is required. In the following, the organization of this effort is described in some detail. Many of these techniques were developed during our studies of LYSO/BaF$_2$ crystals and specialized for the final choice of un-doped CsI. These acceptance criteria are being applied to 72 CsI pre-production crystals from three different producers (Siccas, St.Gobain, Amcrys). Technical specifications for the crystals and QA procedure can be found in dedicated Docdb notes [8-1],[8-2].

### *General inspection and validation of dimensions*

Each crystal is required to satisfy the following:

1.   To be free of cracks, chips and fingerprints. They shall be inclusion-free, bubble-free and homogeneous.
2.   To deviate from a perfect 3-dimensional parallelepiped by less than 100 μm.
3.   Mechanical tolerance of ± 100 μm per side with a 0.3 mm chamfer on all edges.

A visual inspection will be done upon receipt of the crystals and the packaging will be opened in a dedicated humidity controlled room. Crystals will only be handled by experienced personnel wearing latex gloves. Each crystal will then have its dimensions checked using a Coordinate Measuring Machine (CMM). This facility performs, through a touch probe, 20 measurements/face on the long faces and 16 measurements/face on the square faces. Single measurement have a precision of few μm. Several parameters will be evaluated for each crystal to obtain the highest amount of information, such as: (i) the range in which the two transversal dimensions vary along the axis of the crystal; (ii) the flatness for all planes; (iii) the parallelism between opposite planes and (iv) the perpendicularity between intersecting planes. An example of the dimensional check on the flatness and on the transversal dimension of a crystals are reported in Fig. 8.1 and Fig. 8.2, respectively.

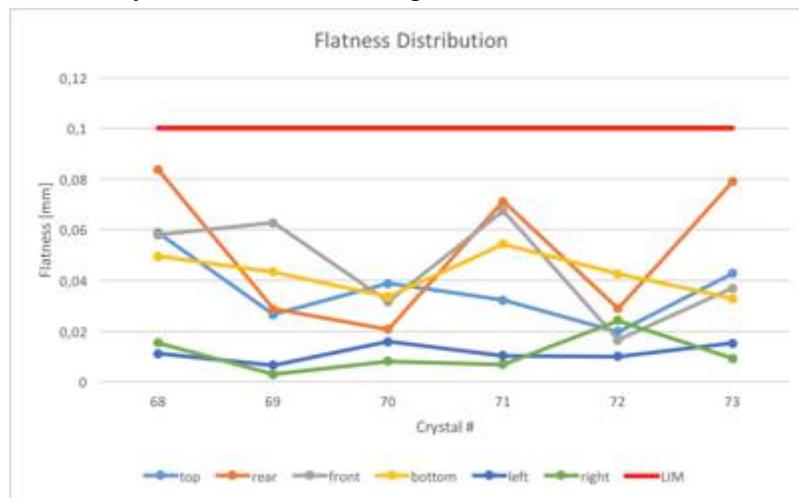

Figure 8.1 *Flatness distribution for few of the pre-production crystals.*

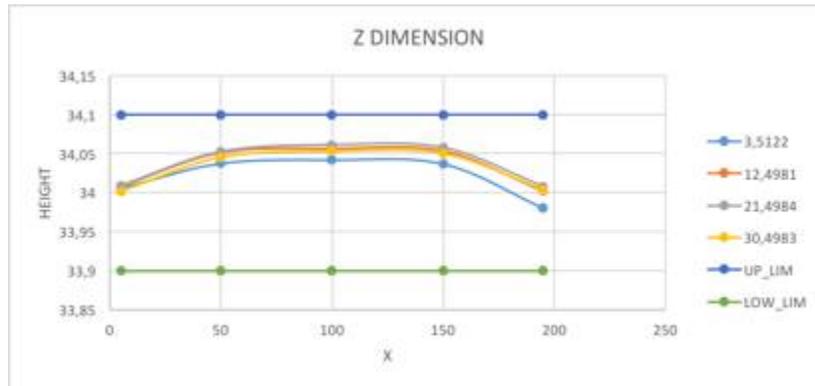

Figure 8.2: *Measurement of the transversal dimension in different longitudinal positions for a pre-production crystal.*

### Measurement of Light Yields and Longitudinal Response Uniformity

The light yield (LY) and Longitudinal Response Uniformity (LRU) tests were first set up using LYSO/BaF$_2$ crystals and then optimized for un-doped CsI [8-3]. The CsI crystals were wrapped in 150 μm thick Tyvek foils and readout in air with a 2" UV-extended PMT, such as ET-9813 QB or similar from Hamamatsu. Our acceptance requirements are:

- a light yield above 100 p.e./MeV;
- an energy resolution (FWHM) less than 45% at 511 keV;
- an LRU, defined as the RMS of the LY measured in 8 or more points along the longitudinal axis, less than 10%;
- a ratio between the fast (200 ns) and the total (3μs) light yield components, R$_{FT}$, above 75%.

All measurements are carried out in a temperature and humidity controlled environment. Both the LNF and Caltech stations use a collimated $^{22}$Na source that illuminates the crystals over a region of a few mm$^2$. Each station is equipped to detect the two 511 keV annihilation γ's produced by this source; one photon is tagged by means of a small monitor system consisting of a LYSO crystal (3x3x10) mm$^3$ readout by a (3x3) mm$^2$ MPPC, while the second γ is used for calibrating the crystals. The tag and test signals are both acquired by means of a CAEN digitizer system running at 1 Gsps. A large effort has been made to make these QA stations user-friendly. In the LNF case, all data taking is done by changing the position of the crystals while running a DAQ program at 500 Hz that allows collection of 20,000 events in 40 sec. A ROOT macro reconstructs, analyzes and fits the data in less than 30 sec.  All analysis and fitting are done automatically. A typical reconstructed 511 keV photon peak for the PMT readout is shown in Fig. 8.3. The LY measured at different longitudinal positions for six crystals is reported in Fig. 8.4.

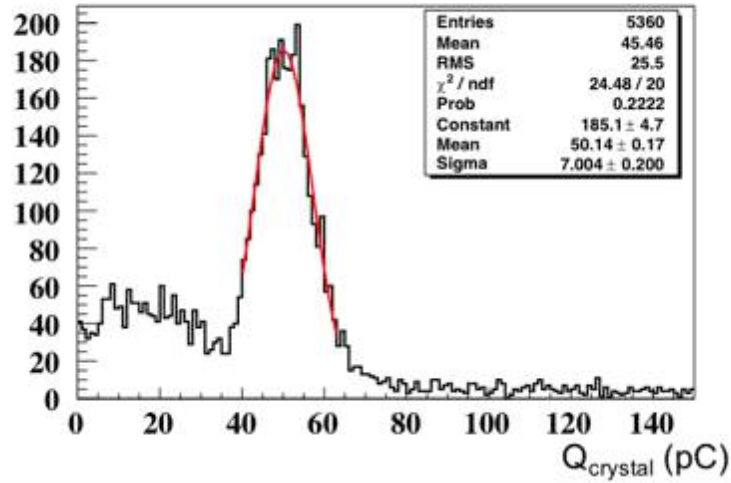

Figure 8.3 : *Response of a CsI crystal to a 511 keV photon from a $^{22}$Na source for the LNF QA station.*

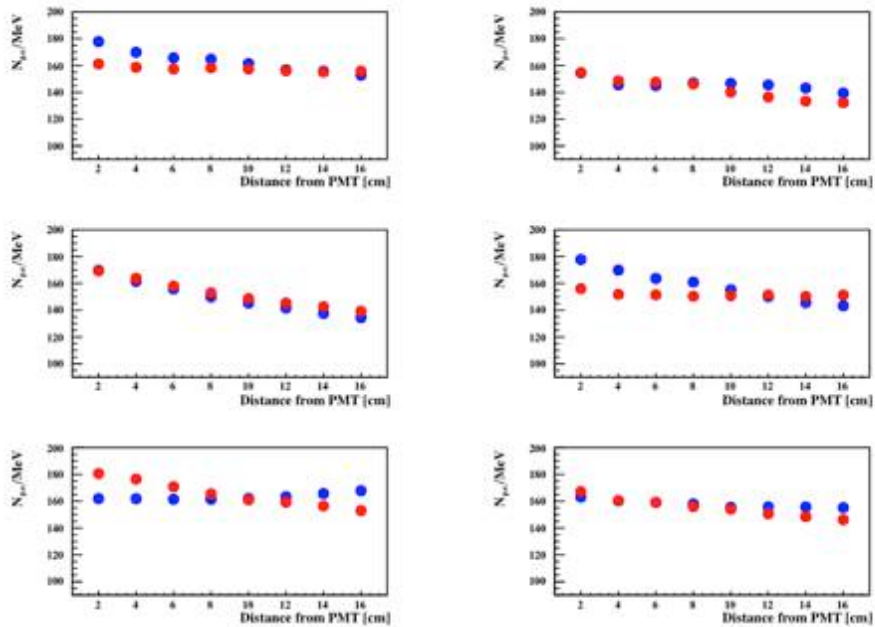

Figure 8.4: *Light Yield for six SICCAS CsI crystals, measured at the LNF station. The measurement is repeated with the source at different positions along the crystal, with a 2 cm step. The crystal response is more uniform when connecting the PMT at one of the two ends (red and blue dots).*

***Acceptance Procedure***

Having described the technique, the acceptance procedures are summarized below.

**1**. *QC at the site of production*

A quality control of the crystals will be carried out from the vendor at the production site before they are shipped to Mu2e. A reference crystal, measured both at Caltech and LNF QA stations, will be provided to the vendor along with the set of specifications that each crystal must satisfy. We expect all parameters of LY, LRU and Rft to be provided from the producers for comparison and optimization purposes. We expect a mechanical test similar to the one performed to our CMM to be done by the producers although with a smaller number of points.

**2.** *QA upon receipt by Mu2e*

Mu2e will perform the QA tests specified above for each delivered crystal. Additional tests will also be performed for radiation hardness as summarized in point.3. ***Final crystal acceptance will be based on the tests made by Mu2e.***

**3.** *Radiation Hardness testing*

A random sample of few % of the production crystals will be tested for radiation hardness. We expect either one small sample from each ingot or at least to select few crystals for production batch. We are exercising this on the pre-production to define the most useful numbers. Gamma irradiation will be performed using a high-intensity $^{137}$Cs source at Caltech where a motorized source mount is already available. Neutron irradiation can be performed at FNG in ENEA, Frascati, Italy and at HZDR, Dresden, Germany. All crystals will be instead tested at a strong $^{137}$Cs, providing a dose of O (rad/h) to estimate the parameter of the radiation induced noise, RIN. As explained in sect. 3.6, our acceptable limit will be to get a RIN < 600 keV.

***Crystal traveler and DB***

A simple excel spread-sheet has been developed for the pre-production to take care of recording all information regarding  dates and location of shipping, crystals barcodes, worker id and the operation/survey done. All the information from QC and QA will then be stored on the official Mu2e Production DB [8-4]. All crystal DB information have been defined (see Fig. 8.5) apart the ones related to the dimensional control that are being finalized at the moment of writing. Scripts for insertion of this information in DB have been prepared and are been exercised. A user-friendly web interface exists to access the stored information in read-only access.

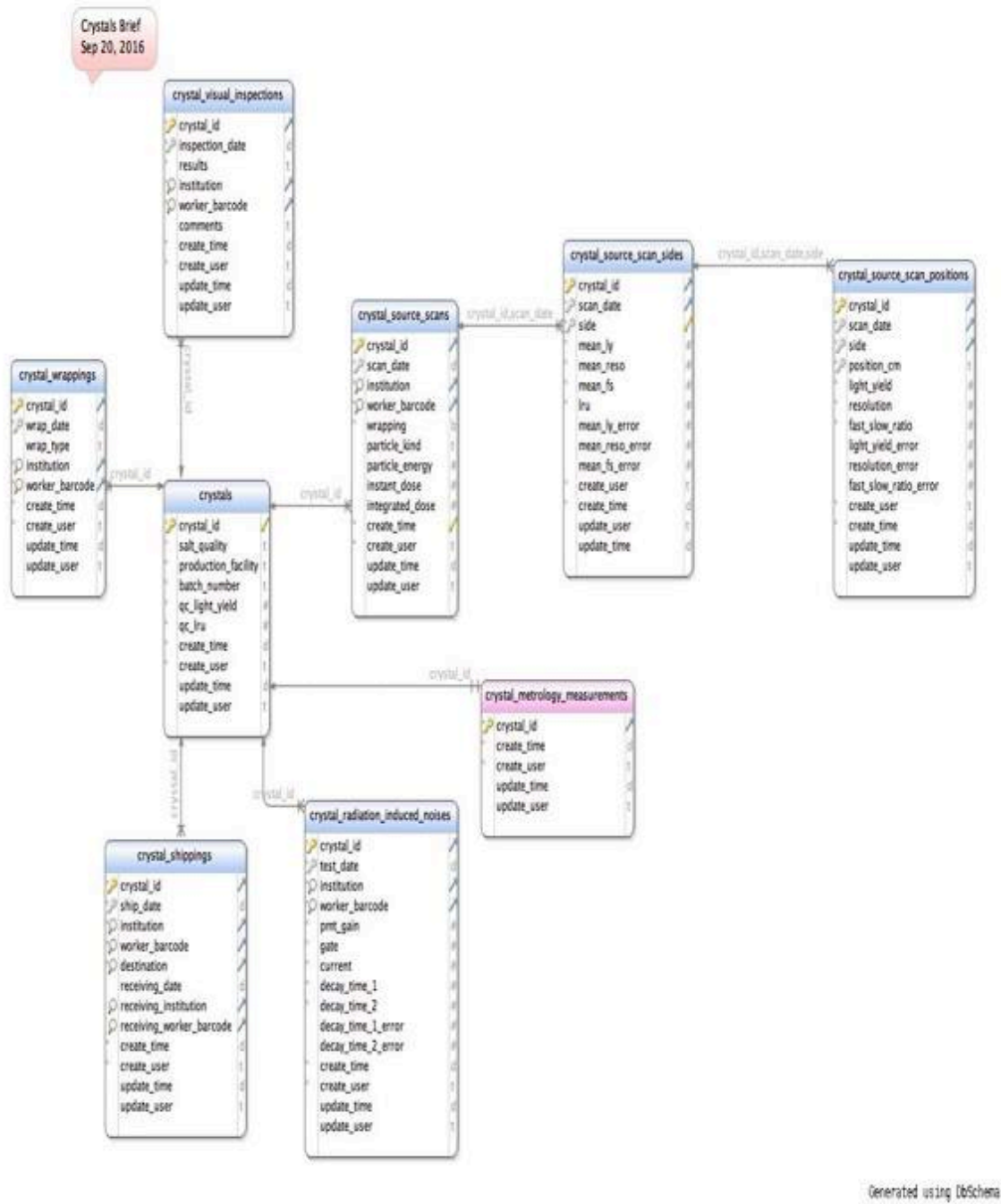

Figure 8.5: *structure of the Hardware DB for the Mu2e crystals*

### 8.3.1    Quality Assurance on SiPMs

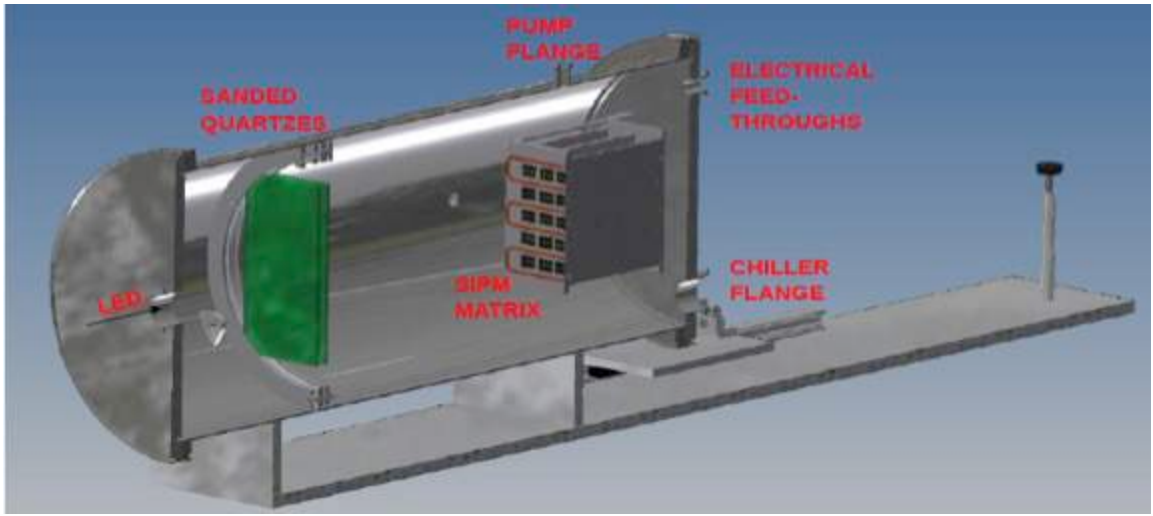

Figure 8.6: *Drawings of the automatized test station for SiPMs.*

In this sub-section, we describe the QA we will apply to the SiPMs. To prevent ambiguity, the whole Mu2e SiPM array is from now on referred to as 'photosensor' or 'sensor', while the single SiPM in the array as 'cell'. As mentioned before, the calorimeter requirements have been translated in a series of technical specification on the sensor.  For accepting a sensor we require [8-5]:

1. A PDE > 20% at 315 nm for each cell of the sensor;
2. A gain greater than $10^6$ for each cell of the sensor;
3. A recovery time $\tau$ < 100 ns on a load greater than 15 $\Omega$;
4. A spread on Vop between the sensor cells < 0.5%;
5. A spread on Idark between the sensor cells < 15%.

The quality control procedure of each sensor is performed in a temperature controlled station that should provide for each cell and for the two series separately: (i) the measurement of the I-V dark curve from which the breakdown voltage is extracted and (ii) the measurement of the Gain at the operation voltage. Moreover, the station should provide, for each cell, also the measurement of the relative PDE with respect to a calibrated reference sensor while illuminating both with a 315 nm light from a pulsed LED. Sensors that do not meet the requirements 1-5 described above will be discarded.

For the SiPM production, an automatized station able to make this QA for 25 SiPMs at the time and at three different running temperatures (namely 0, 10 and 20 °C) will be used. In Fig. 8.6, the schematic drawing of this station is shown. In a vacuum chamber, whose vacuum level is controlled via a vacuum pump, the light of one UV or  green LED is diffused by a couple of sanded quartzes to uniformly illuminate a board with 25 photosensors arranged in a 5x5 matrix. The sensors on the corners and the one on the center are calibrated and kept as reference to monitor the LED intensity.  Each single cell of each sensor can be independently tested thanks to a system of relays controlled by a microcontroller connected to the PC. There is also the possibility to test each series of three cells within a sensor. The temperature is regulated by a chiller  and a cooling coil and is monitored by a set of temperature sensors. The Voltage setting and the current readout are performed by a Keithley picoamperometer connected to a PC. The basic capabilities of this apparatus have been proved with a smaller semi-manual station used only for controlling the pre-production of SiPMs. The test is now underway and involves 105 photosensors from the pre-production, 35 from each of the three vendors: Hamamatsu, SensL and Advansid. At the moment of writing, all the Hamamatsu and SensL sensors have been qualified, while the Advansid are still under test.

**Description of the QA station for pre-production**

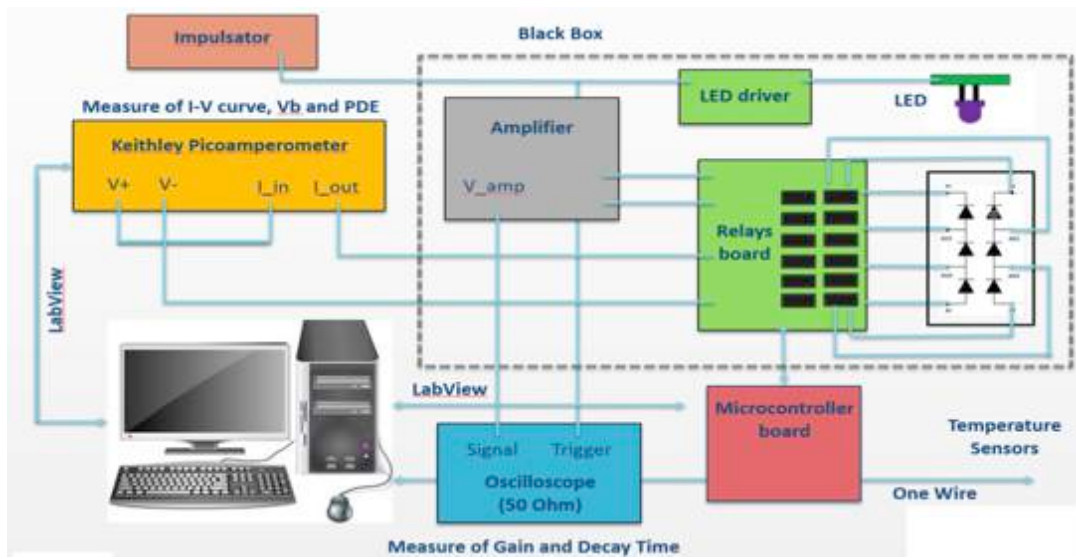

Figure 8.7: *Schematic view of the pre-production QA test station.*

In view of the large number of measurement to perform, we have developed a custom semi-automatized measurement system controlled by computer. The schematic view of the apparatus is shown in Figure 8.6. The Sensor under Test (SUT) is biased by a Keithley picoamperometer, which also measures the pulled current. The basic idea is to select the configuration (a cell or a series) to be biased by means of a custom relays board (Fig. 8.7) photocoupled to a microcontroller. A cascade of Mar-8 amplifiers, with a total gain of 250, can be either excluded or inserted in the circuit to acquire the few photons signal needed for the gain determination by a photon counting method. Via a serial interface, it is possible to control by LABVIEW both the Keithley and the microcontroller to set the bias voltage, select the relays configuration and to program and store the current measurements.

The relays board and the amplifier are located inside a copper box, which acts as a Faraday cage. The ceramic package of the SUT is in thermal contact with the external side of the copper box, while its pins pass through the box in order to be plugged to a custom PC board fixed on the other side, as shown in Figure 8.8. The reference photosensor (RPS) is mounted in the same way and is located closely to the SUT. An UV LED emitting at 315 nm is placed at one end of a metal bar bolted to the copper box and centered such as the LED can illuminate, with similar intensity, both the RPS and the SUT. The temperature is kept stable at 20 ºC, directly on the photosensor package, by using a chiller to refrigerate the copper box. In this way, also the LED is kept cooled. The thermal stability of the system is continuously monitored by a one-wire digital thermometer system (based on the DS18S20) with an accuracy of 0.3 ºC.

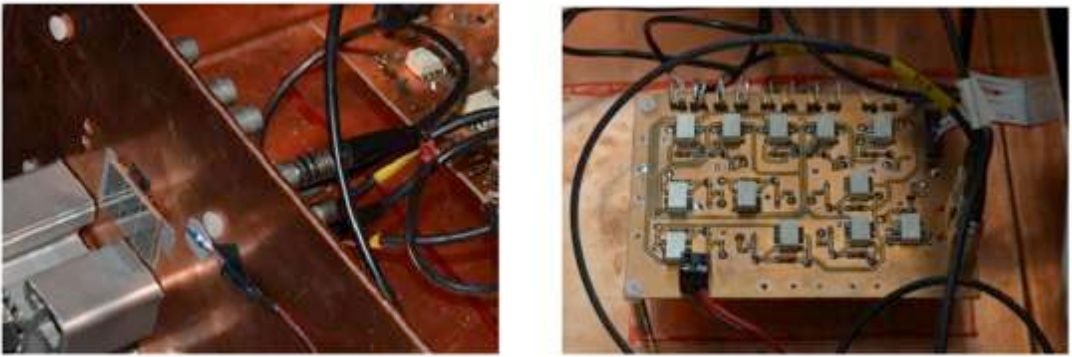

Figure 8.8: *(Left) Picture of a SiPM mounted on the side of the copper box. The reference sensor was not installed. (Right) Relays board controlled by the microcontroller.*

**Determination of the Operative Voltage**

With the LED switched off, a LABVIEW macro drives the Keithley to acquire the I-V curve in an interval starting 1 V below the breakdown voltage, Vbr(firm), as quoted by the producer, up to 4 V above. Our breakdown voltage estimate, Vbr(mu2e), is obtained by fitting the peak position of d(log(I))/dV with a lognormal function. An example of this method is shown in Figure 8.9. The operative voltage is defined as V_br + 3 V.

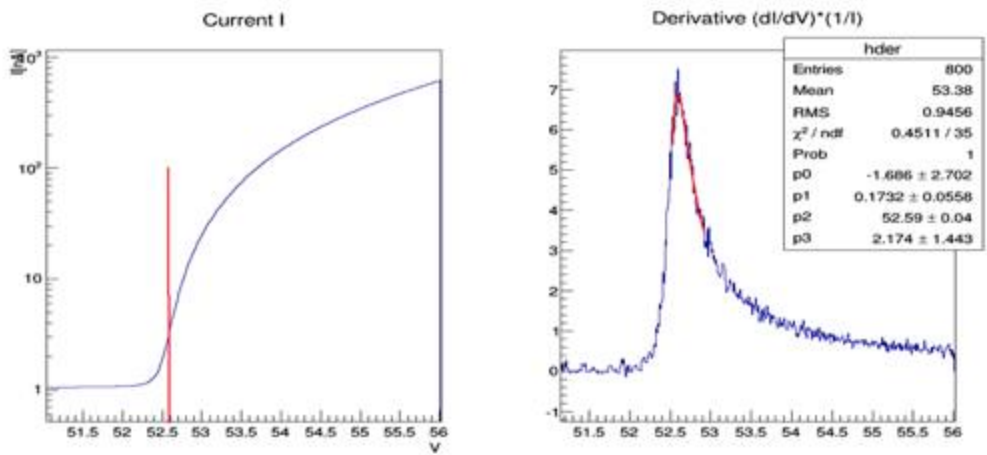

Figure 8.9: *Example of determination of the breakdown voltage.*

**Determination of the Gain at Vop**

The LED is powered by 20 ns wide pulses at a frequency of 100 kHz. The pulse amplitude is finely tuned to let only few photons hitting the photosensor. The SUT is biased at Vop and its signal, amplified by a factor 250, is acquired by a digital scope (on a 50 Ohm resistance) triggering on the light pulse. The charge is obtained by integrating the first 100 ns of the signal and correcting for the 50 Ohm impedance. An example of the resulting charge distribution is reported in Figure 8.10. Each peak corresponds to 0, 1, 2 .. n photons hitting the sensor. The gain is therefore obtained from G = ΔQ/e·Ga, where Δ Q is the charge difference between the 0-photon and the 1-photon peaks, e the fundamental charge of the electron and Ga the amplifier gain.

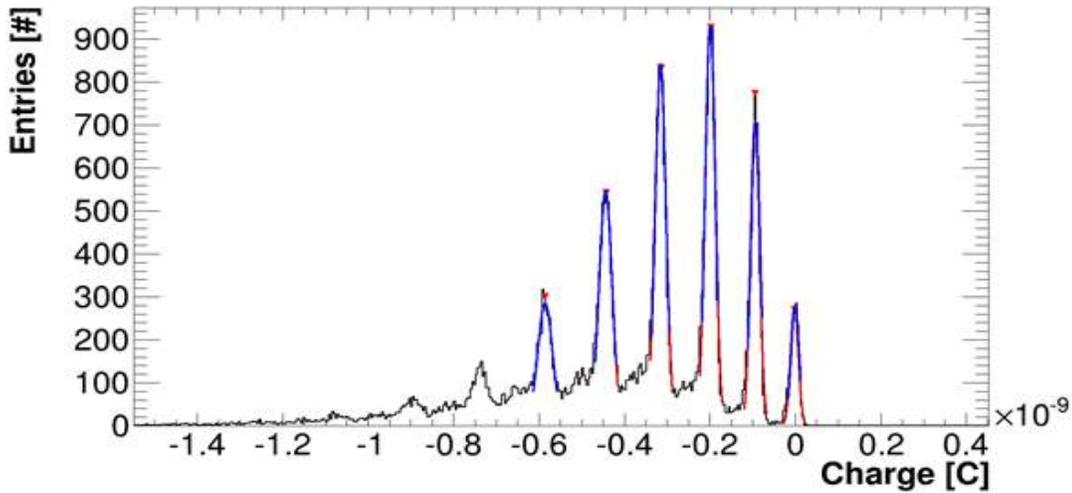

Figure 8.10: *Example of charge distribution for a SiPM.*

**Determination of the Dark Current and the Relative PDE at Vop**

The LED is turned off and the SUT is biased at the operative voltage. The picoamperometer acquires one hundred samples of the dark current and the final measurement is the average. The led is then powered as in the case of the Gain by 20 ns pulses with a 100 kHz frequency, but the pulse amplitude is tuned to have about 10 μA of current pulled from the SUT. The current measurement is then repeated for the SUT and for the reference sensor. Since the photon flux is the same on both the sensors, knowing the gains and the dark currents, the relative PDE is obtained as:

$$PDE_{rel} = \frac{A_{ref}}{A_{sut}} \cdot \frac{G_{ref}}{G_{sut}} \cdot \frac{I_{sut} - I_{sut}^d}{I_{ref} - I_{ref}^d}$$

where  A is the active area of the respective sensor.

**Determination of the recovery time**

The LED is powered by a 20 ns wide pulse at 1 kHz frequency with enough energy to clearly see the pulse at the oscilloscope on a 50 Ohm load without amplifier. The recovery time is defined as the time needed to go from 90% to 10% of the signal peak. The measured value is then rescaled to a 15 Ohm load multiplying by 3/10.

*Acceptance criteria*

As in the case with the crystals, interaction with the photosensor vendors will be important in order to obtain devices that satisfy the Mu2e requirements. Vendors will be expected to provide test data but final acceptance will depend on tests performed by Mu2e. There will be one measurement station in Italy and one measurement station at Caltech to split the characterization of the sensors. A random sample of ~5% of the sensors will be irradiated with neutrons to determine radiation hardness.

### 8.3.3 Other QA activities

The preamplifiers and MB boards will be validated using standard bench test measurements of amplification and noise. A burn-in test of the board will also be performed.

## Chapter 8 References

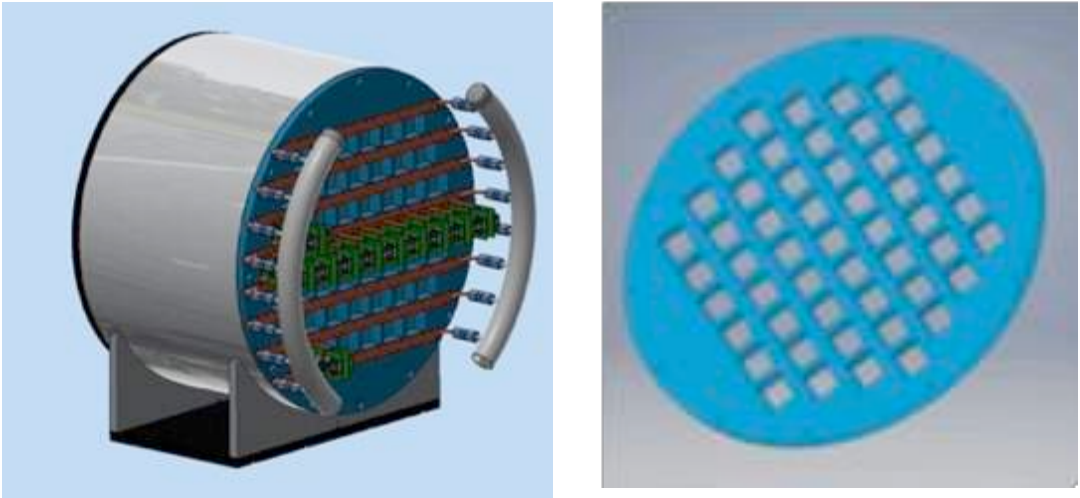

Figure 9.1 *CAD drawings of the Module-0: (left) 3-D view, (right) FEE/cooling back plate .*

# 9 Status of the module-0 and slice tests

Since our calorimeter is not modular by construction, we have defined a structure of staggered crystals, dubbed module-0, with a size large enough to test most of the calorimeter characteristics and resolutions also when performed with electrons beams impinging at 45-50° angles, i.e. similar to that of the CE in the experiment. In its "undressed" version, the module-0 is an array of 51 crystals of final size and shape, connected to 2 Mu2e SiPMs each one inserted in the latest version of SiPM and FEE holders. The sensor holders are attached to a back plate, that apart from its dimension, is similar to the final FEE/back plate thus allowing to test the cooling system and the thermal simulation done with Ansys. In module-0, we expect three different steps for the slice-tests: (step-0)  the FEE version-0 for the AMP-HV chips will be mounted on the back of the SiPM and readout by 5 prototype mezzanine boards in NIM format and 5 WDs with a first version of the input section; the WD will have also a gigabit output for standalone readout; (step-1) the readout will be based on a version 0 of MBs and WDs in final format. The readout will be done via optical fibers to a PC server running a proto-TDAQ. All tests will be done at room temperature. The slice test will be carried out with CR, electron beam and laser running at low rates; (step-2) the pre-production version, v1, of MBs and VDs will be used. A test with the laser system at high rate will also be performed on single MB+WD boards and then for reading the whole module-0.

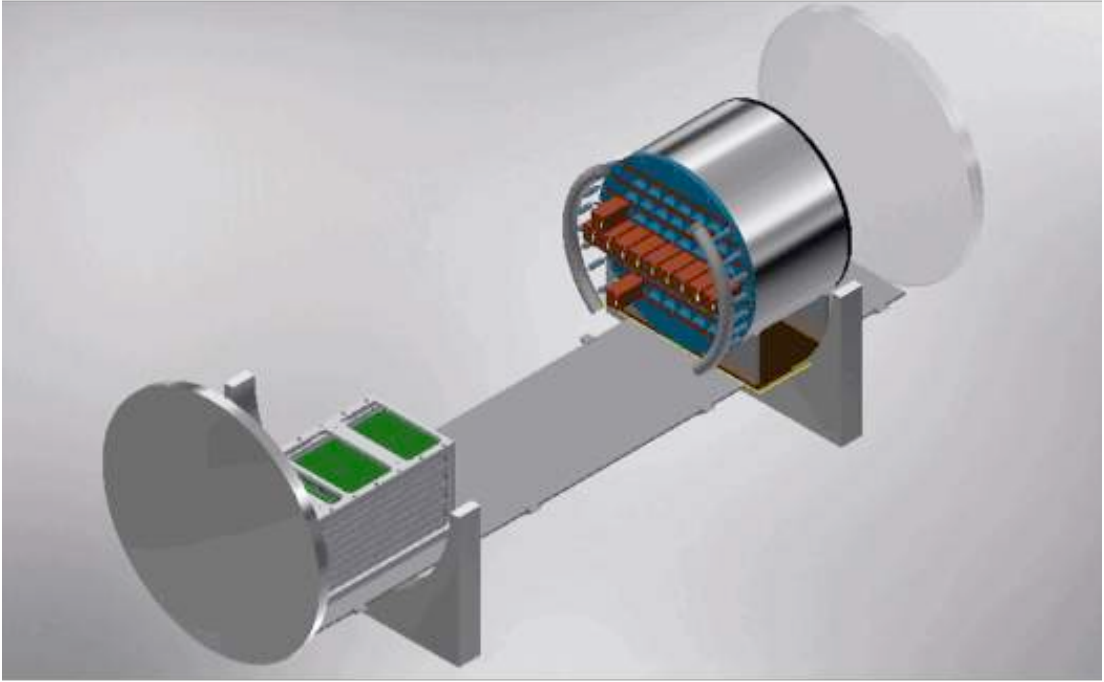

Figure 9.2: *Schematic drawing of the module-0 inserted in a dewar for cooling test.*

The crystals and SiPMs used for the module-0 assembly will be selected from the pre-production samples already in our hands. Mechanical construction is under way for all pieces needed. We expect to start assembly on January 2017. A first test beam with electrons is planned for end of February 2017 with running at room temperature and at a beam frequency of 50 Hz.

Module-0 might result to be very useful also for three other relevant tests: (a) cooling test. after being inserted inside a Dewar (see Fig. 9.2), it can be cool down to 0 °C to check the full functionality of the cooling system; (b) a test of the grounding and shielding scheme as well as the selected feed-throughs  and (c) a radiation hardness test. This last test can be done by exposing it, for a long time (few months) to a neutron flux at a rate 5-6 times larger than expected in the experiment. This will allow to make a complete survey of the full readout chain for a radiation fluence equivalent to at least one year of running.

# 10  Installation and commissioning

After testing the individual components, the two disks will be assembled in a clean room in a controlled temperature and humidity environment. The disks will arrive in the crystal assembly area with the WD crates, the back disk for SiPM and FEE and the cooling lines already mounted and tested for leaks. The assembly will be done with the disks in a vertical position. Special tooling will keep them safely in stable equilibrium.

The insertion of crystals into a disk will start from the front face preparing consecutive rows from the bottom; as soon as each row is assembled, a control of planarity will be performed together with the insertion of special shims to lock the rows of crystals. This will proceed in a stable way until the height of half of the disk will be reached. The inner ring will be then placed and connected to the back disk. Soon after, special filling sectors will be inserted around the inner ring to continue the crystal assembly in the top part. As soon as the last crystal is placed and an optical survey of the crystals' position is done, the already assembled SiPM+FEE holders will be mounted. We will start from the bottom rows up to form "dodecagonal" like sectors. As soon as a given sector will be instrumented, the cabling from FEE chips to MB/WD crates will be carried out. A light-tight box will then cover the disk to allow a test of the completed sector by means of a three-day cosmic ray run using a standalone DAQ system. This allows a first complete debugging of the overall system. A production rate of 160 crystals/month will roughly correspond to the completion of two dodecagonal sector/month. The whole operation would conservatively require 1.5-2 years, assuming a half-year long learning curve.

When the assembly and the cabling of FEE is completed we will conclude the local operation on the disks by closing their ends with light-tight covers on the front face. On each front face the source system and a set of radfets and temperature monitors are mounted. On the fully cabled back disk, we will connect the fiber bundles to the distribution spheres of the laser calibration system and route each fiber to its own crystal. After this operation, the calorimeter is basically ready, needing only to receive the main service cables, the input quartz fibers for the laser calibration system and the fiber optics for the DAQ readout. The laser system will allow a fast debugging of the whole chain, a check of the final cabling and a calibration of the timing offsets for each channel.

Once the two disks are ready, the transportation from the assembly area to the Mu2e building will take place. The insertion on the rails will be carried out one disk at a time, with the FEE mechanics included. Once the two disks are over the rails, the support structure holding each single disk in a vertical position will be taken apart and the two disks joined. The final connection of services and debugging will take place as the final step before a system test with cosmic rays. A 3-4 month cosmic ray run will be carried out with all detector systems to exercise the final DAQ system and confront any system issues.

# ACKNOWLEDGMENTS

This work could not be possible without the help provided by the whole Mu2e group. Collaboration with the simulation and software group, with our friends of the tracking and with the Muon beam line had been excellent. We can not stop thanking G. Ginther for the infinite list of corrections and desiderata/good wishes: we are trying to work on all of them and this is pushing us to get a running detector on feet. We have got a lot of suggestions and directions by our management group lead by R. Ray and J. Whitmore: surely they asked us to pass through so many reviews that we could not do anything if not improve or go back to the starting point. Our spokepersons D. Glenzisnki and J. Miller, were always very keen in following our job and letting us proceed and improve. Finally, our home departments in Caltech, Dubna, LNF, Pisa and Lecce have been extremely useful in helping us on detector tests, engineering and prototype developments. The authors express their sincere thanks to the operating staff of the Beam Test Facility in Frascati (Italy), for providing us a good quality electron beam, and the technical staff of the participating institutions. This work was supported by the US Department of Energy; the Italian Istituto Nazionale di Fisica Nucleare; the US National Science Foundation; the Ministry of Education and Science of the Russian Federation; the Thousand Talents Plan of China; the Helmholtz Association of Germany; and the EU Horizon 2020 Research and Innovation Programunder the Marie Sklodowska-Curie Grant Agreement No.690385.